\documentclass[12pt]{article}
\usepackage{amsmath, amsthm} 
\usepackage{multirow} 
\usepackage{amsfonts}
\usepackage{amssymb,graphics,psfrag}
\usepackage{array,epsfig,multirow,stmaryrd,graphicx}
\usepackage{comment}
\def\hybrid{\topmargin -20pt    \oddsidemargin 0pt
        \headheight 0pt \headsep 0pt
        \textwidth 6.25in       
        \textheight 9.5in       
        \marginparwidth .875in
        \parskip 5pt plus 1pt   \jot = 1.5ex}
\hybrid
\numberwithin{equation}{section}
\numberwithin{table}{section}

\newcommand{\beq}{\begin{equation}}
\newcommand{\eeq}{\end{equation}}
\newcommand{\be}{\begin{equation}}
\newcommand{\ee}{\end{equation}}
\newcommand{\bea}{\begin{eqnarray}}
\newcommand{\eea}{\end{eqnarray}}   
\newcommand{\ben}{\begin{eqnarray*}}
\newcommand{\een}{\end{eqnarray*}}                  
\newcommand{\ba}{\begin{aligned}}
\newcommand{\ea}{\end{aligned}}
\newcommand{\bt}{\begin{tabular}}
\newcommand{\et}{\end{tabular}}
\newcommand{\bc}{\begin{center}}
\newcommand{\ec}{\end{center}}

\newcommand{\cO}{\mathcal{O}}

\newcommand{\cE}{\mathcal{E}}
\newcommand{\cP}{\mathcal{P}}

\newcommand{\cL}{\mathcal{L}}

\newcommand{\cN}{\mathcal{N}}

\newcommand{\cF}{\mathcal{F}}

\newcommand{\cM}{\mathcal M}

\newcommand{\I}{\text{Im}}
\newcommand{\R}{\text{Re}}

\newcommand{\ds}{\displaystyle}

\newcommand{\bi}{{\bar \imath}}

\newcommand{\dd}{d}

\newcommand{\bbZ}{\mathbb{Z}}
\newcommand{\bbR}{\mathbb{R}}
\newcommand{\bbC}{\mathbb{C}}

\newcommand{\half}{\frac12}

\newcommand{\nn}{\nonumber}

\newcommand{\cref}{{\bf [check ref]}}

\newcommand{\rprop}{\oblong}

\newcommand{\CF}{{\cal F}}

\newcommand{\CL}{{\cal L}}
\newcommand{\CM}{{\cal M}}
\newcommand{\CN}{{\cal N}}
\newcommand{\CO}{{\cal O}}
\newcommand{\CP}{{\cal P}}

\def\IZ{{\mathbb Z}}
\def\IR{{\mathbb R}}
\def\IC{{\mathbb C}}
\def\IP{{\mathbb P}}
\def\IT{{\mathbb T}}
\def\IS{{\mathbb S}}
\newcommand{\re}{{\rm e}}
\newcommand{\ri}{{\rm i}}
\newcommand{\rd}{{\rm d}}
\newcommand{\Li}{{\rm Li}}
\psfrag{n1}{$\nu_1$}
\psfrag{n2}{$\nu_2$}
\psfrag{n1'}{$\nu_1'$}
\psfrag{n2'}{$\nu_2'$}
\psfrag{n9}{$\nu_9$}
\psfrag{n10'}{$\nu_{10}'$}
\psfrag{t1}{$\tilde{\nu}_1$}
\psfrag{t2}{$\tilde{\nu}_2$}
\psfrag{t9}{$\tilde{\nu}_9$}
\psfrag{t1'}{$\tilde{\nu}_1'$}
\psfrag{t2'}{$\tilde{\nu}_2'$}
\psfrag{t10'}{$\tilde{\nu}_{10}'$}
\newtheorem{lem14.1}{Lemma}
\newtheorem{cor14.2}[lem14.1]{Corollary}

\newcommand{\tD}{t_D}
\newcommand{\mf}{\Psi}
\newcommand{\coeff}{c}

\newdimen\tableauside\tableauside=1.0ex
\newdimen\tableaurule\tableaurule=0.4pt
\newdimen\tableaustep
\def\phantomhrule#1{\hbox{\vbox to0pt{\hrule height\tableaurule width#1\vss}}}
\def\phantomvrule#1{\vbox{\hbox to0pt{\vrule width\tableaurule height#1\hss}}}
\def\sqr{\vbox{%
  \phantomhrule\tableaustep
  \hbox{\phantomvrule\tableaustep\kern\tableaustep\phantomvrule\tableaustep}%
  \hbox{\vbox{\phantomhrule\tableauside}\kern-\tableaurule}}}
\def\squares#1{\hbox{\count0=#1\noindent\loop\sqr
  \advance\count0 by-1 \ifnum\count0>0\repeat}}
\def\tableau#1{\vcenter{\offinterlineskip
  \tableaustep=\tableauside\advance\tableaustep by-\tableaurule
  \kern\normallineskip\hbox
    {\kern\normallineskip\vbox
      {\gettableau#1 0 }%
     \kern\normallineskip\kern\tableaurule}%
  \kern\normallineskip\kern\tableaurule}}
\def\gettableau#1{\ifnum#1=0\let\next=\null\else
\squares{#1}\let\next=\gettableau\fi\next}

\newcommand{\figref}[1]{Fig.~\protect\ref{#1}}

\def\blfootnote{\xdef\@thefnmark{}\@footnotetext} 
\long\def\symbolfootnote[#1]#2{\begingroup%
\def\thefootnote{\fnsymbol{footnote}}\footnote[#1]{#2}\endgroup}

\begin{document}

\begin{titlepage}

\hfill\vbox{
\hbox{CERN-PH-TH/2007-039}
\hbox{MAD-TH-07-04}
}

\vspace*{ 2cm}

\centerline{\Large \bf  Direct Integration of the Topological String} 

\medskip

\vspace*{4.0ex}

\centerline{\large \rm
Thomas W.~Grimm$^a$,\ Albrecht Klemm$^a$,\ Marcos Mari\~no$^b$ and Marlene Weiss$^{b,c}$\symbolfootnote[0]{\tt \begin{tabular}{ll}$^a$ &grimm@physics.wisc.edu,\ aklemm@physics.wisc.edu\\[.1cm] 
                     $^b$ &marcos@mail.cern.ch,\  marlene.weiss@cern.ch\end{tabular}}}

\vspace*{4.0ex}
\begin{center}
{\em $^a$\ \ Department of Physics, University of Wisconsin, \\[.1cm]
        Madison, WI 53706, USA}

\vspace*{1.8ex}

{\em $^b$ Department of Physics, CERN\\[.1cm]
Geneva 23, CH-1211 Switzerland}

\vspace*{1.8ex}

{\em $^c$ Institut f\"ur Theoretische Physik, ETH H\"onggerberg\\[.1cm]
CH-8093 Z\"urich, Switzerland}

\vskip 0.5cm
\end{center}

\centerline{\bf Abstract}
\medskip
We present a new method to solve the holomorphic anomaly equations governing the free energies of type B topological 
strings. The method is based on direct integration with respect to the non--holomorphic dependence of the amplitudes, 
and relies on the interplay between non--holomorphicity and modularity properties 
of the topological string amplitudes. We develop a formalism valid for any Calabi--Yau manifold and we study 
in detail two examples, providing closed expressions for the amplitudes at low genus, as well as a discussion of 
the boundary conditions that fix the holomorphic ambiguity. The first example is the non-compact Calabi--Yau underlying 
Seiberg--Witten theory and its gravitational corrections. The second example is the Enriques Calabi--Yau, which we 
solve in full generality up to genus six. We discuss various aspects of this model: we obtain a new method to generate 
holomorphic automorphic forms on the Enriques moduli space, we write down a new product formula for the fiber amplitudes 
at all genus, and we analyze in detail the field theory limit. This allows us to uncover the modularity properties of 
$SU(2)$, $\CN=2$ super Yang--Mills theory with four massless hypermultiplets.

\vskip 1cm

\noindent February 2007
\end{titlepage}

\tableofcontents

\section{Introduction}
Topological string theory has played an important role in the quest for a 
better understanding of both physical and mathematical aspects of string theory. 
There are two different topological string theories related to each other by mirror symmetry, 
and known as the A and B-model. 
They are obtained from an $\CN=2$ superconformal field theory, twisted in two distinct 
ways to become type A or type B topological sigma models that are then coupled to gravity.  
The physical relevance of these theories lies in their intimate connection to type II superstring theory. 
In particular, the topological string on a given Calabi-Yau manifold computes higher derivative F-terms 
in the 4d effective action of the corresponding type II theory. 
{}From a mathematical point of view, the topological string partition function provides a generating functional for Gromov-Witten 
invariants in enumerative geometry. 

It is therefore desirable to solve the topological string on a given Calabi-Yau 
manifold, that is to say, to compute all the topological amplitudes $F^{(g)}$ in the genus expansion of the partition 
function. While this problem is completely solved for the case of non-compact toric Calabi-Yau manifolds thanks 
to the techniques of localization and the topological vertex, it remains a challenge for the compact case. 
One of the main tools in solving topological string theory, which also applies to compact Calabi-Yau manifolds, 
is the holomorphic anomaly equations for the B-model found in \cite{bcov}. In this work we present a new approach
to solving these equations. We make use of the fact that for each Calabi-Yau manifold there exists a target space 
symmetry group which provides a symmetry of the topological partition function \cite{abk} and thereby drastically
reduces the space of candidate solutions. The topological string amplitudes 
$F^{(g)}$ turn out to be polynomials in a finite set of generators which transform 
in a particularly simple way under the space-time symmetry group. 
Moreover, it can be shown that all non-holomorphic dependence in these amplitudes arises 
through a very special set of generators that are suitable generalizations of the 
non-holomorphic Eisenstein function $E_2(\tau, \bar \tau)$. The remaining generators are holomorphic. 
Keeping track of these non-holomorphic contributions we will be able to directly integrate the 
holomorphic anomaly equations. This method turns out to be very efficient and gives 
us  rich new information about the remaining holomorphic generators. A similar approach to the 
holomorphic anomaly equations was sketched in \cite{bcov}, in the analysis of toroidal orbifolds. 
For the quintic Calabi-Yau manifold a more complicated method was outlined in~\cite{Yamaguchi:2004bt}.
Other related approaches have been used before in \cite{hosono, hosonorev} to analyze rational elliptic 
surfaces, and in \cite{mnw,mnvw} to study noncritical strings and $\CN=4$ super Yang--Mills theory.

The direct integration of the holomorphic anomaly equations can be performed for 
a generic Calabi-Yau manifold, as we will show in the final section of this 
work. However, in order to fully exploit the interplay of the holomorphic 
anomaly with the space-time symmetry, we will intensively discuss specific examples. 
To illustrate the general ideas we first study the local Calabi-Yau manifold 
associated to the Seiberg-Witten curve. Here the target-space symmetry group is 
a subgroup of $Sl(2,\bbZ)$ and the generating modular functions are well-known. 

Applying these methods to a compact Calabi-Yau manifold is far more 
involved. In the main part of the paper we will focus on the specific example of the Enriques Calabi-Yau \cite{fhsv}, 
arguably the simplest 
Calabi-Yau compactification with nontrivial topological string amplitudes \cite{km,mp}. 
This manifold can be obtained as the free quotient $({\rm K3}\times \mathbb{T}^2)/\bbZ_2$, where $\bbZ_2$ acts as 
the Enriques involution on the K3 fibers. The  
target space duality group of the Enriques Calabi-Yau is shown to be the discrete group $Sl(2,\bbZ)\times O(10,2,\bbZ)$,
with the factors corresponding to the $\mathbb{T}^2$ base and Enriques fiber, respectively. The generating modular forms 
for $Sl(2,\bbZ)$ are well-known, therefore 
we will be particularly concerned with the contributions from the Enriques fiber
and specially their mixing with the $\mathbb{T}^2$ base. 

After integrating the holomorphic anomaly equations the only problem remaining is to fix  
the holomorphic ambiguities, i.e.~the boundary conditions in the integration of the 
equations. These ambiguities are constrained by information coming from boundaries of the moduli space where the $F^{(g)}$ are known 
explicitly. In the Enriques case one can use the fiber limit, where all amplitudes can be determined by heterotic-type II duality \cite{km}, 
and a field theory limit where the manifold degenerates to give rise to $SU(2)$, $N_f=4$ Seiberg-Witten theory. By making use
of these boundary conditions we determine the full topological string amplitudes up to genus $6$, improving in this 
way previous results in \cite{km}. 
As a bonus of our analysis, we clarify the modularity properties of the 
conformal $N_f=4$ theory and its gravitational corrections described in \cite{Nek}. At present the available boundary conditions 
are not enough to completely solve topological string theory on the Enriques Calabi--Yau, but 
we provide efficient tools to 
analyze the amplitudes at all genus with the method of direct integration.

The organization of this paper is as follows. In section \ref{anomaly} we review the derivation 
of the holomorphic anomaly equations. Section \ref{Seiberg-Witten} gives a first simple example 
of the method of direct integration and the fixing of holomorphic ambiguities by application to 
Seiberg-Witten theory. Section \ref{sec:Enriques} reviews what will be our main focus, the Enriques Calabi-Yau. 
We introduce modular and automorphic forms which will be relevant later and discuss the topological amplitudes
on the Enriques fiber. Also an all-genus product formula for the fiber partition function will 
be introduced. 
Section \ref{sec:diE} constitutes the core of this work. We show explicitly how one can solve for $F^{(g)}$
 up to genus six and present the general recursive formalism. Furthermore, boundary conditions and a 
 reduced Enriques model where part of the moduli space is blown down are investigated. 
 In section \ref{sec:ftlim} we analyze the field theory limit corresponding to $N_f=4$ SYM and we relate it in 
detail to the Enriques 
 Calabi--Yau. In section \ref{generic} 
 we present a formalism for direct integration on generic Calabi-Yau manifolds. Section \ref{conclusion} contains 
 conclusions and an outlook on further directions of investigation. Appendix \ref{N=2sp} reviews some special geometry. 
 Appendix \ref{theta} collects some useful formulae for theta functions and modular forms. 
 Appendix \ref{heteroticFg} reviews the heterotic computation of the amplitudes 
in \cite{mm,km} and presents improved formulae for their antiholomorphic dependence. Finally, appendix \ref{Cal_big_Fg} presents the holomorphic anomaly equations on the so-called big moduli space.


\section{The holomorphic anomaly equations \label{anomaly}}

In this section we will briefly recall some basics about topological 
string theory to set the stage for the following sections 
and to fix our conventions. This will force us to introduce some world-sheet notations and techniques.
However, for the rest of this work we will mostly need only the explicit form of the 
holomorphic anomaly equations. For a more detailed 
introduction to topological string theory the reader might want to consult references \cite{mirror,vonk,nv,mbook, klemm}.

Type II string theory on Calabi-Yau threefold $Y$ yields a  
superconformal field theory  with left and right moving $(2,2)$
supersymmetry on the world-sheet. This structure admits two
topological string theories: the A--and the B--model. 
The key quantity in these topological theories is their all genus partition function 
\beq \label{part-function}
    Z=\text{exp}\ \sum^\infty_{g=0} g_s^{2g -2} F^{(g)} \ .
\eeq
This formal expansion in the string coupling $g_s$  contains the 
topological string amplitudes $F^{(g)}$ for maps from genus $g$ Riemann surfaces 
into a target Calabi-Yau manifold. 
The topological string amplitudes of the A-- and B--model are identified by mirror symmetry, which 
maps one theory on $Y$ to its dual on the corresponding
mirror Calabi-Yau. 

We will now briefly recall the B--model definitions of the free energies 
$F^{(g)}$. The B--model describes constant maps from a world-sheet 
Riemann surface $\Sigma_g$ to points in the Calabi-Yau space $Y$. Therefore,
the B--model definition of the $F^{(g)}$ involves only the integration over the
moduli space ${\overline{\cM}_g}$ of the world-sheet  
and not over the moduli space of maps.
More precisely, let us denote by  $({\underline m},{\overline {\underline m}})$ 
coordinates on ${\overline{\cM}_g}$ and abbreviate the correlators of the world-sheet CFT by 
$\big<\cdot\big>_g$. The free energies $F^{(g)}$ are then defined by
\begin{equation}
          F^{(g)}=\langle 1\rangle_g=\int_{\overline{\cM}_g}\ \langle \prod_{k=1}^{3g-3} \beta^k {\bar \beta}^k \rangle_g \ \,
          [\dd {\underline m} \wedge \dd {\overline {\underline m}}]\ .
\end{equation} 
Here we inserted the operators $\beta^k=\int_{\Sigma_g} G^- \mu^k$ and their complex conjugates to 
obtain the correct measure on the moduli space. $\beta^k$ and ${\bar \beta}^k$ contain the 
the world-sheet Beltrami differentials
$\mu^k \in H^1(T\Sigma_g)$ and the world-sheet supersymmetry generators $G^-,\bar G^-$.
The contraction of $[\dd {\underline m} \wedge \dd {\overline {\underline m}}]$ with 
the $(\beta^k,\bar \beta^k)$ factor is
antisymmetric due to the presence of $G^-,\bar G^-$ and  
yields a top form on the complex $3g-3$ dimensional moduli space ${\overline{\cM}_g}$. 
The fact that one has to integrate only over the moduli space of the world-sheet
makes the B-model far simpler to solve than the A-model. Therefore, it is often easier to use the B-model 
and the mirror map to determine A-model quantities.

{}From the point of view of the four-dimensional effective action, one is interested in the dependence 
of the $F^{(g)}$ on the
complex moduli  $t^i,\bar t^i$ in the vector multiplets. These parametrize marginal 
deformations, which in the B-model correspond to complex structure deformation of the Calabi-Yau manifold.
Infinitesimally  the world-sheet action is  
perturbed by the $t^i,\bar t^i$ as follows 
\begin{equation}   
S=S_0+ 
 t^i\ \int_{\Sigma_g} \cO_i^{(2)}+
 {\bar t}^{i}\ \int_{\Sigma_g} 
\bar \cO_{i}^{(2)}\ ,
\end{equation}     
where the sums run over $i=1,\ldots,h^1(Y,TY)=h^{(2,1)}(Y)$.
Here the marginal two-form operators are obtained using the descent 
equations as
\begin{equation} 
\label{descent}
{\cal O}_i^{(2)}= \{G_0^-,[\bar G_0^-,{\cal O}_i^{(0)}]\} \rd z \rd {\bar z}\ , \qquad  
\overline {{\cal O}^{(2)}_{\bar \imath}}=\{G_0^+,[\bar G_0^+,\bar {\cal O}^{(0)}_{\bar \imath}]\}
\rd z \rd {\bar z}\ ,
\end{equation}
where $G_0^+,G_0^-$ are the zero modes of the twisted world-sheet supersymmetries $G^+,G^-$.
In these equations we denoted by ${\cal O}_i^{(0)}$ the zero-form cohomological operators, which are in one-to-one 
correspondence with the $H^1(Y,TY)$ cohomology of the target space. 

{}From the point of view of the target space Calabi-Yau
the complex fields $t^i,\bar t^i$ provide a set of local coordinates on the moduli 
space of complex structure deformations $\cM$. This space is shown to 
be a special K\"ahler manifold with K\"ahler potential 
\beq
   K(t,\bar t)=-\log i\int_Y \Omega(t) \wedge \bar \Omega(\bar t)\ ,
\label{kaehlerpotential}
\eeq
where $\Omega(t)$ is the holomorphic three-form on $Y$ varying holomorphically with a 
change of the complex structure. $\Omega(t)$ is only unique up to rescalings by a 
holomorphic function and hence should be viewed as a section of the line bundle $\cL$ over 
the moduli space $\cM$. 
In appendix \ref{N=2sp} we review how the special geometry of $\cM$ can 
be entirely encoded by a single holomorphic section of ${\cal L}^{2}$, 
the prepotential $F^{(0)}=\cF(t)$. {}From a world-sheet point of view one 
does not obtain $F^{(0)}$ directly, but rather finds the three-point function
\beq
   C^{(0)}_{ijk} =\langle \cO_{i}^{(0)}\cO_{j}^{(0)}\cO_{k}^{(0)} \rangle_g=- \int_Y \Omega(t) \wedge \partial_i \partial_j \partial_k \Omega(t)\ ,
\eeq
where $\partial_i$ are derivatives with respect to $t^i$.

At higher genus a more involved world-sheet analysis can be applied to investigate the 
properties of the higher $F^{(g)}$. It turns out that the  higher genus topological 
string amplitudes $F^{(g)}$ are not holomorphic, but rather fulfill  specific 
holomorphic anomaly equations. These equations are recursive in the genus  and 
determine the anti-holomorphic derivative of $F^{(g)}$.  Therefore, even if the genus zero 
data are given they determine $F^{(g)}$ only up to a holomorphic ambiguity. 
We will now briefly state the essential features and results of the work of
Bershadsky, Cecotti, Ooguri and Vafa \cite{bcov}, who have shown that 

\noindent
{\sl i}.) The $F^{(g)}$ transform as section of ${\cal L}^{2-2g}$ with the connection (\ref{cov_D}).

\noindent
{\sl ii}.) The topological B-model correlation functions 
\beq 
\label{C_prop}
  C^{(g)}_{i_1 \ldots i_n}=
\left\{ \begin{array}{ll}
\langle \int_{\Sigma_g}  \cO_{i_1}^{(2)} \cdots  \int_{\Sigma_g}  \cO_{i_n}^{(2)}\rangle_g= 
D_{i_1}\ldots D_{i_n} F^{(g)} &\ \text{for}\ \ g\ge1\\ [3 mm]
\langle \cO_{i_1}^{(0)}\cO_{i_2}^{(0)}\cO_{i_3}^{(0)}  \int_{\Sigma_g} 
\cO_{i_{4}}^{(2)} \cdots  \int_{\Sigma_g}  \cO_{i_n}^{(2)}\rangle_g=
D_{i_{4}}\ldots D_{i_n} C^{(0)}_{i_1 i_2 i_3} &\ \text{for}\ \ g=0 \end{array}\right.
\eeq
can be obtained using the covariant derivatives \eqref{cov_D} and obey
\beq \label{C_prop_cond}
   C^{(g)}_{i_1 \ldots i_n} = 0 \quad \text{for} \ \ 2g-2 + n \le 0 \ .
\eeq

\noindent
{\sl iii}.) The anti-holomorphic derivative  $\partial_{\bar \imath}=
\frac{\partial}{\partial {\bar t^{i}}}$ of the $F^{(g)}$,
\begin{equation} 
\bar \partial_{\bar \imath} F^{(g)}=
\int_{\overline{\cM}_g} \bar \partial_{\bar \imath} \mu_g=
\int_{\overline{\cM}_g} \partial_m \bar \partial_{\bar m} \lambda_{\bar \imath , g}=
\int_{\partial \overline{\cM}_g} \lambda_{\bar \imath, g},
\end{equation}  
receives only contributions from the complex codimension one locus in the moduli space of Riemann surfaces corresponding
 to world-sheets which are degenerate with lower genus components.  These boundary 
contributions can be worked out and yield recursive equations for the $F^{(g)}$. For $g>1$ one gets
\beq \label{rec_Fg}
   \bar \partial_{\bar \imath} F^{(g)} = \tfrac{1}{2} \bar C^{(0)jk}_{\bar \imath} 
   \Big(D_j D_k F^{(g-1)} + \sum_{r=1}^{g-1}D_j F^{(r)} D_k F^{(g-r)} \Big)  
\eeq 
and for $g=1$ a generalisation of the Quillen anomaly
\beq \label{anomaly_F1}
   \partial_{i} \bar \partial_{\bar \jmath} F^{(1)} = \tfrac{1}{2}     C^{(0)}_{i k l } \bar C^{(0)kl}_{\bar \jmath}  - 
\Big(\frac{\chi}{24} -1 \Big)G_{i\bar \jmath}\ .
\eeq
Here we defined 
\be
\bar C^{(0)kl}_{\bar \jmath}= e^{2K} G^{k\bar k} 
G^{l \bar l} \bar C^{(0)}_{\bar \jmath \bar k \bar l}\ ,
\eeq
where $G_{k \bar k}=\partial_k \bar \partial_{\bar k} K$ is the Weil-Petersson metric of the K\"ahler potential (\ref{kaehlerpotential}).   

These are the recursive holomorphic anomaly equations, which we want to integrate directly  
in this paper. Note that there is no holomorphic anomaly at genus zero. 
$C^{(0)}_{ijk}$ has no world-sheet moduli dependence, 
hence no boundaries, and is therefore holomorphic. 
The genus zero data thus have to be provided from the outset. 
They can be determined from the 
period integrals of the manifold  $Y$.    

It is further shown in ref.~\cite{bcov} that (\ref{rec_Fg}) can be integrated recursively. 
With an iterative procedure of complexity growing exponentially with the genus, 
one rewrites (\ref{rec_Fg}) as 
\begin{equation}
\label{feynman1}
{ \partial}_{\bar k} F^{(g)}(t,\bar t)=\bar \partial_{\bar k} 
\Gamma^{(g)}( \hat \Delta^{ij}, \hat \Delta^{i}, \hat \Delta, C^{(r<g)}_{i_1 \ldots i_n}) \ ,
\end{equation}
and integrates it to 
\begin{equation}
\label{Fgwithf}
F^{(g)}(t, \bar t)= \Gamma^{(g)}( \hat \Delta^{ij}, \hat \Delta^{i}, \hat \Delta, C^{(r<g)}_{i_1 \ldots i_n})+f^{(g)}(t)\ . 
\end{equation}
Here $\Gamma^{(g)}$ is a functional of some propagators $\hat \Delta^{ij}, \hat \Delta^{i}, \hat \Delta$ 
and the lower genus vertices $C^{(r)}_{i_1\ldots i_n}$ with $r<g$.
The holomorphic ambiguity $f^{(g)}(t)$  arises as an integration constant.
To prove that the functional $\Gamma^{(g)}$ exists at every genus,~\cite{bcov} show that it is 
the disconnected Feynman graph expansion of an auxiliary action with the above vertices and propagators, 
whose partition function fulfills a master equation equivalent to (\ref{rec_Fg}) and (\ref{anomaly_F1}).
The propagators can be defined using the genus zero data as follows. Since
\begin{equation} 
\bar D^{\phantom{0}}_{\bar \imath} \bar C^{(0)}_{\bar \jmath \bar k \bar l}=\bar D^{\phantom{0}}_{\bar \jmath} \bar C^{(0)}_{\bar 
\imath \bar k \bar l}
\end{equation}
one can integrate 
\begin{equation}
\bar C^{(0)}_{\bar \jmath \bar k \bar l}=-\tfrac12 e^{-2 K} \bar D_{\bar \imath}\bar D_{\bar \jmath}
\bar \partial_{\bar k} \hat \Delta
\end{equation}
as
\begin{equation}\label{def-small-Delta}
G_{\bi j} \hat \Delta^{j} = \tfrac12 \bar \partial_\bi \hat \Delta\ ,\qquad \quad G_{\bi k}\hat  \Delta^{kj}= 
\bar \partial_{\bi} \hat \Delta^j\ ,\qquad \quad \bar C^{(0) jk}_{\bar \imath}=  \bar \partial_{\bi} \hat \Delta^{jk}\ .
\end{equation} 
Note that the propagators are defined by these equations only up to holomorphic 
ambiguities arising in the integration steps. Fixing these ambiguities directly affects 
the definition of the holomorphic functions $f^{(g)}(t)$ in \eqref{Fgwithf}. 
It turns out that a preferred choice for this ambiguity 
is provided by relating the propagators in a canonical way to $F^{(1)}(t,\bar t)$ \cite{abk}.

The combinatorics of the Feynman graph expansion are useful to establish 
some general properties of the $F^{(g)}$, but its complexity 
grows exponentially with the genus.  However, the $F^{(g)}$ are 
invariant under space-time modular transformations which are 
a symmetry of the full string compactification. As we will discuss 
later, they generically admit a split into a universal factor times a modular 
form. Here the weights of the modular  
forms grow linearly with the genus. Since the ring of modular forms 
is finitely generated, the complexity of modular invariant expressions
grows only polynomially with the genus. The method of direct 
integration that we develop in this paper uses this connection 
with modular forms such that its complexity also 
grows only polynomially with the genus. It has the advantage that 
the modular properties of the amplitudes are manifest 
in all steps of the derivation.


\section{Solving Seiberg-Witten theory by direct integration \label{Seiberg-Witten}}

Local Calabi-Yau geometries provide simple and instructive examples for the interplay
between holomorphicity and modular invariance in topological string theory. 
In this section we will explain the key features using the simplest example, namely 
the local Calabi-Yau corresponding to $SU(2)$ Seiberg-Witten theory with no matter \cite{sw1}. 
In section \ref{SWgeometry} we first recall the geometry of Seiberg-Witten theory.
We show that all genus zero data can be expressed in terms of a finite set of holomorphic 
modular forms. All higher amplitudes $F^{(g)}$ are invariant under the modular 
group. In section \ref{SWintegration}
we directly integrate the holomorphic anomaly equations, determining 
all $F^{(g)}$ up to a holomorphic modular ambiguity. 
Modularity restricts this ambiguity so much 
that simple boundary conditions set by the effective action near 
special points in the moduli space allow one to reconstruct all $F^{(g)}$. We review 
such a convenient  set of boundary conditions in section \ref{SWboundary}.
The general philosophy presented in this section will be later applied to
the more complicated case of compact Calabi-Yau manifolds.

\subsection{The Seiberg-Witten geometry \label{SWgeometry}}

Seiberg-Witten theory  with no matter \cite{sw1} can be obtained in the A--model as a limit of the local 
Calabi-Yau geometry 
${\cal O}(-2,-2)\rightarrow \mathbb{P}^1\times \mathbb{P}^1$ \cite{kkv}. The mirror 
B--model geometry of this limit is the Seiberg-Witten elliptic curve ${\cal E}$
\begin{equation} 
y^2=(x-u)(x-\Lambda^2)(x+\Lambda^2)\ ,
\end{equation}
whose modular group is $\Gamma(2)$. This subgroup of $ Sl(2,\bbZ)$ 
acts on the period integrals
\beq \label{def-ttD}
  t=\int_{a} \lambda\ ,\qquad \qquad t_D=\int_{b} \lambda\ ,  
\eeq
where $\lambda=\frac{\sqrt{2}}{2 \pi }\frac{y}{x^2-1} {\rm d}x$
is the Seiberg-Witten meromorphic differential. In the limit described above, $\lambda$ is 
obtained as a reduction of the holomorphic $(3,0)$ form of the Calabi-Yau manifold. 
Rigid special geometry guarantees 
the existence of a prepotential $F^{(0)}={\cal F}(t)$ with the properties 
\begin{equation} 
t_D=\frac{\partial {\cal F}}{\partial t}\ ,\qquad \qquad 
\tau = -\frac{1}{4 \pi} \frac{\partial^2 {\cal F}}{\partial^2 t} \ .
\end{equation}  
These conditions are obtained as the rigid limit of the
special geometry relations presented in Appendix \ref{N=2sp}. 
Note that $\tau$ is precisely the complex structure parameter 
of the torus and hence parametrizes the 
upper half-plane. In particular, $\I \tau >0$ 
is guaranteed by the Riemann inequality consistent with the fact that 
$\I \tau$ is the gauge kinetic coupling function of Seiberg-Witten theory.
Moreover, a modular transformation acts on $\tau$ as
\beq \label{modulartrans}
   \tau \ \mapsto \ \frac{a\tau + b}{c\tau + d}\ .
\eeq
The genus zero data are functions of $\tau$ and transform 
in a particularly simple way under \eqref{modulartrans}. They 
can be expressed in terms of a finite set of modular generators,
which we will specify in the following.  

A modular function $f(\tau)$ of weight $m$ is defined 
to transform as $f(\tau) \, \mapsto \, (c \tau+d)^m f(\tau)$ under 
\eqref{modulartrans}. Focusing on the modular group of the Seiberg-Witten 
curve, we note that 
the ring of modular functions of $\Gamma(2)$ can be expressed as 
powers of the Jacobi $\theta$-functions. Relevant properties of  the 
Jacobian $\theta$-functions are summarized in Appendix B. We introduce 
two generators 
\begin{equation} 
K_2=\vartheta_3^4+\vartheta_4^4, \qquad  \qquad  K_4=\vartheta_2^8\ , 
\label{gamma2generators} 
\end{equation} 
which are of modular weight two and four respectively. 
The modular transformation properties follow
from (\ref{thetatransformation}).
$K_2, K_4$ generate the graded ring of  holomorphic modular forms 
${\cal M}_*(\Gamma(2))$ of $\Gamma(2)$, which we will also denote by 
$\mathbb{C}[K_2, K_4]$. 
It turns out to be useful to also introduce
\beq \label{ring_hE4}
  h=K_2\ ,\qquad \qquad  E_4=\tfrac{1}{4} (K_2^2 + 3 K_4)\ .
\eeq
As we will see when we develop the method of direct integration, it is natural to take 
$h$, $E_4$ as the generators of the ring ${\cal M}_*(\Gamma(2))$.

Let us now express the genus zero data in terms of modular forms.
The connection with the geometry of the Seiberg-Witten curve is given by the following relation 
\begin{equation}
u(\tau)=\frac{K_2}{\sqrt{K_4}}\ . 
\label{ut}
\end{equation}
The combination $z(\tau)=1/u^2(\tau)$ is modular invariant and can be viewed as the analog of the 
mirror map for this non-compact Calabi-Yau manifold.
The analog of the holomorphic triple coupling is 
\begin{equation}
\label{xisw}
 C \equiv C^{(0)}_{ttt}=
\frac{\partial \tau}{\partial t}=\frac{32 K_4^{1/ 4}}{K_2^2-K_4}\ 
\end{equation}
Note that $C^2$ is a form of weight $-6$ 
under the modular transformations in  $\Gamma(2)$.
The modular group $\Gamma(2)$ also
determines the periods $t,t_D$ as weight $1$ objects \footnote{They 
can be calculated likewise using the Picard-Fuchs equation.}
\begin{equation}
t(\tau)={E_2(\tau)+ K_2(\tau)\over 3 K_4^{1/4}(\tau)}, \qquad \quad 
t_D(\tau_D)=-i{ 2 E_2(\tau_D)- K_2 (\tau_D)- 3K_4^{1/2}(\tau_D) \over 3 \big(2 K_2(\tau_D)- 2 K_4^{1/2}(\tau_D) \big)^{1/2}}\ ,
\label{t-td-period}
\end{equation}
where $\tau_D=-\frac{1}{\tau}$ and $E_2$ is the second Eisenstein series 
defined in \eqref{geneis}. It is natural to give the periods in the above parameters. 
In the electric phase of Seiberg-Witten theory the $q=e^{2 \pi i \tau}$ series converges and
$t$ is the physical expansion parameter, while in the magnetic phase the $q_D=e^{2 \pi i \tau_D}$ 
series converges and $t_D$ is the physical expansion parameter. Of course $t_D(\tau)$ and 
$t(\tau_D)$ can be obtained by performing an $S$-duality transformation on $E_2$ and the Jacobi theta 
functions.

\subsection{Direct integration \label{SWintegration}}

Having discussed the genus zero  geometry, let us now turn to the 
higher genus  free energies $F^{(g)}$ and their holomorphic anomaly. Starting with 
$F^{(1)}$, we note that the holomorphic anomaly equation (\ref{anomaly_F1}) specializes to 
\begin{equation} \label{SWF1anomaly}
\partial_t \partial_{\bar t} F^{(1)}=\tfrac{1}{2}  C^{(0)\, tt}_{\bar t} C^{(0)}_{ttt} \ .
\end{equation}
where the indices are raised with the Weil-Petersson metric $G_{t\bar t}=2 \I \tau$.
This equation integrates immediately to
\begin{equation} \label{SWF1withtau}
F^{(1)}= - \tfrac12 \log \,\I \tau - 
     \log |\Phi(\tau)|\ ,
\end{equation}
where $\partial \tau/\partial t$ is evaluated using \eqref{xisw}.
The holomorphic object $\Phi(\tau)$ is the ambiguity at genus one. It is 
determined from modular constraints and the physical requirement that 
$F^{(1)}$ should only be singular at the discriminant of ${\cal E}$. 
Note that under a modular transformation \eqref{modulartrans} one 
finds that $\I \tau\ \mapsto\ {|c \tau+d|^{-2}}\I \tau$.
Together with the invariance of $F^{(1)}$ this implies that $\Phi(\tau)$ must be a 
modular form of weight $1$.  The only modular 
form of weight $1$ which has only poles at the discriminant of ${\cal E}$
is the square of the $\eta$ function given in \eqref{dede}. This fixes the ambiguity at genus one as $\Phi(\tau) = \eta^2 (\tau)$.

At genus one the non-holomorphic dependence was induced through the appearance of
$\I \tau$. As dictated by the holomorphic anomaly equations, 
all higher $F^{(g)}$ also depend on $\bar t$.
We now show that this dependence arises through the propagator $\hat \Delta^{tt}$ only.
$\hat \Delta^{tt}$ is obtained in the local limit of \eqref{def-small-Delta} and thus obeys
\beq \label{proploc}
    {\partial}_{\bar t} \hat \Delta^{tt}= C^{(0)\, tt}_{\bar t} \ .
\eeq
All other propagators vanish in this limit.
To integrate this condition, we first multiply both sides in \eqref{proploc} by 
$C^{(0)}_{ttt}$. The result is easily compared to the holomorphic 
anomaly equation \eqref{SWF1anomaly} of $F^{(1)}$. 
Changing derivatives 
by inserting ${\partial \tau}/{\partial t}=C^{(0)}_{ttt}$
one evaluates with the help of 
 (\ref{Eta-der})
\begin{equation}
\hat \Delta^{tt}=2 \partial_\tau F^{(1)}(\tau,\bar \tau)=-\tfrac{1}{12} \widehat E_2(\tau,\bar \tau)\ ,\qquad \qquad \partial_\tau = (2\pi i)^{-1} \tfrac{ \partial }{\partial \tau}
\label{swpropagator}
\end{equation} 
The occurrence of the non-holomorphic extension 
of the second Eisenstein series $E_2(\tau)$      
\begin{equation}
\widehat E_2(\tau,\bar \tau)= E_2(\tau)-{3 \over \pi  \I \tau} \ .
\label{hatE2}
\end{equation}
is forced by modular invariance. Since $F^{(1)}(\tau,\bar \tau)$ 
is a modular function of weight zero, its derivative must be a 
modular form of weight $2$ which is not holomorphic. 
The only form with these properties is the almost 
holomorphic form $ \widehat E_2(\tau,\bar \tau)$. This form is 
the canonical, almost holomorphic extension of the second Eisenstein series $E_2$, where $E_2$ is the unique holomorphic quasimodular form of weight 2  
transforming as
\begin{equation}
E_2(\tau) \quad \mapsto\quad (c\tau + d)^2 E_2(\tau) - \tfrac{6}{ \pi} i c (c \tau+d)\ 
\label{E2transformation} 
\end{equation} 
under a modular transformation \eqref{modulartrans}.
The shift in the transformation of the anholomorphic piece in (\ref{hatE2}) cancels 
precisely the shift in (\ref{E2transformation}). More generally the ring 
${\hat {\cal M}}_*$ of almost holomorphic forms of $\Gamma(2)$ is generated 
as $\mathbb{C}[\widehat E_2,h,\Delta]$.

Using  the propagator and general properties of the Feynman graph expansion 
one can extract the fact that the higher genus $F^{(g)}$ are weight 0 forms 
with the structure
\begin{equation}
F^{(g)}(\tau, \bar \tau)=C^{2g-2} \sum_{k=0}^{3g-3}
\widehat E_2^{k}(\tau,\bar \tau) c^{(g)}_k(\tau)\ ,\qquad g>1\ ,
\label{eq:generallocalform}
\end{equation}
where we defined $C=C_{ttt}^{(0)}$.
Modular invariance implies then that the holomorphic forms $c^{(g)}_k(\tau)$ are modular 
 of weight $6(g-1)-2k$ in $\mathbb{C}[h,\Delta]$. We will show 
next that all forms $c^{(g)}_k(\tau)$ with $k>0$ are very easily determined by 
direct integration of the holomorphic anomaly equation. The 
form $c^{(g)}_0(\tau)$ is not determined in this way and  corresponds to
a holomorphic modular ambiguity.        

In order to analyze the holomorphic anomaly equations in the local case, it turns out to be 
very useful to discuss some general properties related to modular transformations. Let us 
first discuss how derivatives transform under the modular transformation \eqref{modulartrans}.
Denoting by $f_k$ a modular form of weight $k$ it is elementary 
to check that its derivative transforms under \eqref{modulartrans} as
\begin{equation} \label{deriv_shift}
\partial_{\tau} f_k \quad \mapsto \quad (c\tau+d)^{k+2}\partial_\tau f_k +
\frac{k}{2 \pi i} c(c \tau +d)^{k+1} f_k \ .
\end{equation}
Similarly, we can evaluate $\partial_{t} f_k = C^{-1} \partial_{\tau} f_k$, where as above 
$C=C^{(0)}_{ttt}$. In order to cancel the shift in \eqref{deriv_shift} 
we will now introduce covariant derivatives. There are two possible 
ways to achieve this\footnote{We thank Don Zagier for explaining us several manipulations involved in the following.}. 
Firstly, one can cancel the shift against the shift of $(\I \tau)^{-1}$ and set
\begin{equation}  \label{der_nonhol}
  D_t f_k=\Big(\partial_t-{ k C \over 4 \pi \I \tau}\Big)f_k \ ,\qquad \quad D_\tau f_k=\Big(\partial_\tau-{ k\over 4 \pi \I \tau}\Big)f_k\ .
\end{equation}
Here $D_t$ is the covariant derivative to the Weil-Petersson metric $G_{t\bar t}$ and $D_\tau$ is the so-called Mass derivative.
$D_t$ maps almost holomorphic forms of $\Gamma(2)$ of weight $k$ 
into almost  holomorphic forms of weight $k-1$, while $D_\tau$ increases the weight
from $k$ to $k+2$. Note that both derivatives in \eqref{der_nonhol} are non-holomorphic 
due to the appearance of $\I \tau$.
There is however  a second possibility to cancel the shift \eqref{deriv_shift} which is manifestly
holomorphic. More precisely, one can cancel the shift against the shift \eqref{E2transformation} of $E_2(\tau)$ and
define
\beq \label{der_hol}
  \hat D_t f_k = \big(\partial_t-\tfrac{1}{12}  k C E_2 \big) f_k\ ,\qquad \quad  \hat D_\tau f_k = \big(\partial_\tau-\tfrac{1}{12}  k E_2 \big) f_k\ .
\eeq  
In this case $\hat D_\tau$ is known as the Serre derivative.
Both $\hat D_t$ and $\hat D_\tau$ are holomorphic. They map holomorphic 
modular forms of weight $k$ to holomorphic modular forms of weight $k-1$ and $k+2$ respectively.
It is easy to check that the following identity holds
\beq \label{splitSW}
    D_t f_k = \hat D_t f_k + \tfrac{1}{12}k C \widehat E_2\, f_k\ ,\qquad D_\tau f_k = \hat D_\tau f_k +\tfrac{1}{12}k  
\widehat E_2\, f_k\ .
\eeq
These equations also imply that whenever $f_k$ is holomorphic all the non-holomorphic 
dependence of $D_t f_k$ and $D_\tau f_k$ lies in a term involving the propagator.
In other words, once again all anti-holomorphic dependence arises through the propagator $\widehat E_2$ only.
The generalizations of the modular derivatives \eqref{der_nonhol} and \eqref{der_hol} will 
reappear in later sections of this work. For the Enriques Calabi-Yau they are given in \eqref{cov_D},\eqref{def-KY}
and \eqref{der_holE}, while in the general discussion of compact Calabi-Yau manifolds they appear in 
\eqref{cov_Dgen1},\eqref{cov_Dgen2} and \eqref{Dhol}.

Here we will us the covariant derivatives  \eqref{der_nonhol} and \eqref{der_hol} 
to rewrite the holomorphic anomaly equations  (\ref{rec_Fg}). Firstly,
we will apply modularity and the fact that all non-holomorphic 
dependence arises through the propagator $\widehat E_2(\tau ,\tau)$ to convert anti-holomorphic 
derivatives into derivatives with respect to $\widehat E_2$. Using \eqref{splitSW} we will be able to 
carefully keep track of the $\widehat E_2$ 
dependence in the holomorphic anomaly equations. Eventually, a solution will be simply obtained by
 direct integration of a polynomial in $\widehat E_2$. 
 
To begin with, note that the holomorphic anomaly equations specialize in the local limit to
\beq
  \partial_{\bar t}   F^{(g)} = \tfrac{1}{2} C^{(0)tt}_{\bar t} \Big(D_t \partial_t F^{(g-1)}+
\sum_{r=1}^{g-1} \partial_t F^{(r)} \partial_t F^{(g-r)}\Big)\ .
\eeq
Using the fact that all non-holomorphic dependence arises only through the 
propagator $\widehat E_2(\tau,\bar \tau)$, this equation can be rewritten as 
\beq \label{SWholan2}
\frac{\partial   F^{(g)}}{\partial \widehat E_2}= \tfrac{1}{48} \Big(D_t \partial_t F^{(g-1)}+
\sum_{r=1}^{g-1} \partial_t F^{(r)} \partial_t F^{(g-r)}\Big)\ .
\eeq
Here we used \eqref{proploc} to substitute $C^{(0)tt}_{\bar t} $ with the derivative $\partial_{\bar t} \widehat E_2$, 
which then cancels with the same factor arising on the left-hand side of this equation. 
Let us now manipulate the right-hand side of \eqref{SWholan-final} and split off the derivative 
of $F^{(1)}$ in the second term
\beq \label{SWholan3}
\frac{\partial   F^{(g)}}{\partial \widehat E_2}= \left\{\begin{array}{ll} \tfrac{1}{48}\Big(D_t \partial_t F^{(1)}+  (\partial_t F^{(1)})^2\Big)  &g=2\ ,\\ 
\ds{\tfrac{1}{48} \Big((D_t + 2 \partial_t F^{(1)}) \partial_t F^{(g-1)}+
\sum_{r=2}^{g-2} \partial_t F^{(r)} \partial_t F^{(g-r)}\Big)}\qquad \qquad& g>2 \ ,\end{array}\right.
\eeq
where the sum now runs from $r=2$ to $r=g-2$. One then notes that $\partial_t F^{(1)}$ can be replaced
by $-\frac{1}{24} C\widehat E_2$ by using \eqref{swpropagator}. Furthermore, we replace
the non-holomorphic derivative $D_t$ with its holomorphic 
counterpart $\hat D_t$ via \eqref{splitSW}. Altogether, one evaluates
\beq \label{F2D}
   \frac{\partial   F^{(2)}}{\partial \widehat E_2}= -\tfrac{1}{48 \cdot 24}\Big(\hat D_t (C \widehat E_2) -  \tfrac18 ( C \widehat E_2)^2\Big) 
\eeq
for genus two and for $g>2$
\beq \label{SWholan-final}
\frac{\partial   F^{(g)}}{\partial \widehat E_2}=\tfrac{1}{48} \Big((\hat D_t -\tfrac16 C \widehat E_2) \partial_t F^{(g-1)}+
\sum_{r=2}^{g-2} \partial_t F^{(r)} \partial_t F^{(g-r)}\Big)\ .
\eeq

We are now in the position to make the dependence on $\widehat E_2$ explicit. This can be done
by rewriting the right-hand side of \eqref{SWholan-final} using \eqref{der_hol}. We also define
$\hat d_t$ and $\hat d_\tau$ as covariant derivatives $D_t,\hat{D}_\tau$ not acting on the propagators $\widehat E_2$, such 
that e.g. $
   \hat d_\tau( \hat E^k_2\, c^{(r)}_k)=   \hat E^k_2\,  \hat D_\tau c^{(r)}_k $.
Applying the chain rule we find 
\beq \label{first_D}
   \partial_t F^{(r)}  = \big[\hat d_t + (\hat D_t \widehat E_2) {\partial}_{\widehat E_2} \big] F^{(r)}
                                = C \big[\hat d_\tau - \tfrac{1}{12} (E_4+ \widehat E_2^2) {\partial}_{\widehat E_2} \big] F^{(r)}\ ,
\eeq
where \eqref{hatE2}, \eqref{der_hol}  and  \eqref{E-der} are applied to evaluate the derivative of $E_2$. 
The Eisenstein series $E_4$ arises naturally in rewriting the derivatives. 
We will therefore work with the ring $\bbC[\widehat E_2, h,E_4]$ introduced in \eqref{ring_hE4}.

Similarly, we rewrite the second derivative 
\bea \label{second_D}
   \hat D_t \partial_t F^{(g-1)}
   &=&\tfrac{1}{12^2}C^2\Big(12^2 \hat d_\tau^2+6^2 h \hat d_\tau + 2   E_4(\widehat E_2{\partial}_{\widehat E_2}  +   \widehat E_2^2 {\partial}_{\widehat E_2}^2)\nn \\
     &&     -( 3 h  + 12 \hat d_\tau )\widehat E_2^2 {\partial}_{\widehat E_2}+  2 \widehat E_2^3{\partial}_{\widehat E_2}+  \widehat E_2^4  {\partial}_{\widehat E_2}^2 \\
    && +  (  -9 E_4 h + 2 h^3 - 12  E_4 \hat d_\tau   ) {\partial}_{\widehat E_2} +E_4^2 {\partial}_{\widehat E_2}^2 \Big)F^{(g-1)}\ ,\nn 
\eea
where we have used that the derivative 
of $C$ is given by $ \hat D_\tau C =\tfrac14  h\, C$.
This is how the holomorphic modular form $h$  defined in \eqref{ring_hE4} arises in the direct integration. 

We can now actually perform the direct integration. This is done by inserting the expressions 
\eqref{first_D} and \eqref{second_D} for $\partial_t F^{(r)}$ and $\hat D_t \partial_t F^{(g-1)}$ into the holomorphic 
anomaly equation \eqref{SWholan-final}. Replacing all $F^{(r)}$ for $1<r<g$ with their propagator expansion 
\eqref{eq:generallocalform}, it is then straightforward to keep track of the number 
of propagators $\widehat E_2$ in each term of the right-hand side of \eqref{SWholan-final}. Finally, $F^{(g)}$ is  
determined up to a $\widehat E_2-$independent ambiguity by integrating the resulting polynomial in $\widehat E_2$.
Without much effort this procedure can be repeated iteratively up 
to the desired genus.

Note that the equation \eqref{F2D} for $F^{(2)}$ is particularly simple to integrate. 
Using  \eqref{der_hol}  and  \eqref{E-der}  one evaluates  
\beq
 \hat D_t (C \widehat E_2) -  \tfrac18 ( C \widehat E_2)^2=\tfrac{1}{24}C^2 \big(-5 \widehat E_2^2+ 6 \widehat E_2 h-2 E_4 \big)\ .
\eeq
Inserted into \eqref{SWholan-final} it is straightforward to integrate this quadratic polynomial in $\widehat E_2$ 
to derive $F^{(2)}$ as
\begin{equation}
{F}^{(2)}(\tau,\bar \tau)= \tfrac{1}{2\cdot 24^3} C^2\left(
\tfrac{5}{3}\, \widehat E_2^3 - 3\,h \widehat E_2^2 + 2\,E_4 \widehat E_2\right) +C^2 c^{(2)}_0,
\label{swF2}
\end{equation}
where $c^{(2)}_0(h,E_4)$ is the holomorphic ambiguity which can be fixed
by additional boundary conditions as we discuss in the next section.
For genus up to $7$ the expressions for $F^{(g)}$ were 
calculated in \cite{hk} using the Feynman graph expansion. 
The direct integration using (\ref{SWholan-final}) provides a far more effective 
method to solve Seiberg-Witten theory and confirms the results of \cite{hk}.
Furthermore, the modular properties of the expressions are 
manifest at each step. As we will discuss in the later sections, similar 
constructions will  provide us with a 
powerful tool to determine the set of candidate modular generators
for more complicated Calabi-Yau manifolds. In particular, holomorphic 
modular forms are needed to parametrize the holomorphic ambiguity. 
In case we know the ring of holomorphic modular forms, fixing the ambiguity
reduces to a determination of a finite set of numerical factors 
at each genus. For Seiberg-Witten theory this can be done systematically, 
as we will discuss in the next section.

\subsection{Boundary conditions \label{SWboundary}}

To systematically fix the  $c^{(g)}_0$ we have to understand the boundary behavior
of the $F^{(g)}$. As it is well known, there are three distinguished regions in the moduli space of pure $SU(2)$
$\cN=2$ SYM which correspond to the geometrical 
singularities of ${\cal E}$. We will parametrize the moduli space by the vacuum expectation 
value  $u=\langle {\rm Tr} \Phi^2\rangle$ of the scalar $\Phi$ in the $\cN=2$ vector
multiplet. The first  region occurs at $u\sim \frac{1}{2} t^2 \rightarrow 
\infty$, and it corresponds physically to the semiclassical regime. 
The monopole region occurs near $u\rightarrow \Lambda^2$, where a magnetic monopole of charge $(e,m)=(0,1)$ becomes 
massless and the electric $SU(2)$ theory with gauge coupling ${\rm Im} \tau$ is 
strongly coupled. At the point $u\rightarrow -\Lambda^2$  a dyon of charge $(e,m)=(-1,1)$
becomes massless. However, this point is identified with the monopole point by a $\mathbb{Z}_2$ exact 
quantum symmetry. For this reason there are no independent boundary 
conditions at $u\rightarrow -\Lambda^2$ and we focus on  
$u\rightarrow \Lambda^2$ and $u\sim \infty$. In both cases the 
elliptic curve acquires a node, i.e. a local singularity of 
the form $\xi^2+\eta^2=(u\pm \Lambda^2)$, where a cycle of $\IS^1$ topology shrinks. 
In string theory, a point in the moduli space where a node in the target 
geometry develops is called a conifold point.

The natural physical parameter in  the magnetic monopole region $u\rightarrow \Lambda^2$
is $t_D$.  We get first a convergent expansion for the  $F^{(g)}$ in the 
variable $q_D=\exp(2 \pi i \tau_D)$ for \mbox{$\tau_D=-{1\over \tau}\rightarrow i \infty$}, which corresponds to 
$t_D\rightarrow 0$.  This is obtained by an $S$- transformation of the modular 
expressions for the $F^{(g)}(\tau,\bar \tau)$ such as (\ref{swF2}), which converge 
in the semiclassical region. The holomorphic magnetic 
expansions ${\cal F}^{(g)}_D(\tau_D)$ can be obtained by formally taking the limit 
$\bar \tau_D\rightarrow \infty$, while keeping $\tau_D$ fixed. Finally we obtain 
the expansion in $t_D$ by inverting (\ref{t-td-period}). In these magnetic expansions, a gap structure was observed near the monopole (or conifold) point \cite{hk}. 
One finds that the leading behavior of ${\cal F}^{(g)}_D(\tau_D)$ is of the form
\begin{equation} 
{\cal F}_D^{(g)}={B_{2g}\over 2 g ( 2g -2) \tilde t_D^{2g-2}}+ k^{(g)}_1 \tilde t_D+
{\cal O}(\tilde t_D^2)\ ,
\label{gap}
\end{equation}
where the $B_{n}$ are the Bernoulli numbers and we used a rescaled variable 
$\tilde t_D=i{t_D\over 2}$. The knowledge of the leading coefficients and the absence 
of the remaining $2g-3$ sub-leading negative powers in the $\tilde t_D$ expansion imposes
$2g-2$ conditions. Since ${\dim }M_{6g-3}(\Gamma(2))=\left[\frac{3 g-1}{2}\right]$ 
this overdetermines the $c^{(g)}_0$, e.g.~for $g=2$ we 
find $c^{(2)}_0=- \frac{1}{2\cdot 24^3}\big(\frac{1}{2}E_4\,h + \frac{1}{30}h^3\big)$. 
It is very easy to integrate (\ref{SWholan-final}) using (\ref{first_D}), (\ref{second_D}) and 
the gap condition, which fixes the ambiguity to arbitrary genus. This solves the 
theory completely. One finds moreover a pattern in the first subleading term in the 
magnetic expansion
\begin{equation}
k^{(g)}_1={((2g-3)!!)^3\over g! 2^{7 g-2}}\ .
\end{equation}

The gap can be explained by using the embedding of Seiberg-Witten theory into type IIA string 
theory compactified on a suitable Calabi--Yau manifold. The most generic singularity of a $d$ complex dimensional 
manifold is a node where an $\IS^d$ shrinks. The 
codimension one locus in the moduli space 
where this happens is called the conifold. It was argued 
in~\cite{strominger,Vafa:1995ta} that at the conifold a 
RR-hypermultiplet becomes massless. This hypermultiplet 
is charged and couples to the $U(1)$ vector multiplets. 
Its one loop effect on the kinetic terms 
of the vector multiplets in the effective action 
is captured by the local expansion of $F^{(0)}$~\cite{strominger}. 
A gravitational one-loop effect yields the moduli dependence 
of the $R_+^2$ term in the effective action and is given by  local 
expansion $F^{(1)}$~\cite{Vafa:1995ta}. Using further one-loop arguments
it was shown that the $F^{(g)}$, which capture the moduli dependence 
of the coupling of the self-dual part of the curvature to the 
self-dual part of the graviphoton $R^2_+ \, F_+^{2g-2}$, have the following 
gap structure
\begin{eqnarray} \label{thegap}
F^{(g)}_{\textrm{conifold}}=\frac{(-1)^{g-1}B_{2g}}{2g(2g-2)t_D^{2g-2}}
+\mathcal{O}(t_D^0),
\end{eqnarray}
where $t_D$ is a suitable coordinate transverse to the conifold 
divisor \cite{hkq}. The Seiberg-Witten gauge theory embedded in type IIA string theory inherits this structure, and the massless 
hypermultiplet at the conifold is identified as a monopole becoming massless at the monopole 
point. In this way, (\ref{thegap}) explains the field theory result (\ref{gap}) and extends it to the full 
supergravity action. 

Once the Seiberg-Witten amplitudes $F^{(g)}$ have been determined in terms of modular 
functions, these can be expanded around every point in the moduli space.
For example, in the semiclassical regime
$\tau\rightarrow i \infty$, $u\rightarrow \infty$ one finds the 
holomorphic amplitudes
\begin{equation} 
{\cal F}^{(g)}={(-1)^gB_{2g}\over  g ( 2g -2) (2 t)^{2g-2}}+ 
{l^{(g)}_{2g+6} \over t^{2g+6}}+{\cal O}(t^{2g+10})\ .
\end{equation}
The higher order terms in this expansion correspond to gauge theory 
instantons and have been computed in \cite{Nek}.


\section{A first look at the Enriques Calabi-Yau}\label{sec:Enriques}

In this section we review some basic properties of topological string theory on
the Enriques Calabi-Yau. We begin by reviewing the $N=2$ special 
geometry of the classical moduli space of K\"ahler and complex structure deformations in 
section \ref{specialE}. 
The first world-sheet instanton corrections arise from genus one Riemann surfaces 
as shown in refs.~\cite{fhsv,hmfhsv,km}. The holomorphic  higher genus free energies, restricted to the K3 fiber,
can be also derived by using heterotic-type II duality \cite{km}. We briefly summarize these results
in section \ref{genus1}. In understanding and deriving the expression for the 
full $F^{(g)}$ an important hint is given by their transformation properties 
under the symmetry group of the full topological string theory on the Enriques Calabi-Yau. 
More precisely, 
generalizing the results of the previous section, one expects 
that all $F^{(g)}$ are built out of functions transforming in 
a particularly simple way under the group $Sl(2,\bbZ) \times O(10,2,\bbZ)$.
In section \ref{aut_prop} we review some essentials about these modular 
and automorphic functions and forms.

\subsection{Special geometry of the classical moduli space \label{specialE}}

The Enriques Calabi-Yau can be viewed as the first non-trivial generalization of 
the product space $\IT^2 \times {\rm K3}$. It is defined as the orbifold 
$(\IT^2\times {\rm K3})/\bbZ_2$, where $\bbZ_2$ acts as a free involution \cite{fhsv}. 
This involution inverts the coordinates of the torus and acts as the Enriques 
involution on the K3 surface. 
The cohomology lattice of $\IT^2\times {\rm K3}$ takes the form \cite{Aspinwall:1996mn}
\begin{equation} 
\Gamma^{6,22}=\Gamma^{2,2}\oplus [\Gamma^{1,1}\oplus E_8(-1)]_1
\oplus[\Gamma^{1,1}\oplus E_8(-1)]_2\oplus \Gamma^{1,1}_g \oplus \Gamma^{1,1}_s\ ,
\label{coveringlattice}
\end{equation}
where the inner products on the sublattices $E_8(-1)$ and $\Gamma^{1,1}$ are given by   
\beq
  (C^{\alpha \beta})=-C_{E_8}\ ,\qquad \qquad (C^{ij})= \left(\begin{array}{cc}0 & 1\\ 1 & 0\end{array} \right)\ . 
\eeq
with $\alpha,\beta=1,\ldots,8$ and $i,j=1,2$. 
Here $C_{E_8}$ is the Cartan matrix of the exceptional group $E_8$.
The lattice \eqref{coveringlattice} splits into $H^1(\mathbb{T}^2)\oplus H_1(\mathbb{T}^2)=\Gamma^{2,2}$ 
and $H^*(K3)=\Gamma^{4,20}$. Under heterotic-type II duality it can be 
identified with the Narain lattice of the heterotic compactification on $\mathbb{T}^6$. 
The $\bbZ_2$ involution on the Enriques Calabi-Yau 
acts on the five terms of the lattice \eqref{coveringlattice} as~\cite{fhsv} \footnote{The  effect of the phase factor 
on the type II side was interpreted as turning on a Wilson line~\cite{fhsv}.}
\begin{equation}
|p_1,p_2,p_3,p_4,p_5\rangle\rightarrow  e^{\pi i \delta\cdot p_5} |-p_1,p_3,p_2,-p_4,p_5\rangle\ ,
\end{equation} 
where $p_i$ is an element of the $i$-th term in \eqref{coveringlattice} and we denoted
$\delta=(1,-1)\in \Gamma^{1,1}_s$.

The Enriques Calabi-Yau has holonomy group $SU(2)\times \bbZ_2$. 
This implies that type II string theory compactified on the Enriques Calabi-Yau
will lead to a four-dimensional theory with $\cN=2$ supersymmetry. 
Nevertheless, due to the fact that it does not have the full $SU(3)$ holonomy
of generic Calabi-Yau threefolds, various special properties related to $\cN=4$ 
compactification on $\IT^2 \times K3$ are inherited. 

As an example of the close relation of the Enriques Calabi-Yau to 
its $\cN=4$ counterpart $\IT^2 \times K3$ one notes that the moduli 
space of K\"ahler and complex structure deformations are simply 
cosets. The complex dimensions of these moduli spaces are given by
the dimensions $h^{(1,1)}$ and $h^{(2,1)}$ of the cohomologies 
$H^{(1,1)}$ and $H^{(2,1)}$. They can be determined constructing 
a basis of $H^{(p,q)}$ of forms of K3 and $\IT^2$ invariant 
under the free involution. One obtains \cite{fhsv}
\beq
    h^{(2,1)}=h^{(1,1)}=11\ ,
\eeq
while $H^{(0,0)},\ H^{(3,3)}$ as well as $H^{(3,0)}$ are one-dimensional. 
Moreover, one can show that the Enriques Calabi-Yau is self-mirror and
that both the K\"ahler and complex structure moduli spaces are given by the coset
\be
\label{coset}
 \cM=\frac{Sl(2,\bbR)}{SO(2)} \times \CN_8\ ,
\ee
where 
\be
\label{nscoset}
\CN_s= 
{O(s+2,2) \over O(s+2) \times O(2)}.
\ee
The actual moduli space is obtained after dividing $\CM$ by the discrete groups $Sl(2,\bbZ)\times O(10,2;\IZ)$. 
$\cM$ is a simple example 
of a special K\"ahler manifold. We will discuss its properties in the following. 

It is a well-known fact that the geometric moduli space of a Calabi-Yau manifold 
consists of two special K\"ahler manifolds corresponding to K\"ahler and 
complex structure deformations. A summary of some of the 
basic definitions and identities of special geometry can be found in 
appendix \ref{N=2sp}. Essentially all information is encoded in one 
holomorphic function, the prepotential $\cF$. Let us for 
concreteness consider the moduli space of K\"ahler structure 
deformations of the Enriques Calabi-Yau which is of the form \eqref{coset}.
Denoting by $\hat \omega$ the harmonic $(1,1)$-form in the $\IT^2$-base
and by $\omega_a$ the $(1,1)$ forms in the Enriques fiber, we 
obtain complex coordinates $S,t^a$ by expanding
  the combination 
  \beq
  \label{kahl}
 J+i B_2=\ S\, \hat \omega +t^a\, \omega_a \ ,\qquad \quad a=1,\ldots,10\ ,
\eeq
where $J$ is the K\"ahler form on the Enriques Calabi-Yau and $B_2$ is the NS-NS two-form.
Note that in our conventions $\R\, S >0$ and $\R\, t^a > 0$ such that the world-sheet  instantons
arise as series in $q_S= e^{-S}$ and $q_{t^a}=e^{-t^a}$ in the large radius expansion. We note that 
these complexified K\"ahler parameters $t^a$ can be regarded as a parametrization of the coset $\CN_8$. The parametrization 
we are using here is the one suitable for the conventional large radius limit and corresponds to what was called in \cite{km} the 
geometric reduction. In terms of (\ref{kahl}), the prepotential takes the form
\beq \label{def-Eprepot}
  \cF = - \tfrac{i}2 C_{ab} t^a t^b S \ .
\eeq
For the Enriques Calabi-Yau the cubic expression for the genus zero free energy $F^{(0)}=\cF$ is exact and
world-sheet instanton corrections will only arise at higher genus. This is precisely the reason 
for the simple form \eqref{coset} of the moduli space.
The symmetric matrix $C_{ab}$ in \eqref{def-Eprepot} 
encodes the intersections in the Enriques fiber $E$ such that 
\be
C_{ab}=\int_E \omega_a \wedge \omega_b\ . 
\ee
The inverse matrix $C^{ab}\equiv C^{-1\, ab}$ can be calculated explicitly and coincide in an appropriate basis 
with the intersection matrix of the $\bbZ_2$ invariant lattice of the second and the third factor in 
(\ref{coveringlattice}), i.e.   
\be \label{def-Gamma_E}
\Gamma_E=\Gamma^{1,1}\oplus E_8 (-1)\ , \qquad \quad (C^{ab})=\left(\begin{array}{cc}0 & 1\\ 1 & 0\end{array} \right) \times (-C_{E_8}).
\ee
Here $C_{E_8}$ is the  Cartan matrix of the exeptional group $E_8$. The
lattice $\Gamma_E$ is identified with the second cohomology group of the Enriques surface. 

The prepotential for the Enriques Calabi-Yau encodes the classical geometry 
of the moduli space \eqref{coset}. The K\"ahler potential is derived using equation
\eqref{KpotII} to be of the form\beq \label{def-KY}
K=-\log \big[  Y (S+\bar S)\big]\ ,\qquad Y=\tfrac12 C_{a b} (t^a+\bar t^a)  (t^b+\bar t^b) \ .
\eeq
Note that $K$ as given in \eqref{KpotII} contains a term 
$-\log |X^0|^2$, with $X^0$ being the fundamental period. Such a term can be removed 
by a K\"ahler transformation $K\rightarrow K-f-\bar f$, where 
$f$ is a holomorphic function, such that our expression \eqref{def-KY} corresponds to 
a certain K\"ahler gauge. In general, all objects we will consider below are sections of 
a line bundle $\cL$ which parametrizes such holomorphic rescalings $V\rightarrow e^{f}V$. 
As an example $e^{-K}$ is a section of $\cL\otimes \bar \cL$.
Such K\"ahler transformations do not change the K\"ahler metric which is 
obtained by evaluating the holomorphic and anti-holomorphic derivative of $K$.
The K\"ahler metric splits into two pieces 
\beq
  G_{S\bar S} = \frac{1}{(S+\bar S)^2}\ ,\qquad G_{a \bar b} = -\frac{C_{a b}}{Y} + \frac{ C_{ac}(t+\bar t)^c C_{bd} (t+\bar t)^c}{Y^2} \ ,
\eeq
with all other components vanishing.
The Christoffel symbols for this metric are easily evaluated to be
\be \label{EnriquesChris}
\Gamma^S_{SS}=2 K_S\ , \qquad
\Gamma^c_{ab}=K_e C^{ed}  \hat \Gamma^c_{ab|d} \ ,
\ee
where $K_S$ and $K_a$ are the first derivatives of the K\"ahler potential \eqref{def-KY} 
and we have defined  
\beq \label{def-hatGamma}
  \hat \Gamma^b_{ac|d} = \big(\delta^b_c C_{ad} +\delta^b_a C_{cd} - \delta^b_d C_{ac} \big)\ .
\eeq
It is also easy to derive the holomorphic Yukawa couplings $ C^{(0)}_{ijk}$ defined in \eqref{def-CC}. 
In coordinates $S,t^a$ one uses the prepotential \eqref{def-Eprepot} to 
show 
\beq
    C^{(0)}_{S ab}= C_{ab}\ .
\eeq

In general $C^{(0)}_{S ab}$ is a section of $\CL^{2}\otimes {\rm Sym}^3(T^*{\CM})$. In the case of the Enriques 
Calabi--Yau it is 
constant in the K\"ahler gauge and coordinates chosen above, and covariantly constant in a general gauge. 
The covariant derivative, acting on a section of 
$\cL^{m}\otimes \bar \cL^{n}$, is \eqref{cov_D} 
\be
D_a=\partial_a+m K_a,\qquad D_{\bar{a}}=\partial_{\bar a}+n K_{\bar a},
\ee
and includes the Christoffel symbols when acting on tensors. 
Applied to $C^{(0)}_{S ab}$ one shows
\be
D_c C^{(0)}_{abS} = -\Gamma_{ca}^d C_{db} -\Gamma_{cb}^d C_{ad} + 2 \partial_c K C_{ab}=0\ , 
\ee
which vanishes by means of the equation \eqref{EnriquesChris} for the Christoffel symbols. 
A similar equation holds for the covariant derivative $D_S C^{(0)}_{abS}$, showing that $C^{(0)}_{abS}$
is indeed covariantly constant. Once again, this special property of the Yukawa couplings is immediately 
traced back to the fact that the prepotential $\cF$ receives no instanton corrections.

\begin{figure}[!ht]
\leavevmode
\begin{center}
\includegraphics[height=5cm]{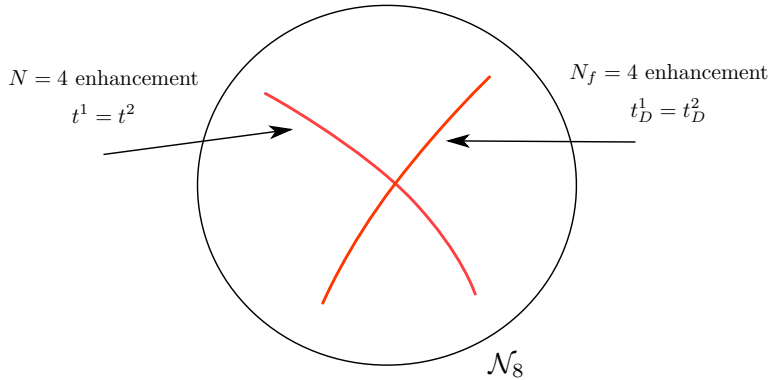} 
\end{center}
\caption{The singular loci in the moduli space $\CN_8$, leading to two different gauge theories in the field theory limit. }
\label{modulispace}
\end{figure}

The space $\CM$ has two different types of singular loci in complex codimension one on the moduli 
space~\cite{fhsv,aspinwall} which lead to conformal field theories in four dimensions. 
The first degeneration comes from the shrinking of a smooth rational
curve $e\in \Gamma_E$ with $e^2=-2$. 
The shrinking $\IP^1$ leads to an $SU(2)$ gauge symmetry enhancement 
together with a massless hypermultiplet, also in
the adjoint representation of the gauge group. We then obtain for
this point the massless spectrum of ${\cal N}=4$ supersymmetric 
gauge theory. In terms of the complexified K\"ahler parameters introduced 
in (\ref{kahl}) this singular locus occurs along  
\be
\label{slocusone}
t^1 =t^2. 
\ee
In order to understand the second singular locus, we first point out that the coset $\CN_8$ can be parametrized in many 
different ways. In \cite{km} it was noticed that there is a parametrization of this coset in terms of some coordinates $\tD^a$, $a=1, \cdots, 10$  
which are related to what was called there the BHM reduction. By using the formulae in \cite{km} it is easy to see that the coordinates $t^a$ and 
$\tD^a$ are related by the following simple projective transformation, 
\be
\label{projtrans}
\ba
t^1&=\tD^1 -{1\over 4 \tD^2}\sum_{i=3}^{10} (\tD^i)^2,\\
t^2&={2 \pi^2 \over \tD^2},\\
t^i &=-\pi \ri { \tD^i \over \tD^2},\quad i=3, \cdots, 10.
\ea
\ee
The second singular locus occurs when 
\be
\label{slocustwo}
\tD^1=\tD^2. 
\ee
On this locus one gets as well an $SU(2)$ gauge symmetry enhancement. In addition
one gets four hypermultiplets in the fundamental representation
of $SU(2)$, and the resulting gauge theory
is ${\cal N}=2$, $SU(2)$ Yang-Mills theory with four massless hypermultiplets. In \figref{modulispace} we represent schematically the two singular 
loci in moduli space, related by the projective transformation  (\ref{projtrans}). In sections \ref{sec:diE} and \ref{sec:ftlim} of this paper we will explore in some detail the 
field theory limit of the topological string amplitudes and we will verify this picture of the moduli space.  

\subsection{Genus one and the free energies on the Enriques fiber\label{genus1}}

So far we have discussed the classical moduli space of the Enriques Calabi-Yau $Y$. 
We introduced the prepotential $\cF$ which is cubic in the K\"ahler structure deformations and 
receives no worldsheet instanton corrections. One expects that such a simple structure will no longer persist at 
higher genus. This is already true at genus one as was shown in \cite{hmfhsv,km}. Heterotic--type II 
duality can also be used to determine all higher genus free energies on the 
K3 fibers of the Enriques Calabi-Yau \cite{km}. In this section we will summarize some results of \cite{km}
and present a closed expression for the fiber free energies also including the anti-holomorphic dependence. 

Let us begin with a brief discussion of the free energies for the Enriques fiber. The fiber limit of the topological 
string amplitudes corresponds to blowing up the volume of the base space by taking 
\beq
S \ \rightarrow \ \infty\ , \qquad \qquad  q_S \equiv e^{-S} \ \rightarrow \ 0\ .
\eeq
In what follows we will need to distinguish the full topological string amplitudes 
$F^{(g)}$ from their fiber limits as well as from their holomorphic limits. We will denote,
\be
F^{(g)}_E(t,\bar{t})=\lim_{S\rightarrow \infty}F^{(g)}(t,\bar{t})
\ee
and
\be
\label{holofiber}
\CF^{(g)}_E(t)=\lim_{\bar{t}\rightarrow \infty}F^{(g)}_E(t,\bar{t}).
\ee
The fiber limit $F^{(g)}_E(t,\bar t)$ can be calculated using heterotic-type II duality \cite{agnt,mm,km}. In the heterotic string 
they are given by a one--loop computation of the form
 \be
 \label{hetint}
F^{(g)}_E (t, \bar t)=\int \rd\tau \, {\overline \Theta}^g_{\Gamma} (\tau, v^+) f_g(\tau, \bar \tau) /Y^{g-1}
\ee
where $Y$ is defined in \eqref{def-KY}, and $\Theta^g_{\Gamma} (\tau, v^+) $ is a theta function with an insertion of $2g-2$ powers of the right--moving 
heterotic momentum. We will not need the precise definitions of $\Theta^g_{\Gamma}$ and $f_{g}$ here. However, 
it is important to note that
these amplitudes can be evaluated in closed form by using standard techniques for one--loop integrals. 
The holomorphic limit (\ref{holofiber}) was determined in \cite{km} and 
it is given by
\beq
\label{gr}
  \cF^{(g)}_E (t) = \sum_{r>0} c_g(r^2) \Big[2^{3-2g} \text{Li}_{3-2g}(e^{-r\cdot t}) -  \text{Li}_{3-2g}(e^{-2r\cdot t}) \Big] \ ,
\eeq
where $\text{Li}_{n}$ is the polylogarithm of index $n$ defined as
\beq
{\rm Li}_n (x) =\sum_{d=1}^{\infty} {x^d\over d^n}\ .
\eeq
In formula \eqref{gr} we have also set $r^2 =C^{ab}r_a r_b$ and $r\cdot t = r_a t^a$. We will sometimes write
\be
r=(n,m, \vec q).
\ee
The restriction $r>0$ means $n>0$, 
 or $n=0, m>0$, or $n=m=0$, $\vec q>0$.
Finally, we need to define the coefficients $c_g(n)$. They can be 
identified as the expansion coefficients of a particular quasi-modular form
\be
\label{geomrmod}
\sum_n c_g(n) q^n =-2 \frac{{\cal P}_{g}(q)}{\eta^{12}(2\tau)},
\ee
with $\CP_g(q)$ given by
\be
\label{defpg}
\biggl( { 2\pi  \eta^3 \lambda \over \vartheta_1(\lambda|\tau)}\biggr)^2=
\sum_{g=0}^{\infty} (2 \pi \lambda)^{2g} {\cal P}_{g}(q).
\ee
The definition of $\eta(\tau)$ and the theta-function $\vartheta_1(\lambda|\tau)$ can be found 
in Appendix \ref{theta}. {}From the definition \eqref{defpg} and the identities summarized in 
Appendix  \ref{theta} one also infers that the $\cP_g$ are quasimodular forms of weight $2g$ 
and can be written as polynomials in the Eisenstein series $E_2, E_4, E_6$. We have for example
\be
\label{casesps}
\CP_1(q)=\tfrac{1}{12} E_2(q)\ , \,\,\,\,\,\, \quad \CP_2(q)=\tfrac{1}{1440} (5 E_2^2 + E_4)\ .
\ee

In general, as we will see in section \ref{sec:diE}, it is very hard to include the $\IT^2$-base in
order to obtain the expressions $\cF^{(g)}$ for the full Enriques Calabi-Yau. It turns out that only $\cF^{(1)}$
factorizes nicely, namely we can write the A--model free energy $\cF^{(1)}$ as \cite{hmfhsv,km}
\beq
   \cF^{(1)}(S,t) = \cF^{(1)}_{\rm base} +\cF^{(1)}_E  \ ,
\eeq  
where $\cF^{(1)}_{\rm base}$ and $\cF^{(1)}_E $ are the contributions from the $\IT^2$ base 
and the K3 fiber. $\cF^{(1)}_{\rm base}$ is the torus free energy given by \cite{bcov1}
\beq
  \cF^{(1)}_{\rm base} = - 12 \log \eta(S)\ ,
\eeq
where $\eta(S)$ is defined in \eqref{dede}, while 
\be
\label{gonehol}
\cF^{(1)}_E=- \tfrac{1}{2} \log \Phi(t), 
\ee
where $\Phi(t)$ is the infinite product
\be \label{def-Phi}
  \Phi(t) =  \prod_{r>0}  \left(\frac{1  -e^{-r\cdot t}}{1 + e^{-r\cdot t}}\right)^{2c_1(r^2)} \ .
\eeq
This infinite product first appeared in the work of Borcherds \cite{borcherds}. As we will discuss in more detail later on, $\Phi(t)$ is the key example of a  
holomorphic automorphic form for the Enriques Calabi-Yau. It is also convenient to introduce, 
\be
\Phi(S,t) = \eta^{24}(S) \Phi(t), 
\ee
so that we can write
\be
  \cF^{(1)}(S,t)=-\tfrac{1}{2} \log \, \Phi(S,t).
  \ee

We presented above formulae for the holomorphic limit of $F^{(g)}_E(t,\bar t)$, but heterotic-type II duality can be 
used as well to obtain the antiholomorphic dependence on $\bar t$. At genus one, one finds \cite{hm,mm} 
\be
\label{feoneanti}
F^{(1)}_E (t, \bar t) =-2\log Y -\log \big|\Phi(t) \big|.
\ee
 The antiholomorphic dependence on $\bar S$ is the usual one for the torus \cite{bcov1} and one has 
\beq \label{F1Enriques}
 F^{(1)}(S, \bar S,t,\bar t) =  F^{(1)}_E(t,\bar t) - 6 \log \Bigl( (S+ \bar S) |\eta^2(S)|^2\Bigr).
\eeq
Equivalently, we can write
\be \label{F1Enriques2}
 F^{(1)}(S, \bar S,t,\bar t)=  - 2\log \big[ (S+\bar S)^3 Y \big] - \log \big|\Phi(S,t)\big|.
\ee
As a consistency check one shows that this anti-holomorphic dependence can also be 
inferred from the holomorphic anomaly equation \eqref{anomaly_F1} for $F^{(1)}$.

The antiholomorphic dependence in the heterotic calculation at higher genus is much more complicated, 
but was written down for the STU model in \cite{mm}. As we show in Appendix \ref{heteroticFg}, this 
computation can be considerably simplified and adapted to the Enriques case. 
We find that the non-holomorphic free energy $F^{(g)}_E(t,\bar t)$ can be cast into the 
form
\bea\label{Fgantihol}
F^{(g)}_E (t, \bar t)
&=&\sum_{l=0}^{g-1}\sum_{C=0}^{\begin{subarray}{c}{\rm min}\\(l,2g-3-l)\end{subarray}}  
\left(\begin{array}{c}2g-3-l \\ C \end{array}\right)
  {(t+\bar{t})^{a_1}\ldots(t+\bar{t})^{a_{l-C}}\partial_{a_1}\ldots \partial_{a_{l-C}}\CF^{(g-l)}_E(t)\over(l-C)! 2^l\ Y^l}\nn\\
  &&-{1\over 2^{g-2}(g-1)Y^{g-1}}\ ,
\eea
where $\cF^{(r)}_E(t)$ is the holomorphic fiber expression given in \eqref{gr}. 
It is easy to check that the $F^{(g)}_E(t, \bar t)$ fulfill the holomorphic anomaly equation on the fiber.

So far we have discussed the heterotic results for the fiber limit by using the K\"ahler parameters (\ref{kahl}) 
appropriate for the large radius limit. As shown in \cite{km}, one can also compute them in the 
coordinates $\tD^a$ introduced in (\ref{projtrans}). This was called the BHM reduction in \cite{km}, and leads to the 
holomorphic couplings,  
\be
\label{finalfgbor}
\CF^{(g)}_E (\tD)=\sum_{r>0} d_g(r^2/2)(-1)^{n+m} {\rm Li}_{3-2g}({\rm e}^{-r \cdot \tD })
\ee
where the coefficients $d_g(n)$ are defined by
\be
\sum_n d_g (n) q^n =\frac{2^{2+g} {\cal P}_{g}(q^4)-2^{2-g} {\cal P}_{g}(q)}{\eta^{12}(2\tau)},
\label{finalmodcoef}
\ee
and in (\ref{finalfgbor}) we regard $r$ as a vector in $\Gamma^{1,1}\oplus E_8(-2)$. Note that
in comparison to \eqref{def-Gamma_E} we now need to include the lattice 
$E_8(-2)$ with inner product given by $-2$ times the Cartan matrix of $E_8$, such that $r^2=2nm-2 \vec q^{~2}$. One has, in particular, 
\be
\CF^{(1)}_E (\tD)=-\tfrac{1}{2} \log \Phi_B(\tD)\ , 
\ee
where 
\be
\Phi_B(\tD)=\prod_{r>0} \Bigl(1- \re^{-r \cdot \tD}\Bigr)^{(-1)^{n+m} c_{B}(r^2/2)}
\ee
with coefficients
\be
\sum_n c_{B} (n) q^n ={\eta(2 \tau)^8 \over \eta(\tau)^8 \eta(4 \tau)^8}.
\ee
This is the modular form introduced by Borcherds in \cite{borcherdsone}, and the above
expression for $F_1$ agrees with that found by Harvey and Moore in \cite{hmfhsv} (up to a factor of $1/2$ due to
different choice of normalizations). 

\subsection{An all--genus product formula on the fiber}

As we have already mentioned, the infinite product (\ref{def-Phi}) was first considered by Borcherds in \cite{borcherds}. Borcherds also noticed 
that (\ref{def-Phi}) is the denominator formula for a generalized Kac--Moody (or Borcherds) superalgebra (see \cite{gebert, hm} for a review of 
Borcherds algebras). The root lattice of this superalgebra 
is $\Gamma^{1,1}\oplus E_8(-1)$ (i.e. the cohomology lattice of the Enriques surface), and 
the simple roots  are the positive, norm $0$ vectors. Each simple root appears also as a superroot, both with multiplicity $8$, and 
this is why the product of (\ref{def-Phi}) has a ``supersymmetric" structure: the numerator is a trace over fermionic degrees of freedom, while the 
denominator traces over bosonic degrees of freedom. Both have the same multiplicity $2c_1 (r^2)$. In addition, the fact that $c_1(-1)=0$ is 
equivalent to the absence of tachyons in the spectrum. 


We will now write down a formula for the total partition function of topological string theory, restricted to the fiber, and we will show that it preserves the 
structure found by Borcherds for (\ref{def-Phi}). As a first step, we define a generating functional $\xi(q,g_s)$ closely related to (\ref{defpg}),
\be
  \xi(q,g_s)= \prod_{n=1}^{\infty} 
 {(1-q^n)^2 \over 1-2 q^n \cos \, g_s + q^{2n}}.
 \ee
We have the identity
\be
 \sum_{g=0}^{\infty} \CP_g(q) g_s^{2g-2} =\biggl( 2 \sin \, {g_s \over 2} \biggr)^{-2} \xi^2(q, g_s), 
 \ee
Let us now define the Enriques degeneracies $\Omega_E(r, \ell)$ as 
\be
\label{omegae}
\sum_{r,\ell} 8 \Omega_E(r,\ell)q^{r^2} q_s^{ \ell} ={2\over  (q_s^{1\over 4} - q_s^{-{1\over 4}})^2} {1\over \eta^{12}(2\tau)} (\xi^2(q, g_s/2)-\xi^2(-q, g_s/2)), 
\ee
where
\be
q_s=\re^{ \ri g_s}
\ee
The r.h.s. of (\ref{omegae}) only involves {\it integer} powers of $q_s^{\pm 1}$. We can collect the Enriques degeneracies in the generating polynomials
\be
\Omega_n(z) =\sum_{r^2=2n,\ell\ge 0} \Omega_E(r,\ell) z^{\ell},
\ee
which are of degree $n$ in $z$. We have for the first few:
\be
\ba
\Omega_0(z)&=1, \\
\Omega_1(z) &=12 + 2z,\\
\Omega_2(z)&= 90 + 24 z + 3 z^2,\\
\Omega_3(z)&=520 + 180 z + 36 z^2 + 4 z^3,\\
\Omega_4(z)&=2538 + 1040 z+ 270 z^2 + 48 z^3+ 5 z^4,\\
\Omega_5(z)&=10944 + 5070 z + 1560 z^2 + 360 z^3 + 60 z^4 + 6 z^5.
\ea
\ee
Notice that the constant terms of $\Omega_n(z)$ are closely related to the Euler characteristics of the Hilbert schemes of the Enriques surface, but there are 
``deviations" which become more and more important as the degree increases.  Finally, notice that 
\be
\sum_{\ell} \Omega_E(r,\ell)q^{r^2} q_s^{  \ell } =\Omega_n(q_s) + \Omega_n(q_s^{-1}) -\Omega_n(0).
\ee
We now define 
\be
F_E=\sum_{g=1}^{\infty} g_s^{2g-2}\CF^{(g)}_E(t), \qquad Z_E =\re^{-2 F_E}.
\ee
Notice that, as $g_s\rightarrow 0$, $Z_E$ is precisely the Borcherds product $\Phi(t)$. It is now an easy exercise to 
evaluate it for finite $g_s$ from (\ref{gr}), and we find
\be
\label{allgproduct}
Z_E(g_s,t)=\prod_{r,\ell} \biggl( { 1- q_s^{\ell} \re^{ -r\cdot t} \over 1+q_s^{\ell} \re^{  -r\cdot t} }\biggr)^{8 \Omega_E(r,\ell)}.
\ee
As in the $g=1$ case, (\ref{allgproduct}) has a supersymmetric structure, with the same degeneracies for 
fermionic and bosonic states. This formula in fact suggests the existence of a superalgebra structure for the 
all--genus result as well. By including $g_s$ we have extended the lattice to
\be
\Gamma^{1,1}\oplus E_8(-1) \rightarrow \Gamma^{1,1}\oplus E_8(-1)\oplus \IZ
\ee
which is reminiscent of the growth of an eleven--dimensional direction associated to the string coupling constant. The fact that 
the all--genus heterotic results seem to lead to an extra direction in the heterotic lattice has been pointed out in \cite{d,kawai}. 
 It would be very interesting to see if there is indeed a superalgebra associated to the all--genus result (\ref{allgproduct}). If this was the case,  the 
 quantities $8 \Omega_E(r,\ell)$ would correspond to root multiplicities. 
 
Finally, we mention that according to the conjecture in \cite{mnop} and the results of \cite{mp}, (\ref{allgproduct}) is 
essentially the generating functional of an infinite family of Donaldson--Thomas invariants on the 
Enriques surface (written already in the right variables). Such product formulas for $Z$ 
exist generically if the latter is expressed in terms of of Gopakumar-Vafa invariants~\cite{Klemm:2004km}.
Our comments above indicate that  the Donaldson--Thomas theory on this manifold has a highly nontrivial algebraic 
structure (see section 3.2.6 in \cite{mp} for a  related observation).

\subsection{Automorphic forms \label{aut_prop}}

The free energies $F^{(g)}_E (t, \bar t)$ on the fiber turn out to be automorphic forms on the coset space $\CN_8$. 
Here we will study in some detail automorphic forms on the space $\CN_s$. We will say that
a function on the moduli space $\CN_s$ is {\it automorphic} 
if it has well--defined transformation properties under the discrete subgroup $ O(s+2,2;\IZ)$.

The transformation properties are easier to understand if we consider explicit generators of 
the symmetry group. We consider the explicit parametrization of the coset space (\ref{coset}) induced by a reduction 
\be
\Gamma^{s+2,2} =\Gamma^{s+1,1} \oplus \Gamma^{1,1}, 
\ee
and let $t \in \IC^{s+1,1}$ be the vector of complex coordinates parametrizing the coset. Our conventions are such that 
$t$ has {\it positive} real part.
For an element $t^a \in \IC^{s+1,1}$ we define the inner product 
\be
t^2=\tfrac{1}{2} C_{ab} t^a t^b, 
\ee
where $C_{ab}$ is the intersection matrix. 

The generators of the symmetry group are taken to be \cite{hm}:
\begin{itemize}
\item $t \mapsto t+2\pi i \lambda$, $\lambda \in \Gamma^{s+1,1}$.

\item $ t \mapsto w(t)$, $w\in O(s+1,1;\IZ)$.

\item The automorphic analog of an S--duality transformation
\be
\label{saut}
t^a \quad  \mapsto\quad \tilde t^a={ t^a \over t^2}.
\ee
\end{itemize}

We say that a function $\mf(t)$ is an {\it automorphic function of weight $k$} if it is invariant under the first two transformations above, and 
if under (\ref{saut}), 
it behaves as follows:
\be
\mf_k(\tilde t)= t^{2 k} \mf_k(t).
\ee
We can also have automorphic forms of weight $(k, \bar k)$ which transform as
\be \label{def-Phi_kk}
\mf_{k,\bar k} (\tilde t)= t^{2k}\ \bar t^{\, 2 \bar k} \mf_{k,\bar k} (t).
\ee
Although we have not indicated it explicitly, these functions might have a non-holomorphic dependence on $\bar t$. 
Automorphic forms are in general non-holomorphic. Some automorphic forms are meromorphic (they have 
poles at divisors). If they do not have poles, they are 
called {\it holomorphic}. 

Notice that (\ref{saut}) transforms the metric $Y=(t+\bar t)^2$ on the ``upper half plane" as follows:
\beq \label{trans-Y}
Y \ \mapsto \  t^{-2}\ \bar t^{-2} Y\ .
\ee
Following the definition \eqref{def-Phi_kk} this identifies $Y$ as an automorphic form of weight $(-1, -1)$.
Recalling the form of the K\"ahler potential for the classical moduli space \eqref{def-KY} this is nothing but a K\"ahler transformation \cite{cardoso}
\beq \label{transK}
   K\ \mapsto\ K+\log t^2 +\log \bar t^2\ .
\eeq
in special coordinates where $X^0=1$.  
Note that, if we keep $X^0$, this shift can be absorbed by the transformation of $X^0$
\beq \label{trans-X0}
   X^0\ \mapsto\ t^2 X^0\ .
\eeq
This can be traced back to the fact that $K$ as given in \eqref{KpotII} is a scalar under the full 
symplectic group.

 In order to understand how the automorphic properties mix with taking derivatives, it is useful to derive the Jacobian $J_a^b$ of the change of 
 coordinates (\ref{saut}). We immediately find, 
 \be \label{def-J}
{\partial \tilde t^a \over \partial t^b} \equiv  (J^{-1})^a_{b} = {1\over t^4} \Bigl( \delta^a_{~b} t^2 - t^a C_{be}t^e \Bigr)\ , \qquad \qquad 
{\partial t^a \over \partial \tilde t^b} =J^a_{b} =\delta^a_{~b} t^2 - t^a C_{be}t^e\ .
 \ee
Notice that $J_a^b$ obeys the following useful identities  
\be
\label{jcont}
 J_a^b = t^4  (J^{-1})^b_{a}\ , \quad  \qquad C_{ab}=t^{-4} C_{cd} J^c_a J^d_b \ ,\quad   \qquad  C^{ab}J_a^c J_b^d=t^4 C^{cd}\ .
\ee
Let us now assume that $\mf$ is an automorphic form of weight $(k,0)$. We want to determine the transformation 
behavior of $D_a\mf$ and $D_a D_b \mf$ under the dualities \eqref{saut}. $D_a$ are here the derivatives covariant both with respect to Christoffel connection and the canonical connection on the vacuum bundle $\CL$, as introduced in section \ref{specialE}. 
Therefore,
\be
D_a \mf =(\partial_a - k K_a)\mf\ . 
\ee
Notice that, since $K$ transforms as given in \eqref{transK}, its first derivative $K_a$ shifts as
\be
    K_a\ \mapsto\ J_a^b \big(K_b + t^{-2} C_{bc} t^c \big)\ .
\ee
Combining this with the transformation of the automorphic form $\mf$ itself 
we conclude 
\beq \label{trans_D_aPhi}
   D_a \mf \ \mapsto \  t^{2k} J^b_a D_b \mf\ .
\eeq
Similarly, we show that the second derivative of $\mf$ transforms as 
\beq \label{trans_D_aD_bPhi}
   D_b D_a \mf\ \mapsto\ t^{2k} J_d^b J_a^c D_b D_c \mf\ ,
\eeq
where we have used that the Christoffel symbols in the second connection 
transform as 
\beq
     J^d_b \partial_d J^c_a - \widetilde \Gamma^d_{ba} J_a^c = \Gamma^d_{ba} J_a^c\ . 
\eeq
Hence, we have shown that the covariant derivatives $D_a$ of $\mf$ 
transform with a factor $t^{2k}$ but are also rotated by the Jacobian $J^a_b$ 
containing another factor of $t^2$. 
Note however, that we can easily obtain automorphic forms 
containing the derivatives $D_a \mf$. More precisely, if $\mf$ and $\mf'$ 
are automorphic forms of weight $(k,0)$ and $(k',0)$ we find by 
using \eqref{jcont} that 
\beq \label{aut-forms}
   C^{ab} D_a D_b \mf\ , \qquad \quad C^{ab} D_a \mf D_b \mf'
\eeq 
are automorphic forms of weight $k+2$ and $k+k'+2$ respectively. Such 
automorphic combinations arise in the derivation of all $F^{(g)}(S, \bar S, t, \bar t)$, $g>1$.
More precisely, we will argue in the next sections that as function of $t, \bar t$, $F^{(g)}(S, \bar S, t, \bar t)$ itself is 
an automorphic form of weight $(2g-2,0)$ such that
\beq \label{trans-Fg}
   F^{(g)}\ \mapsto\ t^{4g-4} F^{(g)}\qquad \text{for} \quad g>1 \ .
\eeq
An important example of an automorphic form is the heterotic integral (\ref{hetint}). It is 
easy to show from the properties of the Narain--Siegel theta function that it has 
weight $(2g-2,0)$. Since this integral gives the fiber limit $F^{(g)}_E$, we obtain a check of the general 
property (\ref{trans-Fg}) from heterotic/type II duality. Note that it is straightforward to define
amplitudes $F^{(g)}$ invariant under automorphic transformations by  
\beq \label{inv_comb}
   (X^{0})^{2-2g}\, F^{(g)}\ .
\eeq
The invariance of this combination is readily checked by using \eqref{trans-X0} and \eqref{trans-Fg}. 
The expressions \eqref{inv_comb} are shown to be invariant under the full target space symmetry group
$Sl(2,\bbZ)\times O(10,2)$. They are the direct analogs of the invariant free energies encountered in the Seiberg-Witten example in 
section \ref{Seiberg-Witten}.

A particularly important and simple example occurs at $g=1$. Since $F^{(1)}_E$ is invariant, one 
deduces from (\ref{feoneanti}) and \eqref{trans-Y} that 
$\Phi(t)$ is an automorphic form of weight $(4,0)$ i.e.
\beq \label{trans-Phi}
  \Phi(\tilde t) = t^{8}\ \Phi(t)\ , \qquad \qquad \tilde t^a = \frac{t^a}{ t^2}\ .
\eeq
One can also show that $\Phi(t)$ is holomorphic. This is proved in \cite{borcherds}, and it 
is in fact a consequence of the regularity of $\CF^{(g)}_E (t)$ at the singular 
locus (\ref{slocusone}), which will be discussed in more detail in section \ref{boundaries}. In addition, $\Phi(t)$ is what is called a 
singular automorphic form (see \cite{binfinite}, section 3, for a definition). Singular automorphic forms are known to 
satisfy a wave equation 
\be
C^{ab} {\partial^2 \over \partial t^a \partial t^b} \Phi(t)=0. 
\ee
Equivalently, they have Fourier expansions involving only vectors of zero norm. It follows that $\CF^{(1)}_E(t)$ satisfies
\be
\label{wavefone}
C^{ab}\partial_a \partial_{b} \CF^{(1)}_E= 2C^{ab} \partial_a \CF^{(1)}_E\, \partial_b \CF^{(1)}_E\ .
\ee
This is equivalent to the recursive relation found in \cite{mp} for genus one invariants on the fiber, and proves that the expression for $ \CF^{(1)}_E(t)$ 
obtained in \cite{km} agrees with the Gromov--Witten calculation of \cite{mp}.


\section{Direct Integration on the Enriques Calabi-Yau}\label{sec:diE}

In this section we illustrate the power of the method of direct integration by 
studying the topological string amplitudes $F^{(g)}$ on the Enriques 
Calabi-Yau. Our approach will follow and generalize the strategy 
developed for the Seiberg-Witten example in section \ref{Seiberg-Witten}.
To begin with, we perform a direct integration along the
$\mathbb{T}^2-$base in section \ref{simple}. Using the fiber results 
obtained in the previous section as additional input, the first 
six free energies $F^{(g)}$ can be determined in a closed form. 
We then present a more general formalism combining direct integration in base and fiber directions. In section \ref{propagators_auto}, we introduce 
the relevant holomorphic and non-holomorphic $O(10,2,\bbZ)$ forms.
A closed recursive expression for $F^{(g)}$ will be derived in section \ref{direct_fiber_base}. 
It determines the $F^{(g)}$ up to a holomorphic ambiguity and we will briefly discuss 
possible boundary conditions in section \ref{boundaries}. Finally, in section 
\ref{red-model} we consider a reduced Enriques  
model with three parameters only, which was already studied in \cite{km}. This model has the advantage that the mirror map 
can be determined explicitly. 
We also study in more 
detail the boundary conditions (such as the gap condition), which 
lead to valuable conclusions also applying to the full model.

\subsection{A simple direct integration and $F^{(g)}$ to genus six \label{simple}}

Let us now perform the direct integration along the $\mathbb{T}^2$ base and derive the first few amplitudes $F^{(g)}$.
In order to do that we carefully keep track of their dependence of on the base direction $S,\bar S$.  
As in the case of Seiberg--Witten theory studied in section 3, it is easy to see from the 
structure of the holomorphic anomaly equations that the only antiholomorphic dependence of 
$F^{(g)}$ on $\bar S$ appears through $\widehat E_2(S, \bar S)$. By taking derivatives with respect to $S$ we will also generate 
in the holomorphic anomaly equations the modular forms $E_4(S), E_6(S)$, and by keeping track of the modular weight one immediately finds 
that $F^{(g)}$ is an element of weight $2g-2$ in the ring 
generated by
\be
\label{genS}
 \widehat E_2(S, \bar S), \quad E_4 (S), \quad E_6(S)\ .
\ee
Our only assumption here is that the holomorphic ambiguity for $F^{(g)}$ is also a modular form of weight $2g-2$ in this ring. This assumption
(as well as the details of the direct integration) 
can be checked in a highly nontrivial way by comparing the resulting expressions to the field theory limit in the $N_f=4$ locus of \figref{modulispace}. This 
check will be performed in section \ref{sec:ftlim}. 

To perform the direct integration 
let us first rewrite the holomorphic anomaly equation for the base direction $\bar S$. 
The general expression (\ref{rec_Fg}) reduces to
\be
\partial_{\bar S}F^{(g)}=- \frac{1}{2}{C^{ab} \over (S+\bar S)^2} \Big(D_a D_b F^{(g-1)}+ \sum_{r=1}^{g-1}D_a F^{(r)}  D_b F^{(g-r)} \Big)\ .
\ee
We now convert the derivative $\partial_{\bar S}$ into a derivative with respect to $\widehat E_2$.
The definition of $\widehat E_2$ was already given in \eqref{hatE2}. Since we now consider an expansion in $q_S=e^{-S}$ 
it takes the form
\be \label{def-widhatE2}
\widehat E_2 (S, \bar S)=-\frac{12}{S+\bar S}+E_2(S)\ .
\ee
Using the above assumption that the dependence of $F^{(g)}$ on $\bar S$ is only through this quantity, we can rewrite the 
anomaly equation as
\be
\label{partialhol}
{\partial F^{(g)} \over \partial \widehat E_2}=- \tfrac{1}{24} C^{ab} \Big(D_a D_b F^{(g-1)}+ \sum_{r=1}^{g-1}D_a F^{(r)}  D_b F^{(g-r)} \Big)\ .
\ee
Here the covariant derivatives $D_a$ are only taken with respect to the fiber directions and do not depend on 
the base due to the simple special geometry of the Enriques Calabi-Yau. This implies that all dependence  
on $\widehat E_2$ arises directly through the $F^{(r)}$.
We thus expand $F^{(g)}$ in powers of $\widehat E_2$ by writing 
\be
    \label{fgexp}
F^{(g)}=\sum_{k=0}^{g-1} \widehat E_2^k (S,\bar S)\ \coeff^{(g)}_{k}\ , \qquad \quad g> 1\ .
\ee
We see that (\ref{partialhol}) determines all the coefficients $c^{(g)}_{ k}$ for $k=1, \ldots, g-1$ in terms of quantities at lower genera. Explicitly, we have 
the solution
\be
\label{cksol}
c^{(g)}_{ k}=- \tfrac{1}{24 k} C^{ab}  \Big(D_a D_b c^{(g-1)}_{ k-1} + \sum_{r=1}^{g-1}\sum_{l+m=k-1} D_a c^{(r)}_{l} D_b c^{(g-r)}_{ m}\Big)\ ,
\ee
where we have set
\be \label{def-F1coeff}
   \coeff^{(1)}_{0}=F^{(1)}\ , \qquad  \qquad \coeff^{(1)}_{i}=0\ ,\qquad i \neq 0\ .
 \ee
The $\widehat E_2$-independent term $c^{(g)}_{0}$ arises as an integration constant and hence 
cannot be determined by the holomorphic anomaly equation. 
However, given our assumptions, we can fix it up to genus 6 as follows. Let us 
denote the coefficients in the fiber limit by
\be
   c^{(g)}_{E|\, k}=\lim_{S, \bar S \rightarrow \infty} c^{(g)}_{k}.
\ee
By also taking the fiber limit of (\ref{fgexp}) we find
\be
\label{sumfiber}
\sum_{k=0}^{g-1}  c^{(g)}_{E|\, k}=F^{(g)}_E(t,\bar t).
\ee
The free energies $F^{(g)}_E(t,\bar t)$ are known from the  heterotic computation and given in \eqref{Fgantihol}. Together with the fact that 
all $c^{(g)}_{E|\, k}$ for $k\ge 1$ are uniquely determined  by the direct integration we can use \eqref{sumfiber} to derive $c^{(g)}_{E|\, 0}$ i.e.~the fiber limit of the integration 
constant. But the condition that $c^{(g)}_{0}$ is a modular form in the ring generated by (\ref{genS}) and does not involve $\widehat E_2$ fixes it uniquely in terms 
of $c^{(g)}_{E|\, 0}$ as 
\be
\ba \label{fix_c}
 c^{(2)}_{0}&= 0\ ,   \qquad \quad &
 c^{(3)}_{0}&=c^{(3)}_{E|\, 0}\  E_4\ ,  \qquad \quad &
 c^{(4)}_{0}&= c^{(4)}_{E|\, 0}\  E_6\ , \\
 c^{(5)}_{0}&= c^{(5)}_{E|\, 0}\ E^2_4\ ,  \qquad \quad & 
 c^{(6)}_{0}&= c^{(6)}_{E|\, 0} \  E_4 \,E_6\ ,
 \ea
 \ee
 where $E_4(S)$ and $E_6(S)$ are the two holomorphic generators in \eqref{genS}. This can be checked 
 by noting that the definition \eqref{geneis} of the Eisenstein series implies that
 \beq \label{Elimit}
    E_2\ ,\ \ E_4\ , \ \ E_6\quad \rightarrow\quad 1\ , 
 \eeq
 in the fiber limit $S,\bar S\rightarrow \infty$.
 For $g\ge 7$, the number of possible modular forms is greater than one and $c^{(g)}_{E|\, 0}$ is no longer uniquely determined 
 in terms of its fiber limit. For example, at genus seven $c^{(7)}_0$ can contain terms proportional to 
 $E_4^3$ as well as $E_6^2$.
 
 Let us now write down some explicit formula for lower genera. For $g=2$ we find, 
\be
F^{(2)}(S, \bar S,t,\bar t) =\widehat E_2(S, \bar S)\ c^{(2)}_{1}\ , 
\ee
where we use \eqref{fix_c} and apply (\ref{cksol}) to derive
\be
\label{f2sol}
 c^{(2)}_{1}= -\tfrac{1}{24} C^{ab}  \left(D_a D_b F_E^{(1)}+ D_a F_E^{(1)} D_b F_E^{(1)} \right). 
\ee
Consistency of the fiber limit requires that
%
$c^{(2)}_{1}=F_E^{(2)}(t, \bar t)$. 
%
This can be checked by using the heterotic expression (\ref{Fgantihol}) for $F_E^{(2)}(t, \bar t)$, the property (\ref{wavefone}), and the identity \cite{km}
\be
 \label{identity}
 \CF^{(2)}_E=-\tfrac{1}{16} C^{ab}\partial_a \partial_b \CF^{(1)}_E\ ,
  \ee
which follows directly from (\ref{hetint}). In the holomorphic limit we find, 
\be 
\CF^{(2)}(S,t)=E_2(S) \CF^{(2)}_E(t)\ , 
\ee
in agreement with the results of \cite{km,mp}. In the following sections we will also need a slightly different 
form of $F^{(2)}$. Namely, it is straightforward to apply \eqref{wavefone} to write
\beq \label{result_F2}
   F^{(2)} = -\tfrac{1}{8} C^{ab} \partial_a F^{(1)} \partial_b F^{(1)}\ .
\eeq

Let us now consider the $g=3$ case. The amplitude $F^{(3)}$ can be expanded by using \eqref{fgexp} and
\eqref{fix_c} as 
\be
F^{(3)}=\widehat E_2^2(S, \bar S)\, c^{(3)}_{2} + E_4(S)\, c^{(3)}_{E|\, 0}\ .
\ee
Using the result of the direct integration (\ref{cksol}) we obtain
\be
  c^{(3)}_{2}=- \tfrac{1}{48} C^{ab} \Big(   D_aD_b F^{(2)}_E + 2 D_a F^{(2)}_E D_b F^{(1)}_E\Big)\ .
\ee
To determine $c^{(3)}_{E|\, 0}$ we use (\ref{sumfiber}), which gives 
\be
 c^{(3)}_{2} +  c^{(3)}_{E|\,0}=F^{(3)}_E(t, \bar t)\ . 
\ee
On the other hand,  one finds that
\be
\label{antiholfthree}
F^{(3)}_E(t, \bar t)=-\tfrac{1}{24} C^{ab} D_a D_b F^{(2)}_E.
\ee
This can be derived in the holomorphic limit by using (\ref{hetint}), and it is similar to (\ref{identity}). The antiholomorphic part can be 
checked with (\ref{Fgantihol}). Using all this, we finally obtain the following simple expression for 
$F^{(3)}(S, \bar S, t, \bar t)$,
\be
F^{(3)}=-\tfrac{1}{24} E_4\, C^{ab}  D_a D_b F^{(2)}_E  
 - \tfrac{1}{48} (\widehat E_2^2-E_4) 
C^{ab} \big(  D_a D_b F^{(2)}_E + 2 D_a F^{(2)}_E D_b F^{(1)}_E \big)\ ,
\ee
with the holomorphic limit
\be
\label{finalfthree}
\CF^{(3)}(S,t)= -\tfrac{1}{24}E_4\, C^{ab}  \partial_a \partial_b \CF^{(2)}_E- \tfrac{1}{48} (E_2^2-E_4)
C^{ab} \big(  \partial_a \partial_b \CF^{(2)}_E + 2 \partial_a \CF^{(2)}_E \partial_b \CF^{(1)}_E\big)\ .
\ee
Note that the second term in these expressions vanishes identically in the fiber limit where $E_2,E_4\rightarrow 1$. 
As we will discuss in more detail in section \ref{boundaries}, this is the first $F^{(g)}$ where the inclusion of the base yields
a behavior near the singular loci that differs significantly from the fiber limit.\\
{}Explicit calculations at genus $4$ proceed in the same way. Modular invariance with respect to $S$ gives
\be
F^{(4)}(S, \bar S, t, \bar t)=\widehat E_2^3 \ c^{(4)}_{E|\, 3} + \widehat E_2 E_4\ c^{(4)}_{E|\, 1} + E_6\ c^{(4)}_{E|\, 0}.
\ee
Once again, the general equation (\ref{gencksol}) allows us to determine the coefficients as
\be \label{c4}
\ba
 c^{(4)}_{E|\, 3}&=- \tfrac{1}{72} C^{ab} \big( D_aD_b c^{(3)}_{E|\,2} + 2 D_a F^{(1)}_E D_b c^{(3)}_{E|\, 2} + D_a F^{(2)}_ED_bF^{(2)}_E \big)\ , \\ 
 c^{(4)}_{E|\, 1}&=-\tfrac{1}{24} C^{ab} \big( D_aD_b  c^{(3)}_{E|\, 0}  + 2 D_a  F^{(1)}_E D_b  c^{(3)}_{E|\, 0} \big)\ .
\ea
\ee
The ambiguity $c^{(4)}_{E|\, 0}$ is again determined by the heterotic computation in the fiber limit. 
More precisely, one specializes \eqref{sumfiber} to 
\be
  c^{(4)}_{E|\, 0}+ c^{(4)}_{E|\, 1}+ c^{(4)}_{E|\,3} = F^{(4)}_E(t, \bar t)\ ,
\ee
and solves for $c^{(4)}_{E|\, 0}$ by inserting the fiber result \eqref{Fgantihol}. This determines 
the free energy $F^{(4)}$. A similar analysis also applies to $g=5,6$. As already discussed 
above, the main obstacle that has to be overcome in order to proceed to higher genus is the difficulty to fix the ambiguities $c^{(g)}_0$. 
We will discuss possible additional boundary conditions in sections \ref{boundaries}, \ref{red-model}
and \ref{sec:ftlim}.

\subsection{Propagators and homolomorphic automorphic forms \label{propagators_auto}}

In the previous section we calculated the first free energies $F^{(g)}$ 
by a direct integration along the base direction. The results were expressed 
in terms of the holomorphic fiber energies $\cF^{(g)}_E$, which are known 
from heterotic-type II duality. Even though the results were rather compact and 
transparent, the information we have extracted is somewhat partial, since 
we have not used the holomorphic anomaly equations for the fiber moduli. In order to exploit the information they contain, we 
will construct building blocks for the automorphic forms in the fiber which enable us to 
perform the direct integration of the remaining holomorphic anomaly equations. Recall that we argued in the previous sections that 
the almost holomorphic modular form 
\beq \label{def-E2}
   \widehat E_2(S,\bar S) = -\frac{12}{S+\bar S}+E_2(S)\ ,\qquad  \qquad E_2(S) = \partial_S \log \Phi\ ,
\eeq
contains all non-holomorphic dependence of $F^{(g)}$ along the base direction $S$. It will be 
the task of this section to introduce the analog of $\widehat E_2$ for the fiber directions $t^a$. 
Furthermore we will define the fiber analogs of the holomorphic modular forms 
$E_4(S)$ and $E_6(S)$. This will lead us to the definition of a new class of holomorphic 
automorphic forms of $O(10,2,\bbZ)$. Eventually, in section \ref{direct_fiber_base} 
we will argue that a direct integration along the fiber direction allows us to 
express all $F^{(g)}$ in terms of these almost holomorphic and holomorphic 
forms of $O(10,2,\bbZ)$.

Let us now introduce the fiber analog of the almost holomorphic modular 
form $\widehat E_2(S,\bar S)$. This can be done by recalling that 
the genus one free energy $F^{(1)}$ is an invariant of the full 
symmetry group $Sl(2,\bbZ)\times O(10,2,\bbZ)$ and hence its
first derivatives transform in a particularly simple way. For the derivative with respect
to $S$ one finds $\partial_S F^{(1)} = \frac12 \widehat E_2$. The derivative with 
respect to $t^a$ we denote by $ \Delta^a = -\tfrac{1}{2} C^{a b} \partial_b F^{(1)}$ and evaluate
\beq \label{def-Deltaa}
     \Delta^a =\frac{t^a+\bar t^a}{Y}  + \epsilon^a(t)=\epsilon^a(t)-K_b(t) C^{ba}\ , \qquad \quad \epsilon^a(t) = \tfrac14  C^{ab} \partial_{t^b} \log \Phi \ ,
\eeq
where $Y=\frac12 C_{ab}(t+\bar t)^b (t+\bar t)^b$ and $\Phi$ is given in \eqref{def-Phi}. The function $\epsilon^a(t)$ is holomorphic in the coordinates 
$t^a$ and is the fiber analog of $E_2(S)$, while $\Delta^a$ plays the role of $\widehat{E}_2$. To see this note that $\epsilon^a$ transforms with a 
shift under the duality $t^a \mapsto t^a /t^2$:
\beq \label{trans-epsilon}
  \epsilon^a \quad \mapsto\quad  t^4 (J^{-1})^a_b (\epsilon^b + t^{-2} t^b)\ .
\eeq
This shift is precisely canceled by the shift of the non-holomorphic term in \eqref{def-Deltaa}
such that $\Delta^a$ simply transforms as
\beq \label{trans-Delta}
   \Delta^a \quad \mapsto\quad  t^{4} (J^{-1})_b^a \Delta^b(t)\ .
\eeq
Note that $\widehat E_2$ and $\Delta^a$ are sufficient to parametrize all propagators $\hat \Delta^{ij},\hat \Delta^{i},\hat \Delta$
introduced in \eqref{def-small-Delta}. Indeed, one has 
\begin{align} \label{Enriques-props}
  \hat \Delta^{a b} &=  -\tfrac{1}{12} C^{a b} \widehat E_2 \ ,\qquad
  &&\hat \Delta^{a S} =  \Delta^a\ ,\\
  \hat \Delta^S& =- \tfrac{1}{2} C_{a b} \Delta^{a} \Delta^{b}\ ,\qquad
  &&\hat \Delta^a = \tfrac{1}{12} \widehat E_2 \Delta^a \ ,\qquad \quad
 \hat \Delta = -\tfrac{1}{12}  \widehat E_2 C_{a b} \Delta^{a} \Delta^{b}\ .\nn
\end{align}
Using the explicit form of $\widehat E_2$ and $\Delta^a$ it is straightforward to check 
that these propagators fulfill the defining conditions \eqref{def-small-Delta}.
The fact that all $\hat \Delta-$propagators can be expressed as polynomials 
in $\widehat E_2$ and $\Delta^a$ will be used in the next section to 
argue that all non-holomorphic dependence of $F^{(g)}$ only arises through $\widehat E_2, \Delta^a$.
However, we also have to extract the non-holomorphic dependence in the 
covariant derivatives $D_a$ defined in \eqref{cov_D}. Following the logic 
of section \ref{Seiberg-Witten} we will show that each derivative can be split into 
a holomorphic covariant  derivative $\hat D_a$ plus holomorphic terms times the 
propagators $\Delta^a$. As an important byproduct, the definition of $\hat D_a$ will
also allow us to find an interesting construction of holomorphic automorphic forms.

Let us now construct a holomorphic covariant derivative $\hat D_a$, which 
has the same properties as $D_a$ under automorphic 
transformations \eqref{saut}. More precisely, given 
an automorphic form $\mf$ of weight $k$
we define its first derivative as 
\beq\label{der_holE}
    \hat D_a \mf \equiv \big(\partial_a - k C_{ab} \epsilon^b\big) \mf\ , 
\eeq 
where $\epsilon^a$ is defined in \eqref{def-Deltaa}, and note that $\hat D_a =D_a-kC_{ab}\Delta^b$. $\hat D_a$ can be viewed 
as the analog of the Serre derivative \eqref{der_hol} for modular forms of subgroups of $Sl(2,\bbZ)$.
It is not hard to check that it transforms under \eqref{saut} exactly as $D_a$. This 
transformation property was given in \eqref{trans_D_aPhi}.
Note however, that $\hat D_a$ maps holomorphic 
forms into holomorphic forms, while $D_a$ contains an anti-holomorphic contribution.
Moreover, by definition of $\epsilon^a$ one has
\beq
   \hat D_a \Phi(t) = 0\ ,
\eeq 
for the automorphic form $\Phi(t)$ given in \eqref{def-Phi}. In order to 
evaluate second derivatives we need to introduce the holomorphic analog of 
the Christoffel symbol in the definition \eqref{cov_D} of $D_k$.
To do that, let us consider a section $ \mf_a$ which transforms as 
$ \mf_a\, \mapsto\, t^{2k} J^b_a \mf_b$ 
under the action \eqref{saut}.
The covariant derivative is then defined to act as
\beq \label{def-hatD}
  \hat D_a  \mf_b = \big(\partial_a - k C_{ac} \epsilon^c\big) \mf_b - \hat \Gamma^{c}_{ab} \mf_c\ .
\eeq
Here we have included the holomorphic Christoffel symbol 
\beq 
   \hat \Gamma_{ab}^c= \hat \Gamma^{c}_{ab|d} \epsilon^d= \tfrac{1}{2} \hat C^{cd}\big(\partial_b \hat C_{da}+\partial_a \hat C_{db}-\partial_d \hat C_{ab} \big)\ ,   
\eeq
where $\hat \Gamma^b_{cd|a}$ is defined in \eqref{def-hatGamma} and related to $\Gamma^b_{cd}$ by $\Gamma^b_{cd}=\hat \Gamma^b_{cd|a}C^{ae}K_e$. We also have introduced the 
holomorphic `metric' $\hat C_{ab}$. Explicitly, $\hat C_{ab}$ is defined as
\beq
  \hat C_{ab} = \Phi^{1/2} C_{ab}\ ,\qquad \hat C_{ab}\ \mapsto\ J^c_a J^d_b \hat C_{cd}\ ,
\eeq
where $\Phi$ is given in \eqref{def-Phi} and  we have also displayed the transformation behavior of $\hat C_{ab}$ under \eqref{saut}
as inferred from \eqref{trans-Phi} and \eqref{jcont}. 
Once again we evaluate the transformation behavior of $\hat D_a \mf_b$ under \eqref{saut}
and finds the holomorphic analog of \eqref{trans_D_aD_bPhi}.
It is now easy to show that every non-holomorphic 
derivative $D_a$ can be split as 
\beq \label{split_Dhol}
   D_a \mf_b = \hat D_a \mf_b +k C_{ac} \Delta^c\, \mf_b+ \hat \Gamma^{c}_{ab|d} \Delta^d\, \mf_c\ .
\eeq
In other words, whenever $\mf_b$ is holomorphic the non-holomorphic 
dependence in $D_a \mf_b$ arises through the propagators $\Delta^a$ 
only.

 Let us now discuss a second interesting application of the holomorphic 
 covariant derivative $\hat D_a$. Namely, we will now show how it can be used to 
 construct new holomorphic automorphic forms. To start with 
 let us note that $\epsilon_a=C_{ab} \epsilon^b$ transforms in \eqref{trans-epsilon} similarly to 
 a vector field. We can use this analogy and define a field strength
 \beq \label{def-epsilon4}
      \epsilon^{4}_{ab} = \partial_{a} \epsilon_b -\tfrac12 \hat \Gamma^{c}_{ab} \epsilon_c=\partial_{a} \epsilon_b -\epsilon_a\epsilon_b+C_{ab}\epsilon^2\ ,\qquad \qquad \epsilon_a=C_{ab} \epsilon^b\ ,
 \eeq
which transforms covariantly, $\epsilon_{ab}^4\, \mapsto\, J^c_a J^d_b \epsilon^4_{cd}$,
under automorphic transformations \eqref{saut}. 
Note that by using the wave-equation \eqref{wavefone} one shows that $\partial_a \epsilon^a = - 4 C_{ab} \epsilon^a \epsilon^b$
 such that 
\beq \label{Cepsilon4=0}
  C^{ab}\epsilon_{ab}^4 = 0\ .
\eeq
Nevertheless, we can use $\epsilon^4_{ab}$ to construct holomorphic automorphic 
forms. To do that, we define
\beq\label{epsilon2k}
  \epsilon^{2k}_{a_1\ldots a_{k}} = \hat D^{\phantom{4}}_{a_k} \ldots \hat D^{\phantom{4}}_{a_3} \epsilon^4_{a_2 a_1}\ ,
\eeq
which is shown to be totally symmetric in the indices.
Holomorphic automorphic forms are now constructed by contraction with 
$C^{ab}$. For example, forms of weight $4$ and $6$ are given by
\bea \label{holfroms}
  \text{weight}\ 4: & \qquad&  C^{ab} C^{cd}  \epsilon^4_{ac} \epsilon^4_{bd}\ ,\\
   \text{weight}\ 6: & \qquad&  C^{ac} C^{be} C^{df}  \epsilon^4_{ab} \epsilon^4_{cd} \epsilon^4_{ef} \ ,\qquad 
                C^{ac} C^{be} C^{df}  \epsilon^{6}_{abd} \epsilon^6_{cef} \ .\nn
\eea 
It is tempting to conjecture that holomorphic automorphic forms of this type are sufficient to parametrize 
the holomorphic ambiguity of $F^{(g)}$. The fact that there is no holomorphic weight $2$
automorphic form of this type due to \eqref{Cepsilon4=0} matches nicely the fact that there is 
no holomorphic ambiguity for $F^{(2)}$. Also the forms in \eqref{holfroms} can be shown to be sufficient 
to parametrize the ambiguities of $F^{(3)}$ and $F^{(4)}$. This will be analyzed in further work.

\subsection{Direct integration of the holomorphic anomaly \label{direct_fiber_base}}

We will now use the material developed in the previous section to perform the direct integration in both fiber 
and base directions. This will allow us to give closed expressions which determine 
the $F^{(g)}$ up to a holomorphic ambiguity. 
To begin with, we show that each 
$F^{(g)}$ can be written as 
\beq \label{Eexp}
  F^{(g)} = \sum_{k=0}^{g-1}\, \sum_{n=0}^{2g-2}\ \widehat E_2^k \Delta^{a_1} \ldots \Delta^{a_n} \coeff^{(g)}_{k\, |\, a_1 \ldots a_n} \ ,\qquad g>1
\eeq
where $\coeff^{(g)}_{k\, |\, a_1 \ldots a_n} $ are holomorphic functions of $S,\, t^a$ and all anti-holomorphic 
dependence arises through the propagators $ \Delta^a$ and $\widehat E_2$ introduced in \eqref{def-E2} and \eqref{def-Deltaa}.
Note that by using the transformation properties of $F^{(g)}$ and $\Delta^a$ given in \eqref{trans-Fg} and \eqref{trans-Delta}
one infers that
\beq
   \coeff^{(g)}_{k\, |\, a_1 \ldots a_n} \ \mapsto\  t^{4g-4-4n} J^{b_1}_{a_1} \ldots J^{b_n}_{a_n}\, \coeff^{(g)}_{k\,|\,b_1 \ldots b_n}
\eeq
under automorphic transformations \eqref{saut}.

Let us now show that each $F^{(g)}$ for $g>1$ can indeed be written as \eqref{Eexp} 
by using induction. 
We first note that $F^{(2)}$ is of the form \eqref{Eexp}, 
\beq \label{trans-k}
  F^{(2)} = -\tfrac{1}2 \widehat E_2 C_{ab} \Delta^a \Delta^b\ ,
\eeq
as is immediately inferred from \eqref{result_F2} and \eqref{def-Deltaa}.
So let us assume that \eqref{Eexp} is true for all $r<g$ and 
show that this implies that \eqref{Eexp} is true for $g$. In order to do that 
we use the Feynman graph expansion \eqref{Fgwithf} of $F^{(g)}$ \cite{bcov}, which states
that each $F^{(g)}$ can be written as an expansion with propagators $\hat \Delta^{ij},\hat \Delta^i,\hat \Delta$
and vertices $C^{(r)}_{i_1\ldots i_n}$ with $r<g$. We have already shown that the $\hat \Delta$-propagators 
are polynomials in $\widehat E_2$ and $\Delta^a$ in \eqref{Enriques-props}. Hence, it remains to 
show that also the vertices $C^{(r)}_{i_1\ldots i_n}$ are polynomials in $\widehat E_2$ and $\Delta^a$.
By definition \eqref{C_prop} and our assertion, the vertices are defined as the covariant derivatives of 
amplitudes $F^{(r)}$ of the form \eqref{Eexp}. Using \eqref{split_Dhol} each of these covariant derivatives 
$D_a$ can be split into a holomorphic covariant derivative $\hat D_a$ and an expansion in $\Delta^a$. 
So we only have to show that $\hat D_a \Delta^b$ admits again an expansion into $\Delta$'s.
A straightforward computation shows that 
\beq \label{der_Delta}
  \hat D_a \Delta^b = C^{bd} \epsilon^4_{da} - \tfrac12 \hat \Gamma^b_{cd|a} \Delta^c \Delta^d\ ,
\eeq
where $\epsilon^4_{ab}$ and $ \hat \Gamma^b_{cd|a}$ are defined in \eqref{def-epsilon4} and \eqref{def-hatGamma}.
Altogether one infers that all vertices and $\hat \Delta$-propagators are polynomial in $\Delta^a$ and 
hence that $F^{(g)}$ is of the form \eqref{Eexp}.

Having shown that every $F^{(g)}$ is of the form \eqref{Eexp} we will now derive a 
closed expression for $F^{(g)}$ by direct integration of the holomorphic 
anomaly equation \eqref{rec_Fg}.
Applying the definition \eqref{def-small-Delta} of the propagators we can write 
the holomorphic anomaly equation as
\beq \label{rec_Fg1}
   \partial_{\bar \imath} F^{(g)} = \tfrac12 \partial_{\bar \imath} \hat \Delta^{ik}
   \Big(D_j D_k F^{(g-1)} + \sum_{r=1}^{g-1}D_j F^{(r)} D_k F^{(g-r)} \Big)  \ .
\eeq
This equation captures the anti-holomorphic derivatives $\partial_{\bar S} F^{(g)}$
along the base as well as the derivative $\partial_{\bar a} F^{(g)}$ along the fiber 
of the Enriques Calabi-Yau. Recall that the only non-vanishing propagators are 
$\hat \Delta^{ab} =-\frac{1}{12} C^{ab}\widehat E_2$ and $\Delta^a =\hat \Delta^{aS}$.
As we have shown, they contain all anti-holomorphic dependence
such that we can rewrite \eqref{rec_Fg1} as
\bea \label{directE}
   \frac{\partial F^{(g)} }{\partial \widehat E_2}&=&  -\tfrac{1}{24}C^{ab}
   \Big(D_a D_b F^{(g-1)} + \sum_{r=1}^{g-1}D_a F^{(r)} D_b F^{(g-r)} \Big)\ ,\\
   \label{directDelta} \frac{\partial F^{(g)} }{\partial \Delta^a}&=& 
   D_a D_S F^{(g-1)} + \sum_{r=1}^{g-1} D_a F^{(r)} D_S F^{(g-r)} \ . 
\eea
As we have seen above, the first equation is already very powerful and can be integrated easily. We can write the 
solution (\ref{cksol}) as 
\beq
\label{gencksol}
  F^{(g)}=  - \tfrac{1}{24} \sum_{k=1}^\infty \tfrac{1}{k} \widehat E_2^k C^{ab}
    \Big(D_a D_b \coeff^{(g-1)}_{k-1}   
   +\sum_{r=1}^{g-1} \sum_{l+m=k-1} D_a \coeff^{(r)}_{l} D_b \coeff^{(g-r)}_{m} \Big)+\coeff^{(g)}_{0}\ ,
\eeq
where $\coeff^{(1)}_m$ is defined in \eqref{def-F1coeff}. Note that $\coeff^{(g)}_{0}(\Delta,S,t)$ 
arises an integration constant of the $\widehat E_2$ integration and hence  
can be a function of the propagators $\Delta^a$ but not 
$\widehat E_2$. 

Let us now determine a second closed expression for $F^{(g)}$ by
integrating the second anomaly equation \eqref{directDelta}. Since $F^{(1)}$
is not of the form \eqref{Eexp} we first split off terms involving $F^{(1)}$.
Inserting the definitions of the propagators $\Delta^a$ and $\widehat E_2$ we find for $g>2$ that
\beq \label{delta-ano1}
  \frac{\partial F^{(g)} }{\partial \Delta^a} = 
  (D_S+ \tfrac{1}{2} \widehat E_2 )D_a  F^{(g-1)} -2C_{ac}\Delta^c D_S F^{(g-1)}
  +\sum_{r=2}^{g-2} D_a F^{(r)} D_S F^{(g-r)}.
\eeq
To make the dependence on the propagators $\Delta^a$ explicit we 
expand the covariant derivative $D_aF^{(g)}$. The covariant 
derivative $D_a$ can be split into a holomorphic derivative $\hat D_a$ defined in \eqref{def-hatD} plus 
a propagator expansion using \eqref{split_Dhol}. 
Moreover, using the chain rule one rewrites 
\beq \label{chain_dec}
    \hat D_a = \hat d_a + (\hat D_a \Delta^b) \partial_{\Delta^b}\ ,
\eeq
where $\hat d_a$ is the covariant holomorphic derivative not acting on the 
propagators, i.e.~we set 
\beq
  \hat d_a \big(\Delta^{a_1}\ldots \Delta^{a_n} \coeff_{a_1\ldots a_n}\big) =  \Delta^{a_1}\ldots \Delta^{a_n} \hat D_a \coeff_{a_1\ldots a_n}\ .
\eeq
Combining \eqref{split_Dhol}, \eqref{chain_dec} and \eqref{der_Delta} we immediately derive
\bea \label{dFa}
  D_a F^{(g)} &=& \Big[\hat d_a + \epsilon^4_{ac}  C^{cb} \partial_{\Delta^b} 
             +(2g-2) C_{ad}\Delta^d - \tfrac{1}{2}  \hat \Gamma^b_{cd|a} \Delta^c \Delta^d \partial_{\Delta^b}\Big]F^{(g)}\ .
\eea
This expansion makes the dependence of $D_a$ on the propagators $\Delta^a$ explicit. 
We note that the $\hat d_a$ term on the right-hand side of this expansion does not 
change the number of propagators. The second term lowers the number of propagators by one, 
while the two last terms raise the number of propagators by one.
Inspecting the holomorphic anomaly equation we note that only 
the first derivative along the fiber direction appears on the right-hand side of  \eqref{delta-ano1}. 
Hence, at least for the integration of  \eqref{delta-ano1} it will not be necessary to evaluate 
the second derivative $D_a D_b F^{(g)}$ as a propagator expansion.

To integrate  expressions such as \eqref{dFa} for $D_a F^{(g)}$ we also need to keep track of the number 
of propagators in the expansion of $F^{(g)}$. Therefore, we introduce the following 
short-hand notation
\beq \label{Fgincg}
  F^{(g)} = \sum_n \coeff^{(g)}_{\ (n)}\ , \qquad \qquad \coeff ^{(g)}_{\ (n)} = \sum_{k=0}^{g-1} \widehat E_2^k \Delta^{a_1} \ldots \Delta^{a_n} \coeff ^{(g)}_{k\,|\,a_1 \ldots a_n}\ ,
\eeq
where each $c^{(g)}_{\ (n)}$ contains $n$ propagators $\Delta^a$.
By counting the number of propagators one finds
\beq
  \int D_a F^{(g)} d\Delta^a =  \sum_n \Big\{ \tfrac{1}{n+1}  \Delta^a \hat d_a 
            + \tfrac{1}{n}\Delta^a \epsilon^4_{ac} C^{cb} \partial_{\Delta^b}     
            +  \tfrac{4g-4-n}{n+2} \Delta^2 \Big\}\, \coeff^{(g)}_{\ (n)}\ ,
\eeq
where as defined above $\Delta^2=\frac12 C_{ab}\Delta^a \Delta^b$.
This integral together  with similar ones for the remaining terms in \eqref{delta-ano1} yields 
a closed expression for $F^{(g)}$ of the form 
\bea \label{general-direct-delta}
F^{(g)} &=&  \big(D_S + \tfrac{1}{2}\widehat E_2) \sum_n \Big\{ \tfrac{1}{n+1} \Delta^a \hat d_a  + \tfrac{1}{n}\Delta^a \epsilon^4_{ac} C^{cb} \partial_{\Delta^b} 
           +  \tfrac{4g-8-n}{n+2} \Delta^2 \Big\}\coeff^{(g-1)}_{\ (n)}\nn \\
   && -  \sum_n \tfrac{4}{n+2} \Delta^2  D_S c^{(g-1)}_{\ (n)} 
         + \sum_{r=2}^{g-2} \sum_n \sum_{k+l=n} D_S \coeff^{(g-r)}_{\ (l)} \Big\{ \tfrac{1}{n+1} \Delta^a \hat d_a\nn \\
&&  + \tfrac{1}{n}\Delta^a \epsilon^4_{ac} C^{cb} \partial_{\Delta^b} 
             +  \tfrac{4r-4-n}{n+2} \Delta^2 
                \Big\} \coeff^{(r)}_{\ (k)} 
               +\coeff^{(g)}_{\ (0)}\ .
\eea
Here $\coeff^{(g)}_{\ (0)}(\widehat E_2,S,t)$ is the integration constant of the $\Delta^a$ integration and hence 
can depend on $\widehat E_2$ but not on $\Delta^a$.

Before turning to the discussion of an explicit example, let us consider the 
fiber limit of \eqref{general-direct-delta}. We therefore 
apply \eqref{E-der} and \eqref{Elimit} to show that
\beq  \label{DSF=0}
 \lim_{S,\bar S\rightarrow \infty}  D_S F^{(g)} =0\ .
\eeq
We also denote by $c^{(g)}_{E\,(k)}$ the fiber limit of the coefficients $c^{(g)}_{\ (k)}$ in \eqref{Fgincg}.
Inserting \eqref{DSF=0} into the formula \eqref{general-direct-delta} for direct integration along the fiber direction
one finds
\beq \label{general-direct-deltaE}
F^{(g)}_E = \tfrac{1}{2} \sum_n \Big( \tfrac{1}{n+1} \Delta^a \hat d_a  + \tfrac{1}{n}\Delta^a \epsilon^4_{ac} C^{cb} \partial_{\Delta^b} 
           +  \tfrac{4g-8-n}{n+2} \Delta^2 \Big)\coeff^{(g-1)}_{E\, (n)} + \coeff^{(g)}_{E\, (0)}\ ,
\eeq
where $c^{(g)}_{E\, (0)}(t)$ is a holomorphic ambiguity in the fiber. Recall that the full expression \eqref{Fgantihol} for
$F^{(g)}_E(t,\bar t)$ is known from heterotic-type II duality. Therefore, verifying that this closed expression 
fulfills the differential equation \eqref{general-direct-deltaE} provides a non-trivial 
check of our derivations.

Let us end this section by presenting the first non-trivial solution to the closed expressions \eqref{gencksol} and \eqref{general-direct-delta}
for $F^{(g)}$. More precisely, one derives that the free energy $F^{(3)}$ admits the 
following propagator expansion
\bea
  F^{(3)} &=& -\tfrac{1}{48}\hat E^2_2 \big( 14 \Delta^4 + 10 \epsilon^4_{ab} \Delta^a \Delta^b
    - \epsilon^4_{ac} \epsilon^4_{bd} C^{ab} C^{cd}   \big) \nn \\
                  &&     -\tfrac{1}{48} E_4  \big( - 2 \Delta^4+2 \epsilon^4_{ab} \Delta^a \Delta^b
                 -  \epsilon^4_{ac} \epsilon^4_{bd} C^{ab} C^{cd}  \big)\ ,
\eea
where $\epsilon^4_{ab}$ is defined in \eqref{def-epsilon4}.
Note that the last term in the first line has to be determined by the direct integration with respect to 
$\widehat E_2$ by using \eqref{gencksol}. Moreover, the purely holomorphic term 
\beq \label{def-f3full}
  f^{(3)}(S,t)= \tfrac{1}{48} E_4 \epsilon^4_{ac} \epsilon^4_{bd} C^{ab} C^{cd}
\eeq
is the holomorphic ambiguity at genus $3$, determined by the fiber limit. 
In other words, applying \eqref{Elimit} one easily derives 
\beq
  F^{(3)}_E = -\tfrac{1}{4} \epsilon^4_{ab} \Delta^a \Delta^b - \tfrac14 \Delta^4
    + \tfrac{1}{24} \epsilon^4_{ac} \epsilon^4_{bd} C^{ab} C^{cd}\ ,
\eeq
which is readily compared with the general expression \eqref{Fgantihol} for 
the fiber free energies. It is straightforward to derive all  $F^{(g)}$ for $g<7$ by
evaluating \eqref{gencksol} and \eqref{general-direct-delta} and fixing the ambiguity by comparison with the 
fiber result \eqref{Fgantihol}. Clearly,  at genus greater than $6$ we will encounter the same difficulties 
as in section \ref{simple}. Only additional boundary conditions can help to 
fix the ambiguities in these cases. In the next section we will summarize possible additional conditions.

\subsection{Boundary conditions}\label{boundaries}

One important feature of the formalism of direct integration is that modular  
and holomorphic properties of the $F^{(g)}$ are manifest. In particular the 
ambiguity is holomorphic, modular invariant and for given genus expressible 
in terms of a modular form of finite weight. This implies that a finite number of 
data will fix it. The latter must be provided from additional 
information at the boundaries of the moduli space of the Calabi-Yau manifold. Let 
us give a short overview over the the nature of these boundary conditions.  

In the large radius limit the holomorphic limit  of the 
$F^{(g)}$ has an expansion in terms of Gromov-Witten invariants $N^{(g)}_\beta$. 
Since the an-holomorphic part is fixed, the $F^{(g)}$ can be completely 
determined by calculating a finite number of Gromov-Witten invariants. 
The reorganisation of the expansion  in terms of Gopakumar-Vafa  
 invariants $n^{(g)}_\beta$ is useful here, because the latter 
vanish if the degree is higher then the maximal degree for which 
a smooth curve exists in a given class.                  

For K3-fibered Calabi-Yau threefolds, the limit of large 
base volume corresponds generically to a perturbative heterotic 
string theory on K3$\times \IT^2$. If the heterotic theory is known one
can calculate the dependence of the $F^{(g)}$ on the fiber moduli  
by calculating a BPS saturated one loop amplitude in the heterotic 
string~\cite{mm,km}. In the Enriques CY case this yields most 
of the information and is the reason that one can tackle an
$11$ parameter model at all. Even if the heterotic dual is not known, 
one may get all the holomorphic $F^{(g)}$ in the fiber from the 
modular properties of the B-model on the K3 and the formula 
for the cohomology of the Hilbert scheme of points on the 
fiber~\cite{Klemm:2004km}.

If the Calabi-Yau admits controllable local limits, e.g. 
to toric Fano varieties with anti-canonical 
bundle, then the $F^{(g)}$ can be 
unambiguously calculated using the topological 
vertex~\cite{vertex}.   

One can also find boundary conditions by looking at the behavior of the 
topological string amplitudes near the conifold point, as we discussed in section \ref{SWboundary}. When there is 
only one hypermultiplet becoming massless at the conifold point, the amplitudes behave like (\ref{thegap}), where 
$t_D$ is a suitable coordinate transverse to the conifold 
divisor. This yields $2g-2$ independent conditions 
on the holomorphic ambiguity.

In contrast to generic $\CN=2$ compactifications, the four dimensional massless 
spectrum at singularities of the Enriques Calabi-Yau is conformal, which 
requires hyper- and vector multiplets to become simultaneously massless. 
The leading behavior of the corresponding effective action is less 
characteristic. We will find a partial gap in the reduced model 
considered in section \ref{red-model}, which is similar to the partial
gap structures that were found in~\cite{hkq} at a point where likewise 
several RR states become massless. 
The determination of the subleading  behavior is possible in the field theory limit and 
yields conditions on the anomaly. We will consider here only the complex codimension 
singularities that we discussed in section \ref{sec:Enriques}. The nontrivial information about the 
$F^{(g)}$ comes from the $N_f=4$ locus: as we will show in section \ref{sec:ftlim}, 
the residue of the leading singularity near (\ref{slocusone}) can be computed using 
instanton counting in field theory. 

Let us now analyze the leading singularity of $\CF^{(g)}$ near the singular loci in the fiber limit. This 
can be done with the heterotic computations of \cite{km} reviewed in section \ref{sec:Enriques}. These 
computations give us expansions around two special regions in moduli space, the large radius limit 
(where $t^a$ are large) and the region appropriate to the BHM reduction (where $t_D^a$ are large). 
As in \cite{agnt,mm}, we can use the computation at large radius to obtain the leading behavior of 
the fiber amplitudes near (\ref{slocusone}), and the computation in the BHM reduction to obtain the 
behavior near (\ref{slocustwo}). 

Let us first look at the behavior near (\ref{slocusone}). A possible singular behavior there must come from the vector 
$r=(1, -1)$ in (\ref{gr}), since this leads to a polylogarithm which, when expanded at the singular locus (\ref{slocusone}),
\be
\label{polyexp}
{\rm Li}_{3-2g}(\re^{-z})={(2g-3)! \over z^{2g-2}} +\CO(z^0), \quad g\ge 2,
\ee
exhibits a pole. Here, $z=t^1-t^2$. However, since $c_g(-2)=0$, the coefficient of this polylogarithm vanishes and 
we conclude that the amplitudes are regular at (\ref{slocusone}). This is indeed 
consistent with the fact that the field theory limit of this model at (\ref{slocusone}) is massless 
$SU(2)$, $\CN=4$ super Yang--Mills theory, which has $\CF^{(g)}=0$ for all $g\ge 2$ \cite{Nek,no,bfmt}.

Let us now look at the behavior near (\ref{slocustwo}). To understand this, we look at the heterotic 
result for the holomorphic couplings in the BHM reduction (\ref{finalfgbor}). Again, 
the singular behavior comes from the vector $r=(1,-1)$. Since the coefficients are defined 
now by (\ref{finalmodcoef}), we find
\be
d_g(-1)={4^g-1 \over 2^{g-2}} {|B_{2g} | \over 2g (2g-2)!}.
\ee
If we set 
\be
\mu=t^1_D -t^2_D,
\ee
and we take into account the behavior of the polylogarithm (\ref{polyexp}), 
we find that the singular behavior of $\CF^{(g)}_E(t_D)$ near (\ref{slocustwo}) is given by
\be
\label{fgsingen}
\CF^{(g)}_{E} (t_D) \rightarrow {4^g-1 \over 2^{g-2}}   {|B_{2g}| \over 2g (2g-2)}{1\over  \mu^{2g-2}} + 
\CO(\mu^0) 
\ee
for $g\ge 2$, while for $g=1$ we have a logarithm singularity %
\be
-{1\over 2} \log \, \mu.
\ee
Since the full $\CF^{(g)}(S,t_D)$ can be written for $g\le 6$ in terms of (\ref{finalfgbor}), as we showed in 
section \ref{simple}, we can compute its leading singular behavior at 
the locus (\ref{slocustwo}). This will be useful in section \ref{sec:ftlim} to compare to the field theory limit. 
The above computation shows that along the fiber direction the topological 
string amplitudes $\CF_E^{(g)}$ show the gap behavior discovered in \cite{hk, hkq}. 
In order to see if the gap also holds in the mixed directions, it is clear from the formulae above 
that we need a precise knowledge of the regular terms in $\mu$ in the expansion of $\CF_E^{(g)}$. 
Unfortunately, this is something we cannot extract from the heterotic expressions. We will however
be able to do this in the reduced model introduced in \cite{km} and studied in more detail below. 
We will see that indeed the strong gap condition obtained for the fiber direction in (\ref{fgsingen}) 
does not hold for the mixed directions.

\subsection{The reduced Enriques model \label{red-model}}

In this section we discuss a reduced model for the Enriques Calabi-Yau introduced in \cite{km}.
The main advantage of this model is that the target symmetry group becomes much simpler, and 
one can easily parametrize the holomorphic functions which appear in the 
expansion of $F^{(g)}$ in the propagators $\Delta^a(t,\bar t)$ 
and $\widehat E_2(S,\bar S)$. In particular, the holomorphic ambiguity can be parametrized in terms of a finite number 
of coefficients at each genus. Also the mirror map is known explicitly and can be used to translate the 
$F^{(g)}$ into a simple polynomial form.
In these aspects, the reduced model is very closely related
to the Seiberg--Witten theory studied in section \ref{Seiberg-Witten}.

\subsubsection{Special geometry and the mirror map}

We begin with a brief discussion of the reduced special geometry and recall the mirror map 
derived in \cite{km}. Out of the eleven special coordinates $S,t^a$ the reduced model 
is only parametrized by three parameters. More precisely, it is
obtained by shrinking $8$ out of the $10$ cycles in the Enriques fiber as
\beq \label{reductionlimit}
   (S,t^a) = (S,t^i ,t^\alpha) \quad \rightarrow \quad (S,t^i,0)\ , \qquad i=1,2\ , \quad \alpha=3,\ldots, 10\ .
\eeq
We denote the reduced moduli space spanned by the remaining coordinates $S,t^1,t^2$ by $\cM_{\rm r}$. 
Explicitly, the full coset \eqref{coset} reduces in this limit to 
\beq \label{cosetr}
   \cM_{\rm r} = \frac{Sl(2,\bbR)}{SO(2)}  \times \left( \frac{Sl(2,\bbR)}{SO(2)} \right)^2\ ,
\eeq 
inducing a split of the full target space symmetry group as
\beq \label{group-reduction}
  Sl(2,\bbZ) \times O(10,2,\bbZ) \quad \rightarrow\quad Sl(2,\bbZ) \times \Gamma(2) \times \Gamma(2)\ .
\eeq
The generators of $Sl(2,\bbZ)$ are precisely the Eisenstein series $\widehat E_2(S,\bar S),\ E_4(S),\ E_6(S)$
as already introduced for the full model in \eqref{genS}. The generators for $\Gamma(2)$ have been introduced 
in the Seiberg-Witten section \ref{Seiberg-Witten}. More precisely, we will generate the ring of almost holomorphic 
modular forms of $\Gamma(2)$ by $\widehat E_2(t,\bar t)$, $K_2(t)$ and $K_4(t)$ explicitly defined in \eqref{def-widhatE2} and \eqref{gamma2generators}.
In the following we will simplify expressions by abbreviating
\begin{align} \label{def-tildeK}
  E_2 &= E_2(t^1)\ , \qquad & K_2&= K_2(t^1)\ , \qquad &  K_4&= K_4(t^1)\ ,\nn \\
  \tilde E_2 & = E_2(t^2)\ , \qquad & \tilde K_2&= K_2(t^2)\ , \qquad& \tilde K_4&= K_4(t^2)\ .
\end{align}
Whenever not stated otherwise, we will keep the $S$-dependence explicit.
Let us also note that the matrix $C_{ab}$ splits as
\beq
   C_{ab} = \left(\begin{array}{cc} C_{ij}& 0\\ 0 & C_{\alpha \beta} \end{array} \right)\ , \qquad C_{ij} =\left(\begin{array}{cc} 0& 1\\ 1 & 0 \end{array} \right)\ ,
\eeq
as already given in \eqref{def-Gamma_E}.
Hence, the holomorphic prepotential \eqref{def-Eprepot} and the fiber K\"ahler potential $Y=(t+\bar t)^2$ reduce to
\beq \label{red-prepot}
  \cF_{\rm r}= iS t^1 t^2\ ,\qquad \qquad Y_{\rm r} = (t^1+\bar t^1)(t^2+\bar t^2)\ .
\eeq
As we have already noted in section \ref{specialE} 
this prepotential and fiber potential are exact and receive no 
instanton corrections.

Let us now turn to a discussion of the mirror map for the reduced Enriques model.
In order to determine this duality map we first note that the reduced Enriques has 
an algebraic realization. Applying standard techniques, one can thus 
derive the three Picard-Fuchs equations for the holomorphic three-form
$\Omega(z)$ as
\beq \label{PF-eqn}
  \cL_1 \Omega(z) =0\ ,\qquad \quad \cL_2 \Omega(z)=0 \ , \qquad \quad \cL_3 \Omega(z)= 0 \ ,
\eeq
where $z^i(t),z^3(S)$ with $i=1,2$ are the mirror coordinates of $t^i,S$ respectively.
The Picard-Fuchs operators are found to be 
\bea
  \cL_i &=& \theta^2_i - 4(4\theta_i +4 \theta_j - 3)(4\theta_i + 4\theta_j-1)z_i\ ,\qquad \quad i,j=1,2\ , \quad i\neq j\ , \\
  \cL_3 &=& 36 (z^3-1)^2  (z^3-2) \theta_3^2 + 36 z^3 (z^3-1) \theta_3 + z^3 \big(8z^3-4(z^3)^2-31\big) \ ,
\eea
where $\theta_i=z^i \frac{\partial }{\partial z^i}$.
The Picard-Fuchs equations \eqref{PF-eqn} can be solved to determine the mirror maps $z^i(t),z^3(S)$. 
This was done in \cite{km} and we will only quote the result here. We first abbreviate 
\beq
   \label{def-zonly}
   z(q_i)= \frac{K_4(t^i)}{ K_2^2(t^i)}\ .
\eeq
Using this shorthand notation the fiber mirror map reads 
\begin{equation} \label{def-mirrorz}
 z^1(t)=z(q_1) \big(1- z(q_2)  \big)\ ,\qquad \qquad 
  z^2(t)= z(q_2)  \big(1- z(q_1)  \big)\ . 
\end{equation}
These coordinates are related by a factor of $64$ to $z_1,z_2$ used in ref.~\cite{km}.
In the base one evaluates
\beq
  z^3(S)= 1-E_4^{-3/2} E_6\ . 
\eeq
Compared to \cite{km} we have rescaled $z^3$ by a factor $864$. Using these
explicit expressions for $z^1,z^2$ and $z^3$, one immediately verifies 
their invariance under the target space symmetry group $Sl(2,\bbZ) \times \Gamma(2) \times \Gamma(2)$.
Also the fundamental period $X^0$ can be obtained from the Picard-Fuchs system 
\eqref{PF-eqn} and reads 
\beq \label{def-X0}
  X^0  = x^0 \hat X^0\ ,\qquad \qquad (\hat X^0)^2 = \tfrac{1}{4} K_2 \tilde K_2\ ,\qquad \qquad (x^0)^4 = E_4\ .
\eeq
We immediately verify that $X^0$ is not invariant under  
the symmetry group $Sl(2,\bbZ) \times \Gamma(2) \times \Gamma(2)$. 
The S-duality transformation \eqref{saut} reads for the reduced model 
$t^1  \mapsto 1/t^2$ and $t^2  \mapsto 1/t^1$. 
Applied to $X^0$ this yields precisely the transformation behavior given in \eqref{trans-X0}.
Before turning to the higher genus amplitudes in the next section let us also
note that the discriminant of the reduced model 
is given by
\beq
   \Delta(z^1,z^2)\ D(z^3)\ , 
\eeq 
where $\Delta(z^1,z^2)$ is the discriminant along the fiber and 
$D(z^3)$ is the discriminant along the base. Explicitly, we find in the 
coordinates \eqref{def-zonly} and \eqref{def-mirrorz} that 
\bea \label{def-disc1}
    \Delta(z^1,z^2) &=& \big(1-z(q_1)-z(q_2)\big)^2  \\ 
                &=&1- 2(z^1 + z^2 + z^1 z^2) + (z^1)^2 + (z^2)^2\ .
\eea 
The second discriminant $D(z^3)$ is given by
\beq \label{def-disc2}
 D(z^3)=\tfrac{1}{2^6 3^3} \big( (z^3)^2-z^3\big) \ .
\eeq
In the next section we will use the mirror coordinates $z^1,z^2$ to express the 
reduced free energies $F^{(g)}_r$. Since along the base direction all equations are 
expressed in terms of simple Eisenstein series $E_{2n}(S)$ we choose to keep this 
$S$-parametrization also in the following discussions.

\subsubsection{Reduced free energies and direct integration}

Let us now discuss the free energies $F^{(g)}_r$  and their holomorphic limits $\cF^{(g)}_r$ 
for the reduced model. In the limit \eqref{reductionlimit} they are simply defined as
\be
F^{(g)}_r(S,t^1, t^2) = F^{(g)}(S,t^1, t^2, t^{\alpha}=0)\ .
\ee
The reduced form of $F^{(1)}$ can be derived by direct computation as was already discussed in \cite{km}. 
Explicitly one finds 
\beq
   F^{(1)}_{\rm r} = - 2\log\big[(S+\bar S)^3 (t^1+\bar t^1)(t^2 +\bar t^2)\big] - \log |\Phi_{\rm r}(S,t)| \ ,
\eeq
where
\beq \label{def-Phir}
    \Phi_{\rm r}(S,t^1,t^2) = \eta^{24}(S) \prod_{m,n} \Big(\frac{1-q^n q^m}{ 1+q^n q^m} \Big)^{c^{\rm r}_1(2mn)}\ .
\eeq
The coefficients $c_1^r(n)$ are given through the modular form 
\beq \label{reduced_c}
  \sum_n c^{\rm r}_1(n) q^n = - \frac{64}{3 \eta^6(q) \vartheta^6_2(q)} E_2(q) E_4(q^2)\ .
\eeq
Note that in comparison with the expression \eqref{geomrmod} for the full Enriques model the 
Eisenstein series $E_4(q^2)$ appears in \eqref{reduced_c}. This extra factor arises due to the summation 
over the $E_8$ vectors in \eqref{def-Phi} and precisely counts their degeneracy.
It was further shown in \cite{km} that the following denominator formula holds
\beq
     \Phi_{\rm r}(S,t^1,t^2) = \tfrac{1}{16} \eta^{24}(S)\ \delta =  \eta^{24}(S) (\hat X^0)^{4} \Delta^{1/2}
\eeq
where 
\beq
   \delta(t^1,t^2) = K^2_2 \tilde K^2_2 - K_4 \tilde K_2^2 - K_2^2 \tilde K_4\ .
\eeq
Here the $\Gamma(2)$ generators $K_2,\tilde K_2 $ as well as $K_4,\tilde K_4$ are defined in \eqref{def-tildeK}, while 
the fundamental period $\hat X^0$ and the discriminant $\Delta$ were given in \eqref{def-X0} and \eqref{def-disc1}.

The holomorphic reduced amplitudes restricted to the Enriques fiber can also be 
computed directly by reducing the heterotic expressions (\ref{gr}) and (\ref{finalfgbor}).
The result reads \cite{km}
\be
\label{red}
\ba
  \cF^{(g)}_{{\rm r},E} (t) &= \sum_{r>0} c^{\rm r}_g(r^2) \Big[2^{3-2g} \text{Li}_{3-2g}(e^{-r\cdot t}) -  \text{Li}_{3-2g}(e^{-2r\cdot t}) \Big]\ , \\
  \CF^{(g)}_{{\rm r},E} (t_D)&=\sum_{r>0} d^{\rm r}_g(r^2/2)(-1)^{n+m} {\rm Li}_{3-2g}({\rm e}^{-r \cdot t_D })\ ,
  \ea
\ee
where the coefficients $c^{\rm r}(n)$, $d^{\rm r}_g(n)$ are defined by
\be
\label{redco}
\ba
\sum_n c^{\rm r}_g(n) q^n& =-2 E_4(q^2) \frac{{\cal P}_{g}(q)}{\eta^{12}(2\tau)}\ ,\\
\sum_n d^{\rm r}_g (n) q^n &=E_4(q^2) \frac{2^{2+g} {\cal P}_{g}(q^4)-2^{2-g} {\cal P}_{g}(q)}{\eta^{12}(2\tau)}\ .
\ea
\ee
Once again we recognize the additional factor $E_4(q^2)$ counting the degeneracies of the $E_8$ 
lattice. Clearly, also the expressions $\cF^{(g)}_{{\rm r},E}(t)$ and $\CF^{(g)}_{{\rm r},E} (t_D)$
can be expressed in terms of the holomorphic generators \eqref{def-tildeK} depending on $t^i$ and $t^i_D$ 
respectively. 

Let us now turn to the discussion of the complete reduced amplitudes including the base and 
the non-holomorphic dependence. In order to do that we describe the direct integration 
for the reduced model focusing on the essential differences to the considerations 
presented in section \ref{direct_fiber_base}.
To begin with, note that the 
propagators of the full model reduce as
\beq \label{def-rprop}
   \Delta^i\quad \rightarrow \quad \rprop^i  \ ,\qquad \qquad
   \Delta^\alpha\quad \rightarrow \quad 0\ ,
\eeq
where $\rprop^i$ is obtained from \eqref{def-Deltaa} by setting $t^\alpha=0$ and using \eqref{def-Phir}.
That $ \Delta^\alpha$ reduces to zero arises from the fact that in
summation over the $E_8$ lattice the vectors cancel pairwise.
In order to perform the direct integration we first have to 
find recursive relations which are valid for the reduced free energies 
$F^{(g)}_{\rm r}$.
Recall that in the full Enriques model we found two sorts of 
recursive relations \eqref{directE} and \eqref{directDelta} capturing the properties $F^{(g)}$
in the base and in the fiber of the Enriques. It turns out that only the second anomaly equation  
 \eqref{directDelta}  admits a simple reduction. More precisely, it can be rewritten for the 
 reduced model as
 \beq \label{recrel_reduced}
   \frac{\partial F^{(g)}_{\rm r} }{\partial \rprop^i}= 
  D_S  D_i F^{(g-1)}_{\rm r} + \sum_{r=1}^{g-1} D_i F^{(r)}_{\rm r} D_S F^{(g-r)}_{\rm r} \ ,
\eeq
since performing the reduction $t^\alpha=0$ interchanges with the differentiation with respect 
to $t^1,t^2$. Note that this is no longer true for derivatives with respect to $t^\alpha$. In particular,
the first equation \eqref{directE} involves a summation over the $\alpha$ indices and one shows 
that the resulting terms do not vanish under the reduction $t^\alpha=0$.
Nevertheless, one can directly integrate \eqref{recrel_reduced} for the reduced free energies
\beq \label{small_propexp}
    F^{(g)}_{\rm r} = \sum_{n=1}^{2g-2} \rprop^{i_1} \ldots \rprop^{i_n} \hat c^{(g)}_{i_1 \ldots i_n} + \hat c^{(g)}\ ,\qquad \quad g>1\ .
\eeq
The function $ \hat c^{(g)}$ is holomorphic in $t^i$ and generally depends on $\widehat E_2(S,\bar S),E_4(S),E_6(S)$. 
Note that due to \eqref{def-rprop} the coefficients of the full and reduced model 
are related by $\hat c^{(g)}_{i_1 \ldots i_n}=c^{(g)}_{i_1 \ldots i_n}(t^\alpha=0)$.
The direct integration is performed in analogy to the integration in the full model 
and results in a closed expression similar to \eqref{general-direct-delta}.
The important difference is that the $\epsilon^4_{\text{r}\, ij}$ as well as the covariant  
derivatives $\hat D^{\rm r}_a$ are not obtained from the full $\epsilon^4_{ab}$ and $\hat D_a$ by simply restricting to 
the $i,j$ indices and setting $t^\alpha=0$. Both $\epsilon^4_{\text{r}\, ij}$
as well as $\hat D^{\rm r}_a$ have to be defined with respect to a new holomorphic 
metric $\hat C_{ij}^{\rm r}=\Phi^{1/2}_{\rm r} C_{ij}$ but otherwise analog to \eqref{def-epsilon4} and \eqref{def-hatD}.
If one had been using the old connection, an additional summation over the $\alpha$ indices would arise and 
yield extra contributions. Applied to the specific free energy $F^{(3)}$ one finds the reduction of 
the holomorphic ambiguity \eqref{def-f3full}
\beq
     f^{(3)}_{\rm r}(S,t)= \tfrac{1}{48} E_4 \big( \epsilon^4_{\text{r}\, ik} \epsilon^4_{\text{r}\, jl} C^{ij} C^{kl} +\tfrac18 (\epsilon^4_{\text{r}\, ij} C^{ij})^2 \big)
\eeq
After these considerations it is not surprising that the contraction of the new $\epsilon^4_{\text{r}\, ij} $ with 
$C^{ij}$ does not vanish as it is the case in the full model \eqref{Cepsilon4=0}.

\subsubsection{The free energies $F^{(g)}$ on the mirror}

So far the reduced free energies $F^{(g)}_{\rm r}$ were expressed as
functions of the variables $t^i,S$ or $t_D^i,S$.
In the reduced model we know the mirror map explicitly and thus will 
be able to translate the expansion  \eqref{small_propexp} 
of $F^{(g)}_{\rm r} $ into a function of the complex coordinates $z^i$. 
We will show that the holomorphic coefficients become polynomials in 
$z^i$ divided by an appropriate power of the discriminant.
Since the dependence of $F^{(g)}_{\rm r}$ is rather transparent we 
chose to keep this variable and do not replace it by its mirror 
counterpart $z^3$. 

The $F^{(g)}$ transform non-trivially 
under the reduced automorphic transformations. We already discussed 
the actually invariant combination in \eqref{inv_comb}. In the coordinates 
$z^i,S$ we thus set 
\beq
    F^{(g)} (z,\bar z,S,\bar S) = (\hat X^0)^{2-2g}\, F^{(g)}(t,\bar t,S,\bar S) \ .
\eeq
This definition is consistent with the fact that the $z^i(t)$ are invariant 
under the target space group \eqref{group-reduction}, while $(\hat X^0)^{2g-2}$ transforms exactly as 
$F^{(g)}(t,S)$. To rewrite the expansion  \eqref{small_propexp} we first note that 
the propagator $\rprop^i$ can be written in the $z^i$ coordinates as
\beq \label{trans-rprop}
    \rprop^i = (\hat X^0)^2 \frac{\partial t^i}{\partial z^j}\rprop^{z^j}\ , \qquad \qquad \rprop^{z^i} = -C^{z^i z^j}\big( \hat K_{z^j} -\tfrac18 \partial_{z^j} \log \Delta \big)\ ,
\eeq
where $\rprop^{z^i}$ is a function of $z^i,\bar z^i$ and we have used 
\beq \label{trans-CC}
  C_{ij}= (\hat X^0)^{-2} C_{z^k z^l} \frac{\partial z^k}{\partial t^i} \frac{\partial z^l}{\partial t^j}\  ,\qquad \quad
  \hat K(z,\bar z) \equiv - \log\big[|\hat X^0|^2 Y_{\rm r}(z,\bar z) \big]  . 
\eeq
It is not hard to use the expressions \eqref{def-mirrorz} for $z^1$ and $z^2$ to evaluate $C_{z^i z^j}$ explicitly as
\beq
  C_{z^1 z^2} = \frac{1}{z^1 z^2 \Delta}(1-z^1-z^2) \ , \qquad C_{z^1 z^1} =  \frac{1}{z^1 z^2 \Delta} 2 z^2\ , \qquad C_{z^2 z^2} = \frac{1}{z^1 z^2\Delta} 2z^1\ . 
\eeq
Once again \eqref{trans-rprop} and \eqref{trans-CC} are in accordance with the transformation behavior of the 
$\rprop^i$ and $C_{ij}$ given in \eqref{trans-Delta} and \eqref{jcont}.
Similarly, we transform the coefficients $\hat c^{(g)}_{i_1\ldots i_n}$ in \eqref{small_propexp} and set 
\beq \label{coeffinz}
    \hat c^{(g)}_{i_1 \ldots i_n } = (\hat X^0)^{2g-2 - 2n} \
    \frac{\partial z^{j_1}}{\partial t^{i_1}}\ldots  \frac{\partial z^{j_n}}{\partial t^{i_n}}\ \hat c^{(g)}_{z^{j_1} \ldots z^{j_n}}(z)\ , 
\eeq
which is consistent with \eqref{trans-k}. It is also straightforward to rewrite the 
direct integration expression for $F^{(g)}_{\rm r}$
by using the $z^i$ coordinates. Let us once again only discuss the appearing
building blocks. We begin by noting that the holomorphic covariant derivative transforms as
\beq
  \hat D_i V_j = (\hat X^0)^{2k} \frac{\partial z^l}{\partial t^i} \frac{\partial z^m}{\partial t^j}  \hat D_{z^l} V_{z^m}\ ,
\eeq 
where the covariant derivative $\hat D_{z^i}$ is given by 
\beq
   \hat D_{z^i} V_{z^j} = \partial_{z^i} V_{z^j} - \tfrac{k}8 (\partial_{z^i}\log \Delta) V_{z^j}+ \hat \Gamma^{z^l}_{z^i z^j} V_{z^l}\ .
\eeq
The holomorphic Christoffel symbol in this expression is defined by 
\beq
   \hat  \Gamma^{z^l}_{z^i z^j} =\tfrac12 \hat C^{z^l z^m} \big(\partial_{z^i} \hat C_{z^m z^j}+\partial_{z^j} \hat C_{z^i z^m}- \partial_{z^m} \hat C_{z^i z^j} \big)\ , \qquad \hat C_{z^i z^j} = \Delta^{1/4} C_{z^i z^j}\ .
\eeq
The second important object in the general equation for the direct integration 
is the automorphic form $\epsilon^4_{\text{r}\, ij}$. One shows that it can be decomposed as 
\beq \label{def-eps4z}
  \epsilon^4_{\text{r}\, ij} =   \frac{1}{z^1 z^2 \Delta^2}\frac{\partial z^l}{\partial t^i} \frac{\partial z^m}{\partial t^j}\ \epsilon^4_{z^l z^m} \ .
\eeq
where for $i=1,2$, $i\neq j$ one finds that
\bea
  \epsilon^4_{z^i z^i}&=&- \tfrac1{16} z^j\,\big( (z^i)^2\,\left( 1 + 3\, z^j \right)  + 
        {\left( -1 +z^j \right) }^2\,\left( 1 + 3\, z^j \right)  - 
        2\, z^i\,\left( 1 - 5\, z^j + 3\, (z^j)^2 \right) \big)\ ,  \nn\\[.2cm]
  \epsilon^4_{z^i z^j}&=&    \tfrac{3}{16}\,z^i\,z^j\,\left( -2 + z^i + (z^i)^2 + z^j + (z^j)^2- 
      2\,z^i\,z^j \right) \ .
\eea
Note that $\epsilon^4_{z^i z^i}$ is polynomial due to the fact that we extracted the denominator $z^1 z^2 \Delta^2$
in \eqref{def-eps4z}. This turns out to be possible for all the coefficients $ \hat c^{(g)}_{z^{i_1} \ldots z^{i_n}}$
appearing in \eqref{coeffinz}. We thus define
\beq
   P^{(g)}_{i_1 \ldots i_n}(z,\widehat E_2,E_4,E_6) = (z^1 z^2 \Delta)^{g-1}\  \hat c^{(g)}_{z^{i_1} \ldots z^{i_n}}\ ,
\eeq
where $P^{(g)}$ are polynomials in $z^i$ as well as $\widehat E_2,E_4,E_6$.
The reduced free energies are thus of the form
\beq
  F^{(g)}_{\rm r}(z,\bar z,S,\bar S) =\frac{1}{ (z^1 z^2 \Delta)^{g-1} } \sum_{n} \rprop^{z^{i_1}} \ldots \rprop^{z^{i_n}}  P^{(g)}_{i_1 \ldots i_n} \ ,\qquad g>1\ .  
\eeq
In particular, this implies that at each genus the holomorphic ambiguity is
parametrized by a polynomial $P^{(g)}(z,E_4,E_6)$ holomorphic 
in $z^i$ and $S$. As it was the case before the coefficients in $P^{(g)}$ 
have to be determined by boundary conditions. For the lower genera this 
can be done explicitly by using the fiber limes. At higher genus 
additional information are needed and we will discuss in the next section 
the possible input from a small gap condition. We believe that essentially all
results on the mirror rewriting can be generalized to the full model in case one 
is able to determine the full mirror map. For the ten parameters along the fiber 
this is however a technically challenging task.

\subsubsection{Boundary conditions and the small gap}
As we have seen in (\ref{group-reduction}), the automorphism group acting on 
the fiber variables is simply 
\be
\Gamma(2) \times \Gamma(2)\ ,
\ee
where these groups act on $t^{1,2}$, respectively,
plus the exchange $t^1 \leftrightarrow t^2$. Moreover, we see from (\ref{projtrans}) that the $\{t^i=0: i=3, \cdots, 10\}$ locus 
maps to the 
$\{t^i_D=0: i=3, \cdots, 10\}$ locus. If we now define 
\be
2\pi \ri \tau^{1,2} =-t^{1,2}, \quad 2\pi \ri \tau_D^{1,2}= -t_D^{1,2} .
\ee
we see that the transformation (\ref{projtrans}) relating the geometric and the BHM expressions 
reduces to 
\be
\label{strans}
\tau^1_D=\tau^1, \quad  \tau_D^2 = -{1\over 2} {1\over \tau^2}.
\ee
By using the explicit expressions for $ \cF^{(g)}_{{\rm r},E} (t)$ in terms of modular forms 
(which can be obtained for example by direct integration), one finds that under (\ref{strans}) 
\be
 \cF^{(g)}_{{\rm r},E} (t) \rightarrow 2^{1-g}  \cF^{(g)}_{{\rm r},E} (t_D),
 \ee
where the factor of $2$ is inherited from the factor of $2$ in (\ref{strans}) and $\cF^{(g)}_{{\rm r},E} (t_D)$ are also given in (\ref{red}). 
Therefore, one can obtain expressions for the amplitudes in the BHM reduction 
in terms of modular forms by simply applying the transformation (\ref{strans}) to the results of the direct integration in the reduced model (which are 
valid for the geometric reduction). 

These expressions for the BHM amplitudes can also be used to study in detail the behavior near the singularity (\ref{slocustwo}), and in particular to 
calculate the subleading terms. One can verify that the discriminant (\ref{def-disc1}) transforms under (\ref{strans}) 
as 
\be
\Delta(t^1, t^2)\quad \mapsto \quad \Delta(t^1_D, t^2_D)= (z(q_D^1)-z(q_D^2))^2,
\ee
which vanishes at the locus (\ref{slocustwo}). This leads to the singular behavior of $\cF^{(g)}_{{\rm r}} (t_D)$, and 
one can now verify the behavior (\ref{fgsingen}) by expanding the expressions in terms of modular forms. One finds, 
\be
\ba
\CF_{{\rm r}, E}^{(1)}(t_D)&=-{1\over 2}\log(\mu)-{1\over 2}\log\Big[\frac{1}{128} K_2K_4(K_2^2-K_4)(q_D^2) \Big]+\CO(\mu), \\
\CF_{{\rm r}, E}^{(2)}(t_D)&={1\over 16\mu^2}-{80E_2^2K_2^2-16K_2^4+3K_2^2K_4+9K_4^2+16E_2(K_2^2+3K_2K_4)\over 9216 K_2^2}(q_D^2)+\CO(\mu),\\
\CF_{{\rm r}, E}^{(3)}(t_D)&={1\over 32\mu^4}+{1\over 53084160 K_2^4}\left(-800E_2^4K_2^4+214K_2^8-726K_2^6K_4+1431K_2^4K_4^2\right.\\
&+405K_4^4-320E_2^3(K_2^5+3K_2^3K_4)+120E_2^2(10K_2^6-15K_2^4K_4+9K_2^2K_4^2)\\
&\left.-540K_2^2K_4-40E_2(14K_2^7-54K_2^5K_4+27K_2^3K_4^2-27K-2K_4^3)\right)(q_D^2)+O(\mu).\\[.2cm]
\ea
\ee
However, if one includes the base directions, the gap is ``partially filled'' starting at genus three (for $\CF^{(2)}(S, t_D)$, the gap property away from the fiber limit is trivially satisfied). 
Indeed, one finds that the term $C^{ab}\partial_a\CF^{(1)}_E(t_D) \partial_b \CF^{(2)}_E (t_D)$, leads, in the reduced model, to the expansion
\be
\ba
&\partial_1\CF^{(1)}_{{\rm r},E} (t_D) \partial_2 \CF^{(2)}_{{\rm r},E} (t_D) + \partial_2\CF^{(1)}_{{\rm r},E} (t_D) \partial_1 \CF^{(2)}_{{\rm r},E} (t_D)=\\
& -{1\over 8\mu^4}-{20E_2^2K_2+17K_2^3+3K_2K_4+4E_2(K_2^2+3K_4)\over 9216 K_2}(q_D^2){1\over \mu^2}+\cdots \ea
\ee
Although there are some nontrivial cancellations (for example, there is no term in $\mu^{-3}$), 
generically one finds, for finite $S$, singular terms in $\mu$ beyond the leading one.


\section{The field theory limit }\label{sec:ftlim}

As we reviewed in section \ref{sec:Enriques}, there is a line of enhanced symmetry in the moduli space of the Enriques Calabi--Yau 
which leads in the field theory limit to $SU(2)$, $\CN=2$ QCD with four massless hypermultiplets. This occurs at the locus (\ref{slocustwo}). 
Similarly to what happens for other K3 fibrations \cite{kklmv}, we expect that near this locus the leading 
singularities of the topological string partition functions become field theory amplitudes of the $N_f=4$ theory. 
At genus zero one should recover the 
prepotential, and at higher genus the gravitational amplitudes introduced by Nekrasov in \cite{Nek} by using instanton counting techniques. In this section 
we will explain this in some detail, and as spin-off we will obtain some new results on the modularity properties of the $N_f=4$ theory and its gravitational 
corrections. 

We first note that the behavior of the amplitudes near (\ref{slocusone}), in the fiber limit, has been already determined with heterotic techniques in (\ref{fgsingen}). 
The results of section \ref{sec:diE} including the base were obtained in principle in the large radius limit, in terms of the ``electric" coordinates $t$. 
However, the calculations of $F^{(g)}$ performed there are also valid in the $t_D$ coordinates, due to general covariance. 
In particular, the holomorphic limit $\CF^{(g)}(S,t_D)$ can be expanded in polynomials in $E_2(S), E_4(S), E_6(S)$ as explained before (\ref{genS}), and we 
can write
\be
\CF^{(g)}(S,t_D)=\sum_k p^g_k (S) f^g_k(t_D). 
\ee
Near the locus (\ref{slocustwo}) the $f^g_k$ should show display a singular behavior of the form 
\be
f^g_k (t_D) = {b_k^g \over \mu^{2g-2}} + \cdots,
\ee
as we checked in the fiber limit in (\ref{fgsingen}). How does this compare to the field theory?

The prepotential and gravitational corrections of the 
massless $N_f=4$, $SU(2)$ $\CN=2$ Yang--Mills theory depend on the vector multiplet variable $a$ and on the 
microscopic coupling $\tau_0$. They can be put together into a generating functional 
\be
\label{genfield}
 \CF^{\rm YM}(a, \tau_0, \hbar) =\sum_{g=0}^{\infty} \hbar^{2g} \CF_g^{\rm YM}(a, \tau_0),
\ee
 where $\CF_0^{\rm YM}(a, \tau_0)$ is the $\CN=2$ prepotential and the higher $g$ amplitudes are the gravitational corrections. 
The statement that the type II theory on the Enriques Calabi--Yau has this gauge theory as its field theory limit near the locus (\ref{slocustwo}) implies that 
the leading singularity of the topological string amplitudes is given by
\be
\label{fscomp}
\CF^{(g)}(S,t_D) \rightarrow {1\over \mu^{2g-2}} \sum_k b^g_k p^g_k(S) =\CF_g^{\rm YM}(a, \tau), 
\ee
where $S$ is related to the coupling constant of the theory $\tau_0$, and 
$\mu$ is related to the $a$ variable of Seiberg and Witten in a way that we will make precise in a moment.
Let us first look at the prepotential. 
While it has been originally assumed \cite{sw} that the prepotential of the self-dual theories with $\mathcal{N}=2$, gauge group $SU(N)$ and $2N$ flavors 
is classically exact, it was found in \cite{dkm1} that it does get instanton corrections.
Those can however be absorbed in the following redefinition of the coupling \cite{dkm2}, 
\be
\tau_0\rightarrow\tau ={1\over 2} {\partial^2\over\partial a^2}\CF^{\rm YM}_0(a, \tau_0)=\tau_0+\sum_{k}c_k q_0^k,
\ee
where
\be
q_0=\exp(2 \pi i \tau_0).
\ee
We then have for the instanton-corrected prepotential 
\be
\CF^{\rm YM}_0(a, \tau_0)={1\over 2} \tau a^2,
\ee
in terms of the renormalized coupling $\tau$.
This is needed in order to match the type II prepotential (\ref{def-Eprepot}), which does not exhibit instanton corrections. 
We will then express the $\CF^{\rm YM}_g$ obtained by instanton computations 
not as functions of $q_0$, but of $q=\re^{2 \pi i \tau}$. 

\begin{figure}[!ht]
\leavevmode
\begin{center}
\includegraphics[height=6.5cm]{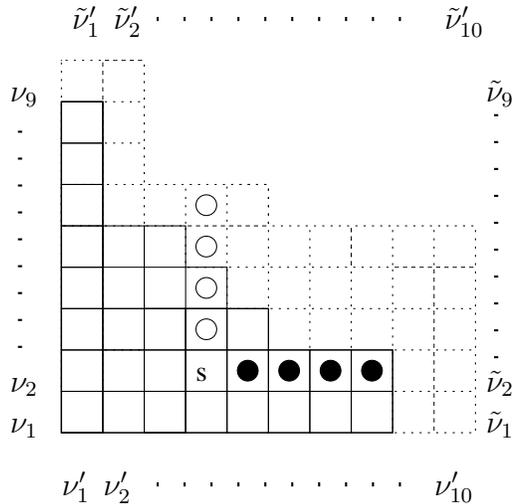}
\end{center}
\caption{A sample pair of Young diagrams $Y_\lambda,Y_{\tilde{\lambda}}$ contributing to \eqref{ZNek}.}
\label{Young}
\end{figure}

The computation of the field theory amplitudes proceeds as follows. The functional (\ref{genfield}) has the structure
\be\label{gf}
 \CF^{\rm YM}(a, \tau_0,  \hbar)=\CF^{\rm YM}_{\rm pert}(a, \hbar) -\hbar ^2 \log Z(a,\tau_0, \hbar), 
\ee
where 
\be
\label{pertfield}
\CF^{\rm YM}_{{\rm pert},g}(a, \hbar)= -{2B_{2g}\over 4^{(g-1)}2g(2g-2)}(1-4^g){1\over a^{2g-2}}
\ee
is the perturbative piece computed in \cite{no}, and 
\be
Z(a,\tau_0, \hbar)=\sum_k Z_k(a,\hbar) q_0^k
\ee
is an instanton sum. Nekrasov's formula for the $k$-instanton contribution to the partition sum $Z_k(a,\hbar)$ can be written as \cite{bfmt}
\be\label{ZNek}
Z_k(a,\hbar)=\sum_{\{Y_\lambda\}}\prod_{\lambda}^{N}\prod_{s\in Y_\lambda}{\varphi_\lambda(s)^4\over \prod_{\tilde{\lambda}} E(s)^2}.
\ee
The sum runs over sets $\{Y_\lambda\}$ of Young diagrams labeled in the $SU(2)$ case by $\lambda=1,2$.
For massless flavors,
\be
\varphi_\lambda(s)=a_{\lambda}-(s_j-s_i)\hbar,
\ee
where $s_i,s_j$ are the coordinates of the cell s inside the Young diagram $Y_\lambda$. We also have
\be
E(s)=a_{\lambda\tilde{\lambda}}-\hbar(h(s)-v(s)-1),\qquad h(s)=\nu_{s_i}-s_j,\qquad v(s)=\tilde{\nu}'_{s_j}-s_i,
\ee
where $\nu_{s_i}$ is the length of row $s_i$ in $Y_\lambda$, $\tilde{\nu}'_{s_j}$ the length of column $s_j$ in $Y_{\tilde{\lambda}}$ and $h(s),v(s)$ are the number of boxes to the right of s inside $Y_{\lambda}$ respectively above s inside $Y_{\tilde{\lambda}}$, see \figref{Young}. The constants $a_\lambda=(a_1,a_2)$ are set to $(-a,a)$.

The relative normalizations between the results in \cite{Nek} and 
the Calabi--Yau case can be obtained from the limit $q \rightarrow 0$, which is the limit $S\rightarrow \infty$. The 
only remaining singularity on the Enriques is then (\ref{fgsingen}), while in the Yang--Mills case we are left with the perturbative piece 
(\ref{pertfield}). Comparing this to (\ref{fgsingen}) and taking into account the relative 
sign in (\ref{fscomp}) we find
\be
(-2)^{g-1}{a^{2g-2}\over \mu^{2g-2}}=1,
\ee 
and one can immediately read off the normalization of a with respect to $\mu=t_D^1-t_D^2$:
\be
a={\mu\over i\sqrt{2}}.
\ee
We notice the following factorization, 
\be
\CF^{\rm YM}_g (q_0, a)={1\over a^{2g-2}} \Xi_g(q_0), 
\ee
where $\Xi_g(q_0)$ is a power series in $q_0$. The relation between $q_0$ and $q$ is defined by
\be
q =q_0 \exp[\Xi_0(q_0)], 
\ee
which can be inverted to obtain the relation between $q_0 $ and $q$. The explicit power series one finds is
 \be
 \label{mmap}
q_0=q-\frac{q^2}{2}+\frac{11 q^3}{64}-\frac{3 q^4}{64}+\frac{359 q^5}{32768}-\frac{75
   q^6}{32768}+\frac{919 q^7}{2097152}-\frac{41 q^8}{524288}+\CO(q^{9}).
\ee
If we now plug this series into $\CF^{\rm YM}_g (a, q_0)$ we find that all gravitational couplings are functions of $q^2$, that is to say, there are no 
odd instanton contributions, as it should be since those are forbidden  by a $\mathbb{Z}_2$--symmetry of the theory \cite{sw}. 
The power series (\ref{mmap}) should be given by a mirror map, corresponding to some algebraic realization of an elliptic curve. Indeed, 
when expressed in terms of 
\be
q = 2^4 q^{1\over 2}_S, \quad q_S=\re^{-S},
\ee
we find
\be
q_0=16 \, q_S^{1\over 2}  - 128 \, q_S + 704 \, q_S^{3\over 2} -3072 \, q_S^2 +\cdots= {\vartheta_2^4(q_S) \over \vartheta_3^4 (q_S)},
\ee
which is (up to an overall factor $16$) the Hauptmodul of $\Gamma_0(4)$. This equality between $q_0$ and the Hauptmodul has only be checked for the 
first few terms of the instanton expansion, and we don't have a general proof. 

We can now express the couplings $\CF^{\rm YM}_g (a, q_0)$, computed from (\ref{ZNek}), in terms of $q_S, \mu$. Due to the connection to the 
Enriques results and the 
field theory limit (\ref{fscomp}), we expect  them to be (up to an overall factor $\mu^{2-2g}$) quasi--modular forms in $q_S$ of weight $2g-2$, and belonging 
to the ring generated by 
$E_2(S)$, $E_4(S)$ and $E_6(S)$. The results obtained with the instanton expansion are in perfect agreement with this. We find at $g=2$
\be
\ba
\mu^2 \CF_2^{\rm YM} &=\frac{1}{16}-\frac{3 q_S}{2}-\frac{9 q_S^2}{2}-6 q_S^3-\frac{21 q_S^4}{2}-9 q_S^5+18 q_S^6+\CO\left(q_S^7\right)\\
&=\frac{1}{2^4} E_2(q_S)\ .
\ea
\ee
Proceeding in the same way we find, 
\be
\ba
\mu^4\CF_3^{\rm YM} &=\frac{1}{2^5}\Big( \frac{2}{3} E_2^2+\frac{1}{3} E_4 \Big)\ ,\\[.2cm]
\mu^6\CF_4^{\rm YM} &=\frac{1}{2^6}\Big(\frac{11}{12} E_2^3+\frac{4}{3} E_2 E_4 +\frac{7}{12} E_6\Big)\ ,\\[.2cm]
\mu^8\CF_5^{\rm YM} &=\frac{1}{2^7}\Big(\frac{17}{9}E_2^4+\frac{97}{18}E_2^2 E_4+\frac{32}{9}E_4^2+\frac{14}{3}E_2 E_6 \Big)\ ,\\[.2cm]
\mu^{10}\CF_6^{\rm YM} &=\frac{1}{2^8}\Big(\frac{619}{120}E_2^5+\frac{218}{9}E_2^3 E_4+\frac{427}{9}E_2 E_4^2+\frac{4501}{144}E_2^2 E_6
    +\frac{4337}{144}E_4 E_6 \Big)\ ,\\[.2cm]
\mu^{12}\CF_7^{\rm YM} &=\frac{1}{2^9}\Big(\frac{1418}{81}E_2^6+\frac{52837}{432}E_2^4 E_4+\frac{12848}{27}E_2^2 E_4^2 +\frac{22631}{108}E_2^3 E_6
+\frac{5423}{9}E_2 E_4 E_6 \\
   &\quad \ \ \ \ \ +\frac{6529}{54}E_6^2 +\frac{352069}{1296}E_4^3 \Big)\ ,
\ea
\ee
We point out that we have not proved these equalities, but rather verified them by 
using the instanton expansion up to high order. It is however highly non--trivial that 
this expansion can be matched to a quasimodular form of the required weight. In addition, one can 
verify that the coefficients of the above combinations agree with the Enriques results. For example, 
if we look at the singular behavior of 
(\ref{finalfthree}) by using (\ref{fgsingen}), one finds, 
\be
\CF^{(3)}(S,t_D) \rightarrow {1\over 32\mu^4} E_4(S) + {1\over 48 \mu^4}(E_2^2(S)-E_4(S)) = {1\over 96}(2 E_2^2(S) + E_4(S)), 
\ee
in agreement with the result above. We have checked that the above polynomials are in accordance with the 
field theory limit of the Enriques model also for $g=4,5,6$. For higher genus the instanton results for the $N_f=4$ theory provide a boundary condition for the 
holomorphic anomaly equation, since they determine the coefficient of the leading singularity near (\ref{slocustwo}) as a function of $S$, and generalize the 
heterotic result (\ref{fgsingen}) away from the fiber. 

In summary, we have verified with the instanton computations of \cite{Nek} our general results about the structure of the topological string amplitudes in the 
Enriques Calabi--Yau (in particular our assumption after (\ref{genS}) about the modular properties of the holomorphic ambiguity). Conversely, the 
results on the Enriques side have been instrumental in clarifying the modularity structure of the massless $N_f=4$ theory.


\section{Direct integration on generic Calabi-Yau manifolds}\label{generic}

In this section we present a general formalism which allows for direct integration of the 
holomorphic anomaly equation \eqref{rec_Fg} for a generic Calabi-Yau manifold.
In order to do that we will first have to rewrite these equations by using new
coordinates and introduce the so-called big moduli space $\widehat \cM$ in 
section \ref{sec:bigrecanomaly}. The holomorphic anomaly equations on the big
moduli space have been also discussed in~\cite{Dijkgraaf:2002ac,Verlinde:2004ck}.
The target space symmetry group acts naturally on the coordinates 
of this extended moduli space and we will briefly discuss modular forms 
on $\widehat \cM$ in section \ref{symp_group}. This has been also studied in~\cite{abk}, 
see also~\cite{Gunaydin:2006bz}. Alternatively to the direct integration, the higher 
genus amplitudes can be derived using a Feynman graph expansion in generalization 
of \cite{bcov,Verlinde:2004ck,abk}. We introduce the appropriate propagators and vertices in section \ref{sec:vertprop}.
Finally, in section \ref{direct-int} we derive a closed expression for the $F^{(g)}$ 
using direct integration. This can be viewed as the generalization of the discussion 
of the Seiberg-Witten example in  section \ref{Seiberg-Witten} for compact Calabi-Yau manifolds with 
an arbitrary number of moduli.

\subsection{The recursive anomaly for $F^{(g)}$ \label{sec:bigrecanomaly}}

In this section  we rewrite the holomorphic anomaly equations \eqref{rec_Fg}  for an enlarged moduli space in which the $2h^{(2,1)}$
coordinates $t^i,\bar t^i$ on $\cM$ are promoted to $2h^{(2,1)}+2$ coordinates $Y^K,\bar Y^K$. 
{}From a geometric point of view, this amounts to working on the moduli space of 
complex normalized $(3,0)$-forms $\Omega$ on the Calabi-Yau manifold under consideration.
We denote this moduli space by $\widehat \cM$ and call it the big moduli space. 
The coordinates $Y^K$ on $\widehat \cM$ are defined as functions 
of $t^k$ by using the homogeneous coordinates $X^K(t)$ arising in the expansion \eqref{exp_Omega} of the holomorphic three-form $\Omega$.
Explicitly, we define 
\beq
   Y^I = \lambda^{-1} X^I(t)\ ,\qquad I = 0,\ldots,h^{(2,1)}\ ,
\eeq
where $\lambda$ is the complex string coupling.
The big moduli space $\widehat \cM$ 
is shown to be a rigid special K\"ahler manifold with K\"ahler potential $\widehat K $ and 
K\"ahler metric $\widehat K_{I\bar J}$ given by
\beq \label{def-Khat}
  \widehat K =\tfrac{i}{2}\big( Y^K \bar \cF_K(\bar Y) - \bar Y^K \cF_K(Y)\big)\ , \qquad 
  \widehat K_{I\bar J} = \partial_{I}\partial_{\bar J}\widehat K = \I\tau_{IJ} \ ,
\eeq
where $\partial_I \equiv \partial_{Y^I}$ and $\partial_{\bar I} \equiv \partial_{\bar Y^I}$
are the derivatives with respect to the coordinates on $\widehat \cM$ and $\tau_{IJ}=\partial_I \partial_J \cF$
is the second derivative of the prepotential.
Note that the K\"ahler metric $\I \tau_{IJ}$ is not positive definite, but rather has complex signature $(h^{2,1},1)$, 
i.e.~has one complex negative direction.   
The metric connection is shown to be 
\beq \label{biggamma}
   \Gamma^I_{JK} = \widehat K^{I \bar M} \partial_J \widehat K_{K \bar M} = -\tfrac{i}{2} C_{JK}^{\ \ I}\ ,
\eeq
where $C_{IJK}(Y)=\partial_I \partial_J \partial_K \cF$ is the third derivative of the prepotential $\cF$.
This implies that the covariant derivative of a tensor $V_K$ on $\widehat M$ is given by
\beq\label{cov_Dgen1}
   D_I V_K \equiv \partial_{I} V_K -\Gamma^J_{IK}V_J =  \partial_{I} V_K + \tfrac{i}{2} C_{IK}^{\ \ J} V_J \ .
\eeq
Here and in the following, we will raise and lower indices using the metric $\widehat K_{I\bar J} = \I\tau_{IJ} $.
For a more exhaustive discussion of rigid 
special geometry we refer to the existing literature \cite{N=2rev,Craps:1997gp,Freed}.

Let us now lift the holomorphic anomaly equations  \eqref{rec_Fg} for the free energies $F^{(g)}$ 
to the big moduli space $\widehat \cM$. In order to do that, we evaluate $F^{(g)}(t,\bar t)$ as functions of the 
homogeneous coordinates $X^K$. As reviewed in section \ref{anomaly}, they transform as sections of $\cL^{2-2g}$ such that 
\beq \label{Fg_Y}
   F^{(g)}(Y,\bar Y) = \lambda^{2g-2} F^{(g)}(X,\bar X)\ ,\qquad \qquad Y^K \partial_K F^{(g)} = (2-2g) F^{(g)}\ . 
\eeq
Rewriting the holomorphic anomaly equations \eqref{rec_Fg} using the $Y^K$ coordinates and the functions $F^{(g)}(Y,\bar Y)$
we find  
\beq \label{big_Fg}
       \partial_{\bar I} F^{(g)} = -\tfrac{i}{8} \bar C_{I}^{\ JK} \Big( 
       D_J\partial_{K} F^{(g-1)} +\sum_{r=1}^{g-1} \partial_{J} F^{(r)} \partial_{K} F^{(g-r)} \Big)\ .
\eeq
A detailed derivation of \eqref{big_Fg} can be found in appendix \ref{Cal_big_Fg1}.
 We can also lift the equation \eqref{anomaly_F1} for $F^{(1)}$ to the 
big moduli space $\widehat \cM$ in a way similar to the lift of the holomorphic anomaly equations for $g>1$. 
First recall that $F^{(1)}$ is a section of $\cL^0$ and hence 
as a function of the homogeneous coordinates $X^K(t)$ homogeneous of degree 
$0$ as seen in \eqref{Fg_Y}. This implies that 
\beq \label{F1_hom}
  Y^K \partial_{Y^K} \partial_{\bar Y^M}F^{(1)} = \bar Y^K \partial_{\bar Y^K} \partial_{Y^M}F^{(1)} = 0\ .
\eeq
Using this property and the special geometry identities summarized in appendix \ref{N=2sp}, one derives 
the holomorphic anomaly for $F^{(1)}(Y,Y)$ on the big moduli space 
(see appendix \ref{Cal_big_F1})
\beq \label{big_F1}
  \partial_{I} \partial_{\bar J} F^{(1)} =  \tfrac18 C_{ILM} \bar C_{J}^{\ LM} 
   - \Big(\frac{\chi}{24} -1 \Big) K_{I \bar J}(Y,\bar Y) \ ,
\eeq
where the second derivative of the K\"ahler potential \eqref{kaehlerpotential} is shown to be
\beq \label{deriv_KY}
  K_{I \bar J}\equiv \partial_{Y^I} \partial_{\bar Y^J} K(Y,\bar Y) = 2e^{K} \widehat K_{IJ}+ 4 e^{2K} \bar Y_{I} Y_J\ ,
\eeq
and indices were lowered by contraction with the metric \eqref{def-Khat}.
Note that the last term in the expression for $ K_{I \bar J}$
ensures that the holomorphic anomaly  \eqref{big_F1}
also implies \eqref{F1_hom}.
In this big moduli space formulation, it is straightforward to integrate the holomorphic 
anomaly equation \eqref{big_F1} for $F^{(1)}$. One thus shows that the genus one 
free energy is locally of the form 
\beq \label{F_1solution}
  F^{(1)}(Y,\bar Y) = - \tfrac{1}{2}\log \det(\I\tau_{IJ}) - \Big(\frac{\chi}{24}-1\Big)K(Y,\bar Y) - \ln |\Phi|+f^{(1)}+\bar f^{(1)}\ ,
\eeq
where $\Phi(Y)$ and $f^{(1)}(Y)$ are holomorphic functions arising as integration constants. 
For reasons which will become clear later, we introduced the seemingly artificial split of the 
holomorphic ambiguity into $\Phi$ and $f^{(1)}$.  
The expression \eqref{F_1solution} provides the direct generalization for $F^{(1)}$ in Seiberg-Witten theory \eqref{SWF1withtau} and 
also reduces to the Enriques result \eqref{F1Enriques2}. Clearly, the holomorphic anomaly does not determine $\Phi,\ f^{(1)}$
which were derived in the Seiberg-Witten and Enriques example by using additional information due to modularity
and string dualities. In the next section we will briefly discuss modularity from the point of 
view of the big moduli space $\widehat \cM$.

\subsection{Monodromy, symplectic group and modular forms \label{symp_group}}

In this section we discuss the action of the target space symmetry group on 
the coordinates of the big moduli space $\widehat \cM$ and introduce some 
basic modular forms and modular derivatives.
To begin with, let us note that there is a natural symplectic
action on the periods $(\cF_J,Y^I)$ of the holomorphic three-form given by
\beq \label{period_rot}
   \left(\begin{array}{cc}a & b \\ c & d \end{array}\right) \left(\begin{array}{c}\cF \\ Y \end{array}\right)=\left(\begin{array}{c}\cF' \\ Y' \end{array}\right)\ ,
\eeq
where $a,b,c$ and $d$ are real integer-valued matrices obeying 
\beq \label{def-M}
  a^T c=c^T a\ ,\qquad \quad b^T d = d^T b\ , \qquad \quad a^T d - c^T b = 1\ .
\eeq
These transformations change the basis of the third cohomology of the Calabi-Yau manifold 
and form the symplectic  group $Sp(H^3,\mathbb{Z})$. Note that in general only 
a subgroup $\Gamma_M$ of $Sp(H^3,\mathbb{Z})$ provides a true 
symmetry of the topological string theory. $\Gamma_M$ is the monodromy 
group. We encountered specific examples for $\Gamma_M$ in the sections on Seiberg-Witten 
theory and the Enriques Calabi-Yau: $\Gamma_M(\text{SW})=\Gamma(2)$ and $\Gamma_M(\text{E})=Sl(2,\bbZ)\times O(10,2,\bbZ)$.
The monodromy group $\Gamma_M$ is a symmetry of all higher genus amplitudes 
$F^{(g)}(Y,\bar Y)$.

Given the action of $Sp(H^3,\mathbb{Z})$ on the periods, we can also investigate 
its induced action on the geometrical objects on $\widehat \cM$. 
First note that both K\"ahler potentials $K$ and $\widehat K$ are 
invariant under \eqref{period_rot} since they contain the symplectic 
scalars $Y^K \bar \cF_K - \bar Y^K \cF_K$. The second derivative 
$\tau_{IJ}=\partial_{Y^I} \cF_J$ of the prepotential transforms as
\beq
  \tau \quad \mapsto \quad (a\tau +b)(c\tau +d)^{-1}\ .
\label{tau_{IJ}transformation}
\eeq
This implies that $\tau_{IJ}$ transforms as a modular 
parameter and is the higher-dimensional analog of \eqref{modulartrans}.
Once again one easily shows that the inverse of $\I \tau_{IJ}$ transforms 
with a shift
\beq
   \I \tau^{IJ}\quad \mapsto \quad (c\tau + d)_{\ K}^I (c\tau +d)_{\ L}^J \I \tau^{KL} -2i c^{IK} (c\tau +d)^J_K\ .
\label{imtau^{IJ}transformation}
\eeq
On the other hand, the third derivative $C_{IJK}$ of the prepotential $\cF$ 
transforms without such a shift
\beq
  C_{IJK}\quad \mapsto \quad (c\tau + d)_I^{-1\, M}(c\tau +d)_J^{-1\, N}(c\tau + d)^{-1\, P}_K C_{MNP} 
\eeq
This is precisely the transformation property of a modular form of weight $-3$.  
In general, we say that a modular form is of weight $-n$ if it transforms as
\beq
   M_{I_1 \ldots I_n}\quad \mapsto \quad (c\tau + d)_{I_1}^{-1\, J_1} \ldots (c\tau + d)_{I_n}^{-1\, J_n} M_{J_1 \ldots J_n}\ .
\eeq
The holomorphic form $\Phi$ appearing in \eqref{F_1solution} has no indices, but nevertheless transforms 
under the modular group $\Gamma_M$. It is chosen such that $F^{(1)}$ as well as $f^{(1)}$ are invariant.
This implies that it has to transform as 
\beq
   \Phi\quad \mapsto \quad \det (c \tau +d)\ \Phi
\eeq
to compensate the transformation  of $\det(\I \tau_{IJ})$ in \eqref{F_1solution}. A major challenge is to find the appropriate 
$\Phi$ for a given Calabi-Yau manifold and show that it can be expressed as a function of $\tau_{IJ}$ 
only. In order to do that one can change $f^{(1)}$ by holomorphic modular invariant combinations.
$\Phi(\tau_{IJ})$ can be explicitly derived for the Enriques Calabi-Yau. It is desirable 
to explore further examples such as the quintic Calabi-Yau.

As we have seen before, the derivative $\partial_{I_0} M_{I_1 \ldots I_n}$
of a modular form $M$ is no longer a modular form, since the derivative also acts on the matrices $(C\tau + D)_{I_i}^{-1\, J_i}$. 
However, this action can be compensated by using covariant derivatives on $\widehat \cM$. One easily shows 
that the Christoffel symbols \eqref{biggamma} shift under \eqref{period_rot} such that the covariant 
derivative $D_{I_0} M_{I_1 \ldots I_n}$ of a modular form is again a modular form of weight reduced by one.
If we express $M_{K}$ as a function of $\tau_{IJ}$, we can also take derivatives 
\beq\label{cov_Dgen2}
   D^{IJ}M_K\equiv  \partial_{\tau_{IJ}}M_K -  \tfrac{i}{2}\delta_K^{\{J} \I \tau^{I \}L} M_L\ ,\qquad \quad D_I = C_{IJK} D^{JK}\ ,
\eeq
where $\{IJ\}$ indicates symmetrization in the indices $I$ and $J$ with symmetry factor $\frac12$. In order to relate 
$D_I$ and $D^{JK}$ we have used that $\partial_I \tau_{KL}=C_{IKL}$. Since $C_{IKL}$ has weight $-3$
this also implies that $D^{IJ}$ raises the weight of the modular form by $2$.
 Let us note that $D_I$ and 
$D^{IJ}$ are the higher-dimensional analogs of the derivatives $D_t$ and $D_\tau$ displayed in \eqref{der_nonhol}.

\subsection{Feynman rules for $F^{(g)}$: vertices and the propagators \label{sec:vertprop}}

Let us now come back to the discussion of the holomorphic anomaly equations \eqref{big_Fg}.
 As argued in \cite{bcov} and briefly recalled in section \ref{anomaly}, the traditional way of finding a solution to equations 
of the form \eqref{big_Fg} is via a Feynman 
graph expansion.
In this section we will derive the vertices and propagators to describe 
such an expansion in the large moduli space. This can be done by first directly solving 
\eqref{big_Fg} for the smallest possible genus $g=2$. The resulting section $F^{(2)}$
can be identified as a sum over Feynman graphs counting degeneracies of Riemann surfaces.
This example allows us to identify the vertices and propagators, which can be used to 
systematically construct every solution $F^{(g)}$. The generating functional encoding 
these Feynman rules is then derived and can be shown to be equivalent to the generating 
functional of Bershadsky, Cecotti, Ooguri and Vafa \cite{bcov}.
  
In order to extract the solutions for the free energies $F^{(g)}$ we first define  
complex tensors  
\beq \label{def-C^g}
  C^{(g)}_{I_1 \ldots I_n}(Y,\bar Y) \equiv  \left\{ \begin{array}{ll}D_{I_1}\ldots D_{I_n} F^{(g)}(Y,\bar Y) &\ \text{for}\ \ g\ge1\\
                                                                          i D_{I_1}\ldots D_{I_n} C_{I_{n-2} I_{n-1} I_{n}} &\ \text{for}\ \ g=0 \end{array}\right.\ .
\eeq
and demand
\beq\label{def-C^g2}
   C^{(g)}_{I_1 \ldots I_n} = 0 \quad \text{for} \ \ 2g-2 + n \le 0 \ .
\eeq
These two equations are the big moduli space equivalents of \eqref{C_prop} and \eqref{C_prop_cond}. 
They imply that $C^{(g)}_{I_1 \ldots I_n}$ is a section of  $\cL^{2-2g-n}$ such that
we can infer the homogeneity relation
\beq \label{proj_2}
  Y^K C^{(g)}_{K I_1 \ldots I_n} = (2-2g-n) C^{(g)}_{I_1 \ldots I_n}\ .
\eeq
 
In equation \eqref{F_1solution} we already displayed the general local form of
solutions for the free energy $F^{(1)}$. The next function to determine is 
$F^{(2)}(Y,\bar Y)$. Evaluating \eqref{big_Fg} for $g=2$ one obtains 
\beq \label{F_2big}
    \partial_{\bar Y^I} F^{(2)} = -\tfrac{i}{8} \bar C_{I}^{\ JK} \Big( 
    D_J \partial_{K} F^{(1)} + \partial_{J} F^{(1)} \partial_{K} F^{(1)} \Big) \ ,
\eeq
As we discuss in appendix \ref{Cal_big_Fg1} such an equation can be solved by 
an integration by parts method. This amounts 
to writing the right-hand side of \eqref{F_2big} as an anti-holomorphic 
derivative of some expression $\Gamma^{(2)}(Y,\bar Y)$. The solution 
to \eqref{F_2big} is then given by $F^{(2)} = \Gamma^{(2)}(Y,\bar Y) + f^{(2)}(Y)$,
where  $f^{(2)}$ is the holomorphic ambiguity at genus two.
This method of solving \eqref{F_2big} is equivalent to the one used in ref.~\cite{bcov} 
to solve the holomorphic anomaly equations \eqref{rec_Fg} for $F^{(g)}(t,\bar t)$.
However, in contrast to \cite{bcov} it will be sufficient to introduce one type of propagator 
denoted by $\Delta^{IJ}$. The propagator $\Delta^{IJ}$ has to be chosen such that
\beq \label{d_prop}
   \partial_{\bar K} \Delta^{IJ} = \tfrac{i}4 \bar C_K^{\ IJ}\ .
\eeq
Clearly, this fixes the form of $\Delta^{IJ}$ only up to an holomorphic function. 
As for the examples discussed in the previous sections this ambiguity can be fixed by modular invariance and
compatibility with the solution $F^{(1)}$. 

In order to derive an explicit expression for the propagator $\Delta^{IJ}$ we note that a solution to \eqref{d_prop} is
always of the form 
\beq
\label{DIJanholomorphic}
  \Delta^{IJ} =-\tfrac12 \I \tau^{IJ} + {\cal E}^{IJ}(Y)\ ,
\eeq
where ${\cal E}^{IJ}(Y)$ is a holomorphic function, which compensates 
the shift transformation \eqref{imtau^{IJ}transformation} of $\I \tau^{IJ}$.
As in the Seiberg-Witten and Enriques example we want to express $\cE^{IJ}$
as a derivative of the holomorphic part of $F^{(1)}$ given in \eqref{F_1solution}.
In order to do that, let us assume that 
we can express $\Phi(Y)$ as a function of $\tau_{IJ}$ itself. To achieve this,
it might be necessary to appropriately split the holomorphic ambiguity of 
$F^{(1)}$ into $\Phi(\tau)$ and an additional function $f^{(1)}(Y)$. 
$f^{(1)}$ is a modular invariant function which might not be 
expressible as a function of $\tau_{IJ}$.
We then identify the 
holomorphic part in \eqref{DIJanholomorphic} to be
\beq
  \cE^{IJ}=  -\frac{i}{\Phi} \frac{\partial \Phi(\tau)}{\partial \tau_{IJ}}\ .
\eeq
{}From this definition one can immediately conclude that $\Delta^{IJ}$ is a modular form of weight 
$2$ under the target space symmetry group $\Gamma_M$.
To see this, note that since $F^{(1)}$ and $K$ are invariant under $\Gamma_M$   
also the section
\beq \label{def-F1tilde}
   \tilde F^{(1)} =- \tfrac{1}{2}\log \det(\I\tau_{IJ}) - \ln |\Phi(\tau)|+f^{(1)}+\bar f^{(1)}\ ,
\eeq
is trivially invariant under $\Gamma_M$. But evaluating the first derivative on the weight 
zero forms $\tilde F^{(1)}$ and $f^{(1)}$ one finds
\beq \label{Deltaasder}
  \partial_I \tilde F^{(1)}=-\tfrac12 C^{(0)}_{IJK} \Delta^{JK}+\partial_I f^{(1)}\ ,
\eeq
and infers from the discussion of section \ref{symp_group} that $\Delta^{IJ}$ is of weight $2$ and does not shift under
$\Gamma_M$.

\begin{figure}[!ht]
\leavevmode
\begin{center}
\includegraphics[height=2.4cm]{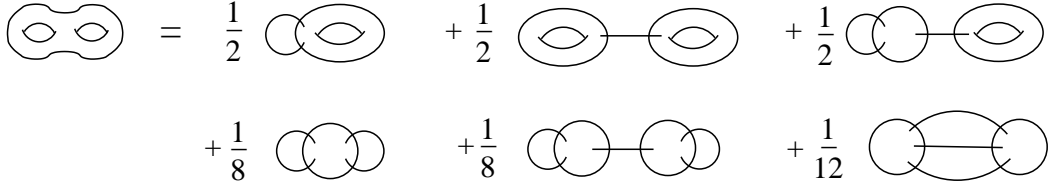}
\end{center}
\caption{The Feynman graph expansion for $F^{(2)}$.}
\label{F_2pic}
\end{figure}

Now that we have discussed the propagator $\Delta^{IJ}$, let us turn to the definition of the 
vertices. We do that by continuing the evaluation of the $F^{(2)}$ example.
In appendix \ref{Cal_big_Fg1} we determine by the partial integration method of \cite{bcov} that 
$F^{(2)}(Y,\bar Y)$ to be
\bea \label{F_2_big}
   F^{(2)}(Y,\bar Y)&=& f^{(2)} - \Delta^{JK} \Big( 
   \tfrac12 \tilde C^{(1)}_{JK} + \tfrac12 \tilde C^{(1)}_J \tilde C^{(1)}_K \Big)
   -\Delta^{JK}\Delta^{LM} \Big(\tfrac18 C^{(0)}_{KLMJ} + \tfrac12 C^{(0)}_{JLM}   \tilde C_K^{(1)} \Big)  \nn \\
         &&- \Delta^{JK} \Delta^{LM} \Delta^{QP}\Big(\tfrac{1}{12} C^{(0)}_{KMQ} C^{(0)}_{PLJ} + \tfrac18 C^{(0)}_{KJQ} C^{(0)}_{PML} \Big)  \ ,
\eea
where  $f^{(2)}(Y)$ is the holomorphic ambiguity.
Note that in this expansion we introduced the shifted $F^{(1)}$ vertices 
\beq
    \tilde C^{(1)}_{JK} = C^{(1)}_{JK} + \big(\tfrac{\chi}{24} -1 \big) K_{J} K_K\ ,\qquad \qquad  \tilde C^{(1)}_K = C^{(1)}_K + \big(\tfrac{\chi}{24} -1 \big)  K_{K} \ .
\eeq
It is not hard to interpret the resulting $F^{(2)}$ as being obtained from a Feynman graph expansion. Each term 
in \eqref{F_2_big} corresponds to one Feynman diagram representing a degeneration of a genus $2$ Riemann surface.
The vertices are $C^{(0)}_{IJK},\ C^{(0)}_{IJKL}$ and $\tilde C^{(1)}_{I},\ \tilde C^{(1)}_{IJ}$ which are connected 
by propagators $\Delta^{IJ}$. The whole Feynman sum is shown in Figure \ref{F_2pic}.

{}From this example we can also infer the general Feynman rules which generate the solutions $F^{(g)}(Y,\bar Y)$
to the holomorphic anomaly equation \eqref{big_Fg}. The propagator is defined in \eqref{DIJanholomorphic} and \eqref{Deltaasder} as 
the derivative of $\tilde F^{(1)}$. The vertices take the form 
\bea \label{def-tildeC}
  \tilde C^{(g)}_{I_1 \ldots I_n} &=& C^{(g)}_{I_1 \ldots I_n} \qquad \text{for}\quad g \neq 1 \ ,\qquad \qquad  \tilde C^{(1)} = 0\ .\\
  \tilde C^{(1)}_{I_1 \ldots I_n} &=& C^{(1)}_{I_1 \ldots I_n} + (n-1)! \big(\tfrac{\chi}{24} -1 \big) K_{I_1}\ldots K_{I_n}\qquad  \text{for}\quad n >0 \ ,
\eea
where $K_I = \partial_{Y^I}K(Y,\bar Y)$ are the first derivatives of the K\"ahler potential \eqref{kaehlerpotential}.
It is straightforward to check that by using $K_I = - \widehat K_I /\widehat K$ and $D_I \widehat K_J=0$ 
one finds 
\beq
 \tilde C^{(1)}_{I_1 \ldots I_n}= D_{I_1} \ldots D_{I_n}  \tilde F^{(1)} \ ,
\eeq
where $\tilde  F^{(1)} $ is defined in \eqref{def-F1tilde}.
 This Feynman graph expansion
can be obtained as a saddle point expansion of the formal integral
\beq \label{Z[Y]}
    \hat Z\big[Y \big] = \int dZ\ \sqrt{\det \Delta}\ \text{exp}\big(-\tfrac12 g_s^{-2}\ \Delta^{-1}_{IJ}\, Z^I Z^J  +  W[Z;\, Y,\bar Y] \big)\ ,
\eeq
where $g_s$ is the expansion constant playing the role of $\hbar$.
Here $W[Z;\, Y,\bar Y]$ contains the vertices \eqref{def-tildeC} and reads
\bea \label{big_W}
  W[Z;\, Y,\bar Y]&=& \sum_{g=0}^{\infty} \sum_{n=0}^{\infty} \tfrac{1}{n!}\, g_s^{2g-2}\, \tilde C^{(g)}_{I_1\ldots I_n} Z^{I_1} \ldots  Z^{I_n}\\
    &=&  \sum_{g=0}^{\infty} \sum_{n=0}^{\infty} \tfrac{1}{n!}\, g_s^{2g-2}\, C^{(g)}_{I_1\ldots I_n} Z^{I_1} \ldots  Z^{I_n} - \big(\tfrac{\chi}{24} -1 \big) \ln\big(1-Z^I K_I \big)\ .\nn
\eea 
Note that the holomorphic anomaly equations on the big moduli space together with (\ref{def-C^g}) 
and (\ref{def-C^g2}) imply that the integrand of (\ref{Z[Y]}) transforms as a wavefunction 
\cite{Witten:1993ed,Dijkgraaf:2002ac,Verlinde:2004ck,abk}. Moreover, following~\cite{bcov,Verlinde:2004ck} 
one shows that $\hat Z[Y]$ is actually holomorphic 
in $Y^I$. One thus concludes that each coefficient of $g_s^{2g-2}$ in the saddle point 
expansion of $\log \hat Z$ is a holomorphic ambiguity $f^{(g)}(Y)$. On the other hand, 
each coefficient is of the form $F^{(g)}(Y,\bar Y)- \Gamma^{(g)}(Y,\bar Y)$, 
where $\Gamma^{(g)}$ are the Feynman graphs described above.  
We thus solve for $F^{(g)}= \Gamma^{(g)}(Y,\bar Y)+f^{(g)}(Y)$ and find the desired result. 
In the remainder of this section, we will argue that 
the big moduli space formulation is indeed completely 
equivalent to the results obtained in \cite{bcov}.

Let us now turn to the comparison of the big moduli space formalism with the
standard results of \cite{bcov} reviewed in section \ref{anomaly}.
Firstly, note that we only needed one type of propagator $\Delta^{IJ}$. 
This propagator is related to the propagators $ \hat  \Delta^{ij}, \hat\Delta^i $ and $\hat \Delta $
introduced in \eqref{def-small-Delta} by 
\bea
\label{DIJ}
  \Delta^{IJ}  &=&  \chi^I_i \hat  \Delta^{ij} \chi^J_j -  \chi^I_i \hat\Delta^i X^J -  X^I \hat \Delta^i \chi^J_i + X^I \hat\Delta X^J\ ,\nn \\
    &=&  \left(\begin{array}{cc}X^I  &\chi_i^I  \end{array} \right)
          \left(\begin{array}{cc} \hat \Delta & -\hat \Delta^{j}\\ -\hat \Delta^{i} &\hat \Delta^{ij} \end{array} \right)
          \left(\begin{array}{c} X^J \\  \chi_j^J\end{array} \right)\ ,
\label{DIJM}
\eea
where $\chi_i^I$ is defined in \eqref{exp_Omega}.
To check this identity, we can evaluate the $\bar t^i$-derivative of $\Delta^{IJ}$. Clearly,
from the form \eqref{DIJanholomorphic} we find 
\beq
   \partial_{\bar t^i} \Delta^{IJ}= \tfrac{i}{4} \lambda^{-1} \bar \chi^K_{\bi} \bar C_{K}^{\ \ IJ} \ .
\eeq
Precisely the same equation is obtained by using the identification \eqref{DIJ} the special geometry identities 
\eqref{special_geom} and the derivatives \eqref{def-small-Delta} of the small propagators $ \hat  \Delta^{ij}, \hat\Delta^i $ and $\hat \Delta $.
In other words, we found a non-holomorphic lift of the small propagators $\hat \Delta$ to 
$\widehat \cM$ such that $\Delta^{IJ}$ takes the simple form \eqref{DIJanholomorphic}.
Even though we did not completely specify the holomorphic dependence of $\Delta^{IJ}$
we already notice that all non-holomorphic dependence entirely arises through the inverse 
of $\I \tau_{IJ}$. This already hints to the fact that in the formulation on $\widehat \cM$
we have much better control over the $\bar Y^I$ dependence of each $F^{(g)}(Y,\bar Y)$. In section 
\ref{direct-int} we will show that this fact can be used to directly integrate the holomorphic anomaly 
equations, which provides an efficient and direct method to find $F^{(g)}$. In order to show
that expressions such as \eqref{F_2_big} are completely equivalent to the ones of \cite{bcov}, 
we also need the projection of the vertices $\tilde C^{(g)}_{I_1 \ldots I_n}$.
These vertices are related to the correlation functions 
$C^{(g)}_{i_1 \ldots i_n}$ defined in \eqref{C_prop}  by
\beq \label{proj_1}
  C^{(g)}_{i_1 \ldots i_n}(t,\bar t) = \lambda^{2-2g-n} \chi^{I_1}_{i_1} \ldots \chi^{I_n}_{i_n} \tilde C^{(g)}_{I_1 \ldots I_n}(Y,\bar Y) \ .
\eeq
In order to derive this equation we have used the special geometry relations 
\eqref{special_geom} as well as the scaling behavior of $C^{(g)}_{I_1 \ldots I_n}$ when inserting 
$Y^K ={ \lambda}^{-1} X^{K}(t)$. This equation also holds for $\tilde C^{(1)}_{I_1\ldots I_n}$
since due to \eqref{K_Iproj} the additional terms are zero under the contraction with $\chi^I_{i}$. They 
are however of importance once one contracts $\tilde C^{(g)}_{I_1\ldots I_n}$ by $Y^K$ yielding
\beq
   Y^K \tilde C^{(g)}_{KI_1\ldots I_n} = 
                                                            (2-2g -n)\, \tilde C^{(g)}_{I_1 \ldots I_n}\ .
\eeq
By using these identities and the identification \eqref{DIJ} of the propagators 
the expansion \eqref{F_2_big} on $\widehat \cM$ gets transformed into the 
known result of \cite{bcov}. Moreover, also the generating function \eqref{big_W}
reduces to the one found in \cite{bcov} if we identify 
\beq
   Z^I = -\varphi Y^I +  x^i \chi^I_i(Y,\bar Y) \ .
\eeq
This proves that the solutions for $F^{(g)}$ are actually 
identical in the formulation of section \ref{anomaly} and the big
moduli space formalism presented here. As already mentioned above,
the advantage of this new formulation is that all non-holomorphic 
dependence arises entirely through $\I \tau^{IJ}$ in the unified 
propagator $\Delta^{IJ}$ and the covariant derivatives $D_I$.
We will use this fact in the next section to perform a direct integration 
and to derive a closed expression for $F^{(g)}$.

\subsection{Direct integration of the holomorphic anomaly \label{direct-int}}

In this section we make use of the special properties of the big moduli space formulation 
to directly integrate the holomorphic anomaly equations \eqref{big_Fg}.
To begin with, we will argue that every $F^{(g)}$ for $g>1$ can be expressed as 
a finite power series in the propagators $\Delta^{IJ}$ as
\beq \label{Fgexp}
    F^{(g)}(Y,\bar Y) = \sum^{3g-3}_{k=0}  \Delta^{I_1 J_1} \ldots \Delta^{I_k J_k}\ c^{(g)}_{I_1 J_1 \ldots I_k J_k}\ ,
\eeq
where $c^{(g)}$ without indices is the holomorphic ambiguity at genus $g$.
Due to modular invariance of $F^{(g)}$ the coefficients $c^{(g)}_{I_1 J_1 \ldots I_k J_k}(Y)$ are shown to be 
holomorphic modular forms of weight $-2k$ on the big moduli space $\widehat \cM$. All non-holomorphic dependence of 
$F^{(g)}$ arises entirely through $\I \tau^{IJ}$ appearing in the propagators $\Delta^{IJ}$ defined in \eqref{DIJanholomorphic},
It is this fact which will allow us to directly integrate the holomorphic anomaly equations \eqref{big_Fg}.

First of all, we have to show that indeed each $F^{(g)}$ for $g>1$ can be written as a power series in 
the propagators $\Delta^{IJ}$ with holomorphic coefficients. We check this for $F^{(2)}$ first. 
$F^{(2)}$ was expressed in \eqref{F_2_big} as a power series in $\Delta^{IJ}$ with coefficients 
containing $\tilde C^{(0)}_{IJKL},\ \tilde C^{(1)}_I$ and  $\tilde C^{(1)}_{IJ}$. {}From their definitions 
\eqref{def-tildeC}, it is clear that these three quantities are not holomorphic. Hence, in order to establish that \eqref{Fgexp} is true for $F^{(2)}$,
they have to be written as power series in $\Delta^{IJ}$. For $\tilde C^{(0)}_{IJKL} \equiv D_I C^{(0)}_{JKL}$ this requires 
that we have to expand the connection $D_I$. Note that $D_I$ contains the Christoffel symbol
$\Gamma^{K}_{IJ}=-\frac{i}{2} C_{IJL}\I \tau^{LK}$ and is only non-holomorphic due to the appearance of $\I \tau^{IJ}$.
However, by using \eqref{DIJanholomorphic} one 
can replace $\I \tau^{LK}$ and split the connection as
\beq \label{Dsplit}
   D_I V_J = \check D_I V_J - C^{(0)}_{IJK} \Delta^{KL} V_L\ ,\qquad \qquad \Gamma^{K}_{IJ}=\check \Gamma^K_{IJ} + C^{(0)}_{IJM} \Delta^{MK}\ ,
\eeq
where we introduced the holomorphic connection
\beq\label{Dhol}
  \check D_I V_J = \partial_I V_J - \check \Gamma^{M}_{IJ} V_M= \partial_I V_J + i C_{IJK} \cE^{KM} V_M\ .
\eeq
The holomorphic connection $\check D_I$ maps holomorphic sections 
$V_K(Y)$ into holomorphic sections $\check D_K V_L(Y)$. Moreover, it maps modular
forms into modular forms, decreasing the weight of the modular form by one. 
$\check D_I$ are the generalizations of the holomorphic covariant 
derivatives \eqref{der_hol} and \eqref{def-hatD} for the Seiberg-Witten example and the Enriques Calabi-Yau.
We can now split $\tilde C^{(0)}_{IJKL}$ into a holomorphic part and a term linear in the propagator
\beq \label{tildeC04}
   \tilde C^{(0)}_{IJKL} = \check D_I  C^{(0)}_{JKL} - \Delta^{MN} \big(C^{(0)}_{IJM} C^{(0)}_{NKL}+  C^{(0)}_{IKM}  C^{(0)}_{NJL}+ C^{(0)}_{ILM} C^{(0)}_{NJK}\big) \ .
\eeq
Clearly, due to the holomorphicity of $C^{(0)}_{IJK}$ both $ \check D_I  C^{(0)}_{JKL}$ and the coefficient of $\Delta^{MN}$ are holomorphic 
functions.

Let us now evaluate $\tilde C^{(1)}_I$ and $\tilde C^{(1)}_{IJ}$.
The first derivative $\tilde C^{(1)}_I =\partial_I \tilde F^{(1)}$ was already given in \eqref{Deltaasder}
and shown to have an expansion in the propagator $\Delta^{IJ}$ with 
holomorphic coefficients.
In order to also evaluate the remaining vertices, we will need to take derivatives of $\cE^{IJ}$. 
As in the examples, 
note that the first derivative $\partial_K \cE^{IJ}$ is not a modular form, but rather transforms with a 
shift. These shift transformations can be compensated by adding another 
term quadratic in $\cE^{IJ}$. Indeed, we find that the linear 
combination 
\beq \label{E(Phi)}
  \cE^{KL}_{\ \ \ I} \equiv \partial_I \cE^{KL} - \cE^{KM} \cE^{LN} C^{(0)}_{IMN}\ ,
\eeq 
transforms as a modular form without an additional shift. Not surprisingly, $\cE^{KL}_{\ \ \ I}$ is not the same as 
$\check D_I \cE^{KL}$ but rather the field strength of $\cE^{IJ}$. However, there is another important 
representation of $\cE^{KL}_{\ \ \ I}$ in terms of derivatives of $\Phi(\tau)$. Using \eqref{E(Phi)}
one finds 
\beq
   \cE^{KL}_{\ \ \ I} = \cE^{KLMN}_4 C^{(0)}_{MNI}\ ,
\eeq
where 
\beq
  \cE^{I_1 J_1 \ldots I_k J_k}_{2k} = (-i)^{k}\frac{1}{\Phi} \frac{\partial \Phi(\tau)}{\partial \tau_{I_1 J_1} \ldots \partial \tau_{I_k J_k}}\ .
\eeq
These holomorphic modular forms of weight $2k$ are the direct generalizations of the forms $\epsilon^{2k}_{a_1\ldots a_k}$ introduced 
in \eqref{epsilon2k}. A direct calculation shows that we can also express the holomorphic 
modular derivative of $\Delta^{IJ}$ as a propagator expansion,
\beq \label{dDelta}
 \check D_I \Delta^{KL} =- \Delta^{KM} \Delta^{LN} C^{(0)}_{MNI} 
                                           + \cE^{KLMN}_4 C^{(0)}_{MNI}\ .
\eeq

We are now in the position to evaluate the vertex $\tilde C^{(1)}_{IJ} \equiv D_J \partial_I \tilde \cF^{(1)}$.
Using the derivatives \eqref{dDelta} of the propagators together 
with \eqref{Dsplit}, one easily derives 
\bea \label{tildeC^1_IJ}
    \tilde C^{(1)}_{IJ}  &=&-\tfrac12 \cE^{KLMN}_4 C^{(0)}_{MNJ} C^{(0)}_{IKL} + \check D_J\partial_I f^{(1)}
    -\tfrac12 \Delta^{KL}\big( \check D_J C^{(0)}_{IKL} + 2 C^{(0)}_{IJK} \partial_L f^{(1)}\big)
      \nn \\
     &&\qquad 
     + \tfrac{1}{2}\Delta^{KL} \Delta^{MN} \big(C^{(0)}_{JI M} C^{(0)}_{NKL} +  C^{(0)}_{J K M} C^{(0)}_{NIL}\big)\ .
\eea
Inserting \eqref{Deltaasder},  \eqref{tildeC04} and \eqref{tildeC^1_IJ} into the expansion \eqref{F_2_big} for 
$F^{(2)}$ one finds 
\bea \label{F2exp}
   F^{(2)}& =&  \Delta^{I_1 J_1} \Delta^{I_2 J_2} \Delta^{I_3 J_3}\big( \tfrac{1}{24} C^{(0)}_{I_1 I_2 I_3} C^{(0)}_{J_1 J_2 J_3} +  \tfrac{1}{16} C^{(0)}_{I_1 J_1 I_2} C^{(0)}_{J_2 I_3 J_3}  \big) \\
    && - \tfrac18 \Delta^{I_1 J_1} \Delta^{I_2 J_2}\big(\check D_{I_1} C^{(0)}_{J_1 I_2 J_2}
   + 4 C^{(0)}_{I_1 J_1 I_2} \partial_{J_2} f^{(1)} \big)\nn \\
     && - \tfrac14 \Delta^{I_1 J_1} \cE_4^{KLMN} C^{(0)}_{I_1 MN} C^{(0)}_{J_1KL}  
      + \tfrac12 \check D_{I_1} \partial_{J_1} f^{(1)} + \tfrac{1}2 \partial_{I_1} f^{(1)} \partial_{J_1} f^{(1)}+c^{(2)}\ .\nn
\eea
This shows that the calculation of $F^{(2)}$ using the partial integration and the expansion of the non-holomorphic
coefficients yields the desired expansion \eqref{Fgexp} of $F^{(2)}$. 
We will now use an inductive argument to show that every $F^{(g)}$ can be 
written in the form \eqref{Fgexp} and derive a recursive expression 
by direct integration.

Let us now go one step further and show that if all $F^{(r)}$  for $1<r<g$ 
can be written in the form \eqref{Fgexp} also $F^{(g)}$ itself 
admits this expansion. To do that, we use the Feynman graph expansion 
introduced in section \ref{sec:vertprop}. It was shown there that each 
$F^{(g)}(Y,\bar Y)$ can be obtained from vertices $\tilde C^{(r)}_{I_1\ldots I_k}$, $r<g$
connected with propagators $\Delta^{IJ}$. But it is not hard to see that $\tilde C^{(r)}_{I_1\ldots I_k}$
is actually an expansion in $\Delta^{IJ}$ with holomorphic coefficients. More precisely, note that 
the vertices are obtained 
by taking covariant derivatives $D_I$ of $F^{(r)}$ and 
we can apply \eqref{Dsplit} to rewrite these into holomorphic 
covariant derivatives $\check D_I$ and a propagator contribution. But since by our
induction assumption $F^{(r)}$ is of the form \eqref{Fgexp} for $r<g$ we can apply \eqref{dDelta} 
to show that this is equally true for $F^{(g)}$ itself. This proves that \eqref{Fgexp} is true for all $g>1$.
It is also straightforward to 
count  the number of propagators arising in the expansion \eqref{Fgexp}. 
One simply notes that the term in the Feynman graph expansion 
with coefficients $C^{(0)}_{I JK}$ only is already an
expansion in $\Delta^{IJ}$ with holomorphic coefficients. It has the 
maximal number of propagators, namely $3g-3$.

Having shown that $F^{(g)}$ can be always brought to the form
\eqref{Fgexp}, let us now determine a closed expression by direct integration.
Since all non-holomorphic dependence arises through $\Delta^{IJ}$, the 
holomorphic anomaly equation can be rewritten as 
\beq \label{reg_Fgtointegrate}
  \frac{\partial F^{(g)}}{\partial \Delta^{IJ}}=\tfrac12 D_I \partial_J F^{(g-1)} + \tfrac12 \sum^{g-1}_{r=1} \partial_I F^{(r)} \partial_J F^{(g-r)}\ . 
\eeq
To integrate this expression we introduce the following shorthand notation
\beq \label{m_kex}
     F^{(g)}(Y,\bar Y) = \sum^{3g-3}_{k=0}  c^{(g)}_{\ (k)} \ , \qquad \quad  
     c^{(r)}_{\ (k)} = \Delta^{I_1 J_1} \ldots \Delta^{I_k J_k}\,  c^{(r)}_{I_1 J_1 \ldots I_k J_k}\ ,
\eeq
where $c^{(g)}_{\ (k)}$ is the term containing $k$ propagators $\Delta^{IJ}$.
We also rewrite the right-hand side of the holomorphic anomaly equation 
as 
\beq
  D_I \partial_J F^{(g-1)} + \partial_I \tilde F^{(1)} \partial_J F^{(g-1)} + \partial_I F^{(g-1)} \partial_J \tilde F^{(1)}
                       + \sum_{r=2}^{g-2} \partial_I F^{(r)} \partial_J F^{(g-r)}\ .
\eeq 
Here the first three terms can be rewritten as 
\bea \label{DDexpression}
  && D_I \partial_J F^{(g-1)} + \partial_I \tilde F^{(1)} \partial_J F^{(g-1)} + \partial_I F^{(g-1)} \partial_J \tilde F^{(1)} \\
  &&\ = \check D_I \partial_J F^{(g-1)} 
         -\Delta^{KL} C^{(0)}_{IJK} \partial_{L} F^{(g-1)} -\Delta^{KL} C^{(0)}_{KL\{I} \partial_{J\} } F^{(g-1)} + 2 \partial_{\{ I} f^{(1)} \partial_{J\} } F^{(g-1)} \ , \nn
\eea
where we have applied \eqref{Dsplit} and inserted \eqref{Deltaasder}.
 We also introduce the derivative $\check d$,
which acts on the coefficients of the $\Delta$-expansion as the holomorphic covariant derivative 
$\check D_I$ but leaving $\Delta^{IJ}$ invariant. For $c^{(g)}_{\ (k)}$ given in \eqref{m_kex} we thus set
\beq
  \check d_{I} c^{(g)}_{\ (k)}=\Delta^{I_1 J_1} \ldots \Delta^{I_k J_k}\,  \check D_I c^{(g)}_{I_1 J_1 \ldots I_k J_k}(Y) 
\eeq
Using this definition, we calculate
\bea \label{firstbig_der}
  \partial_I c^{(g)}_{\ (k)}&=&  \check d_I c^{(g)}_{\ (k)}  + \big(\check D_I \Delta^{MN} \big) \frac{\partial }{\partial \Delta^{MN}} c^{(g)}_{\ (k)}   \\
                   &=& \check d_I c^{(g)}_{\ (k)}
                   + C^{(0)}_{IPQ}\big(\cE^{PQMN}_4  - \Delta^{PM} \Delta^{QN} \big)    \frac{\partial }{\partial \Delta^{MN}} c^{(g)}_{\ (k)}\ . 
\eea
Note that the first term is homogeneous of degree $k$ in $\Delta$, the second is homogeneous of degree $k-1$, while the last is 
homogeneous of degree $k+1$. 
We also evaluate the second derivative
\bea \label{secondbig_der}
 \check D_I \partial_J c^{(g-1)}_{\ (k)} &=& 
 \Big[\cE_4^{MNQP} \cE_4^{TURS} C^{(0)}_{IMN} C^{(0)}_{JTU} \Big] \frac{\partial^2}{\partial \Delta^{QP} \partial \Delta^{RS}} c^{(g-1)}_{\ (k)}  \\
&+&\Big[\cE^{CDKLMN}_6 C^{(0)}_{KLI} C^{(0)}_{MNJ}  + \cE^{CDFG}_4 \check  C^{(0)}_{JFGI}+ 2  \cE^{CDFG}_4 C^{(0)}_{FG\{I}\check d^{\phantom{(}}_{J\}} \Big] \frac{\partial}{\partial \Delta^{CD}} c^{(g-1)}_{\ (k)}\nn \\
&+&\Big[\check d_I  \check d_J - 2 \cE^{QCFG}_4  C^{(0)}_{FG\{ I} C^{(0)}_{J \}QB} \Delta^{BD} \frac{\partial}{\partial \Delta^{CD}} \nn \\
  &&    - 2 \cE^{CDEF}_4 C^{(0)}_{MN\{I} C^{(0)}_{J\}EF} \Delta^{MQ} \Delta^{NP}  \frac{\partial^2}{\partial \Delta^{QP} \partial \Delta^{CD}} \Big] c^{(g-1)}_{\ (k)} \nn \\
  &-& \Big[\check C^{(0)}_{IJAB}+2 C^{(0)}_{AB \{I} \check d^{\phantom{(}}_{J\} } \Big] \Delta^{AC} \Delta^{BD}  \frac{\partial}{\partial \Delta^{CD}} c^{(g-1)}_{\ (k)}\nn\\
  &+& \Big[ 2 C^{(0)}_{IMN} C^{(0)}_{JQB} \Delta^{QM} \Delta^{NC} \Delta^{BD}  \frac{\partial}{\partial \Delta^{CD}}  \nn \\
  &&       +   C^{(0)}_{IMN} C^{(0)}_{JAB}  \Delta^{QM} \Delta^{NP} \Delta^{AC} \Delta^{BD}   \frac{\partial^2}{\partial \Delta^{QP} \partial \Delta^{CD}}  \Big] c^{(g-1)}_{\ (k)}\nn\ ,
\eea
where $\{IJ\}$ indicates the symmetrization of the indices and we abbreviated 
\beq
   \check C^{(0)}_{IJKL}= \check D_I C^{(0)}_{JKL}\ .
\eeq
Once again, we can specify the $\Delta$-homogeneity of the terms: first line $k-2$, second line $k-1$, third and fourth line $k$,
fifth line $k+1$, sixth and seventh line $k+2$. In performing the direct integration of the holomorphic anomaly equation
we keep track of the number of propagators on the right-hand side of \eqref{reg_Fgtointegrate}. We can do this explicitly 
by inserting \eqref{DDexpression} together with \eqref{firstbig_der} and \eqref{secondbig_der} 
into \eqref{reg_Fgtointegrate}. Due to the vast number of indices 
the result looks rather complicated and will be presented in the following.

Performing the direct integration one finds
\bea
  F^{(g)} &=& \tfrac12 \sum_{k=0}^{3g-6}\Big[ \tfrac{1}{k-1}
  \cE_4^{MNQP} \cE_4^{TURS} C^{(0)}_{IMN} C^{(0)}_{JTU} \Delta^{IJ} \frac{\partial^2}{\partial \Delta^{QP} \partial \Delta^{RS}}  \nn \\
&+&\tfrac{1}{k}\Big(\cE^{CDKLMN}_6 C^{(0)}_{KLI} C^{(0)}_{MNJ}  + \cE^{CDFG}_4 \check C^{(0)}_{JFGI}
       + 2  \cE^{CDFG}_4 C^{(0)}_{FGI}\big(\check d^{\phantom{(}}_{J} +  \partial_{J} f^{(1)}\big) \Big) 
         \Delta^{IJ} \frac{\partial}{\partial \Delta^{CD}} \nn \\
&+&\tfrac{1}{k+1}\Big( \big(\check d_I + \partial_{I} f^{(1)} \big) \big(\check d_J + \partial_{J} f^{(1)} \big) \Delta^{IJ} - \partial_{I} f^{(1)}   \partial_{J} f^{(1)} \Delta^{IJ} \nn \\
&&- 2 \big( \cE^{QCFG}_4  C^{(0)}_{FGI} C^{(0)}_{J QB}  \Delta^{IJ} \Delta^{BD}+\cE^{PQCD}_4  C^{(0)}_{IJK}  C^{(0)}_{LPQ} \Delta^{IJ} \Delta^{KL} \big) \frac{\partial}{\partial \Delta^{CD}} \nn \\
  && \qquad\   - 2 \cE^{CDEF}_4 C^{(0)}_{MNI} C^{(0)}_{JEF}  \Delta^{IJ}\Delta^{MQ} \Delta^{NP}  \frac{\partial^2}{\partial \Delta^{QP} \partial \Delta^{CD}} \Big)  \nn \\
  &-& \tfrac{1}{k+2}\Big(2 C^{(0)}_{IJK}      \check d_L  \Delta^{IJ} \Delta^{KL}  + \big(\check C^{(0)}_{IJAB}+2 C^{(0)}_{ABI}( \check d^{\phantom{(}}_{J} + \partial_J f^{(1)})\big) \Delta^{IJ} \Delta^{AC} \Delta^{BD}  \frac{\partial}{\partial \Delta^{CD}} \Big) \nn\\
  &+&\tfrac{1}{k+3} \Big( 2 \big( C^{(0)}_{IJK}     C^{(0)}_{LPQ}\Delta^{IJ}  \Delta^{KL} \Delta^{PC} \Delta^{QD}  + C^{(0)}_{IMN} C^{(0)}_{JQB} \Delta^{IJ} \Delta^{QM} \Delta^{NC} \Delta^{BD}\big)\frac{\partial }{\partial \Delta^{CD}} \nn \\
  &&    \qquad\     +   C^{(0)}_{IMN} C^{(0)}_{JAB} \Delta^{IJ} \Delta^{QM} \Delta^{NP} \Delta^{AC} \Delta^{BD}   \frac{\partial^2}{\partial \Delta^{QP} \partial \Delta^{CD}}  \Big) \Big] c^{(g-1)}_{\ (k)}\nn \\
  &+&\tfrac12 \sum_{r=2}^{g-2} \sum_{k=0}^{3g-6} \sum_{m+n=k} \Big[  \tfrac{1}{k+1} \Delta^{IJ} \big(\check d_I   c^{(g-r)}_{\ (m)} \big) \big(\check d_J c^{(r)}_{\ (n)}\big)\nn \\
  &+&\tfrac{1}{k-1}\cE^{PQMN}_4 \cE^{RSTU}_4  C^{(0)}_{IPQ}  C^{(0)}_{JRS} \Delta^{IJ} \Big(    \frac{\partial }{\partial \Delta^{MN}}   c^{(g-r)}_{\ (m)} \Big) 
   \Big(    \frac{\partial }{\partial \Delta^{TU}}   c^{(r)}_{\ (n)} \Big)\nn\\
  &+&  \tfrac{1}{k+3}  C^{(0)}_{IPQ}  C^{(0)}_{JRS}  \Delta^{IJ} \Delta^{PM} \Delta^{QN}\Delta^{PT}\Delta^{SU} \Big(    \frac{\partial }{\partial \Delta^{MN}}   c^{(g-r)}_{\ (m)} \Big) 
   \Big(    \frac{\partial }{\partial \Delta^{TU}}   c^{(r)}_{\ (n)} \Big) \nn\\
   &+&\tfrac{1}{k} \cE^{RSTU}_4 C^{(0)}_{JRS} \Delta^{IJ}   \Big\{ \big(\check d_I   c^{(g-r)}_{\ (m)} \big)    \Big( \frac{\partial }{\partial \Delta^{TU}}   c^{(r)}_{\ (n)} \Big)  +
                               \big(  \frac{\partial }{\partial \Delta^{TU}}   c^{(g-r)}_{\ (m)} \big)    \Big(\check d_I  c^{(r)}_{\ (n)} \Big)    \Big\} \nn \\
   &-&\tfrac{1}{k+2} C^{(0)}_{JRS} \Delta^{IJ} \Delta^{PT} \Delta^{SU}  \Big\{ \big(\check d_I   c^{(g-r)}_{\ (m)} \big)    \Big( \frac{\partial }{\partial \Delta^{TU}}   c^{(r)}_{\ (n)} \Big)  +
                               \big(  \frac{\partial }{\partial \Delta^{TU}}   c^{(g-r)}_{\ (m)} \big)    \Big(\check d_I  c^{(r)}_{\ (n)} \Big)    \Big\}                                \Big]\nn \\
   &-&\tfrac{1}{k+1}  \cE^{PQMN}_4 C^{(0)}_{IPQ} C^{(0)}_{JRS} \Delta^{IJ} \Delta^{PT} \Delta^{SU} \Big\{
          \Big(    \frac{\partial }{\partial \Delta^{MN}}   c^{(g-r)}_{\ (m)} \Big)  \Big(    \frac{\partial }{\partial \Delta^{TU}}   c^{(r)}_{\ (n)} \Big) \nn\\
          &&\qquad +
          \Big(     \frac{\partial }{\partial \Delta^{TU}}   c^{(g-r)}_{\ (m)} \Big)  \Big(   \frac{\partial }{\partial \Delta^{MN}}   c^{(r)}_{\ (n)} \Big)     \Big\} \Big] + c^{(g)}_{\ (0)}\ .
\eea
Let us end with some brief remarks about the properties of the direct integration in the big moduli space.
Firstly, we note that the building blocks of $F^{(g)}$ are the propagators $\Delta^{IJ}$ as well as the holomorphic modular forms 
\beq \label{hol_mod_forms}
   \cE^{I_1\ldots J_k}_{2k}\ ,\qquad \check D_{I_1}\ldots \check D_{I_k} f^{(1)}\ ,\qquad \check D_{I_1}\ldots \check D_{I_k} C^{(0)}_{KLM}\ ,
\eeq
induced by $F^{(1)}$ and $F^{(0)}$. It seems likely that also the holomorphic 
ambiguity can be parametrized by \eqref{hol_mod_forms}. To determine these forms it is essential to 
find $\Phi(\tau)$, which will be harder for examples other than the Enriques Calabi-Yau. 
Moreover, in order to efficiently derive all $F^{(g)}$ one also needs to show that the 
forms \eqref{hol_mod_forms} are generated by a finite number of holomorphic modular forms of $\Gamma_M$.
Clearly, the most challenging task is then to fix the ambiguity by appropriate boundary conditions.
To explore these issues for other interesting examples will be left for further work.

\section{Conclusion and Outlook}\label{conclusion}

In this paper we have developed a new approach to solving the holomorphic anomaly equations of \cite{bcov}, 
based on the interplay between modularity and non-holomorphicity, which makes possible to perform 
a direct integration of the equations at each genus. This approach is more efficient than the diagram 
expansion of \cite{bcov} and leads to closed expressions for the topological string amplitudes, 
once the ambiguities are fixed by appropriate boundary conditions. The amplitudes obtained with this procedure  
can be written as polynomials in a finite set of generators that transform in a particularly simple way under 
the space-time symmetry group, making the modularity properties manifest. 
There are many open questions and possible avenues for future work. We list here some of them. 

 Although we have been able to improve the results of \cite{km} for the Enriques Calabi--Yau manifold, 
it would be interesting to push further the formalism developed in this paper. In section \ref{sec:diE} we have introduced 
a set of holomorphic automorphic forms on the Enriques moduli space which might be enough to parametrize the holomorphic 
ambiguity. Using these forms, the boundary conditions obtained from the field theory and the fiber limits, and some extra 
information coming for example from Gromov--Witten theory, one might be able to obtain the topological string amplitudes 
at higher genus. 

As explained in \cite{kkvbh}, Gopakumar--Vafa invariants should provide a microscopic counting of degrees of freedom for 
5d spinning black holes, although in order to make contact with the macroscopic Bekenstein--Hawking entropy one typically 
needs a knowledge of these invariants (therefore of the topological string amplitudes $F^{(g)}$) at arbitrary high genus. 
Some of the results of this paper might be useful in studies of these black holes. For example, the all--genus fiber result 
for the Enriques Calabi--Yau manifold should give a detailed microscopic counting for small 5d black holes obtained by 
wrapping M2 branes in the Enriques fiber.

Vast progress has been made in the understanding of compactifications which allow to stabilize many or
all moduli in $\cN=1$ supersymmetric vacua \cite{Douglas:2006es}. 
These vacua often rely on the inclusion of background fluxes 
and D brane instanton effects. Orientifolds of the Enriques Calabi-Yau might serve as 
very controllable examples in which certain corrections to the $\cN=1$ 
low energy effective theory can be derived. In particular, it is an interesting 
task to identify and compute corrections to the four-dimensional super- and K\"ahler potentials 
encoded by the higher genus amplitudes.

As shown in  \cite{hk,emo} the free energies of matrix models satisfy the holomorphic 
anomaly conditions. Hence, the techniques of this paper could lead to a useful method to analyze matrix models. Using 
matrix model technology one can also write down holomorphic anomaly equations for open string amplitudes in local Calabi--Yau 
manifolds \cite{emo}, and it would be interesting to study them using the methods of this paper. In view of the results of 
\cite{mmopen}, this could lead to a powerful  approach to compute open string amplitudes on toric Calabi--Yau manifolds.

\section*{Acknowledgments}
We would like to thank Mina Aganagic, Richard Borcherds, Vincent Bouchard, Robbert Dijkgraaf, 
Ian Ellwood, Eberhard Freitag, Min-xin Huang, Peter Mayr, Nikita Nekrasov and Don Zagier, 
for very  valuable discussions and correspondences. 
M.M.~would like to thank Greg Moore for conversations on the field theory limit of the 
FHSV model in 1998. A.K.~and T.G.~are supported by DOE Grant 
DE-FG-02-95ER40896. M.W.~is supported by the Marie Curie EST program. This work has been initiated during a stay of A.K., M.M.~and T.G.~at 
the MSRI in Berkeley. 

\bigskip \bigskip

\appendix

\noindent {\bf \Large Appendices}

\section{$\cN=2$ special geometry \label{N=2sp}}

In this appendix we summarize some basics about
$\cN=2$ special geometry~\cite{CandelasdellaOssa,N=2rev,Craps:1997gp,Freed}.
Let $Y$ be a Calabi-Yau threefold, i.e.~a complex three-dimensional K\"ahler 
manifold with $SU(3)$ or $SU(2)\times \bbZ_2$, but no smaller, holonomy 
group.  In particular $Y$ has a 
no-where vanishing holomorphic three-form $\Omega$, which is unique 
 up to complex rescaling. $\Omega$ depends on the complex structure 
of $Y$ and hence varies over the space of complex structure deformations $\cM$.
Local  coordinates on $\cM$ are denoted by $t^i,\bar t^i$. $\Omega(t)$ can be used to define a 
K\"ahler potential 
\beq \label{Kpot}
K(t,\bar t)= - \log\big[i \int_Y \Omega \wedge \bar \Omega \big]
\eeq
$K$ induces the following K\"ahler metric structures on $\cM$ 
\begin{equation}
\begin{array}{rl}
G_{i\bar \jmath}&=\partial_i\partial_{\bar \jmath} K, 
\quad \ 
\Gamma^{k}_{ij}=G^{k\bar l}\partial_i G_{j\bar l},
\quad  \ 
\Gamma^{\bar k}_{\bar \imath \bar \jmath}=G^{l\bar k}{\bar \partial}_{\bar \imath} G_{l\bar \jmath}\\[0.2cm]
R_{i \bar \jmath k \bar l}&=-\partial_i \bar \partial_{\bar \jmath} G_{k\bar l}+
G^{m\bar n} (\partial_i G_{k\bar n})(\bar \partial_{\bar \jmath} G_{m\bar l}), 
\quad \ \ 
R_{i\bar j k}^{\phantom{i \bar \jmath k} l}=-\bar \partial_{\bar \jmath} \Gamma^l_{ik}
\\[0.2cm]
R_{i\bar \jmath}&\equiv G^{k\bar l}R_{i\bar \jmath k\bar l}=-\partial_i 
\bar \partial_{\bar \jmath} \log \det(G_{i\bar \jmath})\ .
\end{array}
\label{kaehlersimplifications}
\end{equation}
$\Omega$ and $\bar \Omega$ are sections of holomorphic and anti-holomorphic 
lines bundles ${\cal L}$ and ${\overline  {\cal L}}$ over ${\cal M}$ 
respectively and holomorphic gauge transformations 
$\Omega\rightarrow e^f\Omega$  in ${\cal L}$ correspond to 
K\"ahler transformations, i.e. $e^{-K}\in {\cL}\otimes {\overline {\cL}}$.  
The derivatives $\partial_i$ are with respect to coordinates $t_i$ of ${\cal M}$, 
 and sections like $V_{j\bar \jmath}$ in 
$T\cM^*_{(1,0)}\otimes T\cM^*_{(0,1)} 
\otimes {\cal L}^m  \otimes {\overline  {\cal L}}^{n}$ have a natural 
connection with respect to the Weil-Petersson metric $G_{i\bar \jmath}$ and the 
line bundle $K_i=\partial_i K$, $K_{\bar \imath} =\partial_{\bar \imath } K$
\begin{equation}
\label{cov_D}
D_i V_{j\bar\jmath}=\partial_i- \Gamma_{ij}^l V_{l\bar \jmath}+ m K_i V_{j\bar \jmath},\quad
D_{\bar \imath} V_{j\bar\jmath}=\partial_{\bar \imath}- 
\Gamma_{\bar \imath \bar \jmath}^{\bar l} V_{j \bar l}+ n K_{\bar \imath} V_{j\bar \jmath}\ .
\end{equation}
For a given complex structure $\Omega$ defines a Hodge decomposition 
\beq
  H^{3}(Y,\bbC) = H^{(3,0)} \oplus H^{(2,1)} \oplus H^{(1,2)} \oplus   H^{(0,3)}\ .
\label{hodgedecomposition}
\eeq
The forms $\Omega$, $\chi_i\equiv D_i\Omega$, ${\overline\chi}_{\bar \imath} 
\equiv D_{\bar \imath} \bar \Omega$ and $\bar \Omega$ provide a basis which 
spans the above cohomology groups over $\bbC$. Since it depends on the complex
structure we call it the moving basis.  
By Kodaira theory, infinitesimal deformations of the complex structure  
are elements of  $H^1(Y,TY)$. $\Omega$ induces an isomorphism 
$H^1(Y,TY)\sim H^{(2,1)}(Y)$.  Hence the harmonic  
$(2,1)$-forms $\chi_i$, $i=1,\ldots,h^{21}$ can be identified as (co)tangent 
vectors to ${\cal M}$ and these deformations are unobstructed on a CY 
manifold \cite{TianTodorov}.

We also introduce a fixed integer symplectic basis $(\alpha_K,\beta^L)$ of $H^{3}(Y,\bbZ)$ with 
\begin{equation} \label{integralsymplecticbasis}
\int_Y \alpha_K \wedge \beta^L =- \int_Y \beta^L \wedge \alpha_K = \delta_K^L\ , 
\quad \qquad
\int_Y \alpha_K \wedge \alpha_L = \int_Y \beta^K \wedge \beta^L = 0\ ,
\end{equation} 
which is independent of the complex structure. We can expand the moving 
basis in terms of the fixed basis 
\beq \label{exp_Omega}
  \Omega = X^I \alpha_I - \cF_I \beta^I \ , \qquad \chi_i = 
\chi_i^I \alpha_I - \chi_{Ii} \beta^I\ , \qquad \text{etc\ .}
\eeq
The  expansion coefficients are the period integrals 
\begin{equation}
\label{periods}
X^I=\int_{A^I} \Omega, \quad \quad F_I=\int_{B_I}\Omega,\quad \quad \chi^I_i=\int_{A^I}\chi_i, 
\quad \quad \chi_{I\, i}=\int_{B_I}\chi_i\ ,
\end{equation}  
where $(A^K,B_I)$ is a basis  of $H_{3}(Y,\bbZ)$ dual to $(\alpha_K,\beta^L)$.
Using (\ref{integralsymplecticbasis}) and (\ref{periods}) we can express (\ref{Kpot}) in terms of the periods
\beq 
\label{KpotII}
  K=- \log i \big[ \bar X^K \cF_K  -  X^K \bar \cF_K\big] \ .
\eeq
Note that $X^I\in \cL$, ${\overline \cF}_I\in {\overline \cL}$, 
$\chi^I_i \in T^*_{(1,0)}\cM \otimes \cL$ etc. 
Obviously the periods carry the information about 
the complex structure deformations. The  $X^I$, $I=0,\ldots h^{21}$ 
can serve locally as homogeneous coordinates on ${\cal M}$. 
Local special coordinates on ${\cal M}$ are 
defined by $t^i=X^i/X^0$, $i=1,\ldots h^{(2,1)}$. 
The $\cF_I$ on the other hand are not independent. It follows rather 
from 
\be
\int_Y\Omega \wedge {\partial \over \partial X^I}\Omega=0
\ee
that there is a holomorphic section $\cF$ of ${\cL}^{2}$   
called prepotential obeying
\be
\cF=\tfrac{1}{2}  X^I F_I\ ,\qquad \quad \cF_I = \partial_{X^I} \cF\ .
\ee
This also implies that $\cF(X)$ is homogeneous of degree two in $X^I$.
In special coordinates $t^i$ one also writes $\cF(t)=
(X^0)^{-2}{\cF(X)}$.
 It turns out to be useful to introduce the 
second and third derivative of the prepotential as
\beq
   \tau_{IJ} = \partial_I \partial_J \cF\ ,\qquad \quad C_{IJK} = \partial_I \partial_J \partial_K \cF\ ,
\eeq
which are homogeneous of degree zero and minus one respectively.

Special K\"ahler geometry describes the relation between the metric structure 
and the Yukawa coupling 
\beq
\label{def-CC}
  C^{(0)}_{ijk} \equiv iC_{ijk} \equiv - \int_Y \Omega \wedge 
\partial_i \partial_j \partial_k \Omega= - \int_Y \Omega \wedge D_i D_j D_k \Omega\ ,
\eeq
a section of $C_{ijk}\in {\rm Sym}^3 (T^*_{(1,0)})\otimes \cL^2$.
Using $\langle \chi_i ,\bar \chi_{\bar \imath} \rangle =G_{i\bar \jmath} e^{-K}$ 
and transversality of $\langle,\rangle$ under the the decomposition 
(\ref{hodgedecomposition}), i.e. $\langle \gamma_{(k,l)},\gamma_{(m,n)} \rangle=0$ unless
$k+m=l+n=3$  one gets the special geometry identities \cite{CandelasdellaOssa} 
\beq \label{special_geom}
  D_i X^I \equiv \chi^I_i \ ,\qquad 
  D_i \chi_j^I  = i C_{ijk} G^{k\bar k} \bar \chi^I_{\bar k} e^K\ , \qquad 
  D_i \bar \chi_{\bar j}^I = G_{i\bar j} \bar X^I\ .
\eeq
{}From (\ref{cov_D}) and (\ref{kaehlersimplifications}) follows 
$[D_i,D_{\bar \jmath}]\chi_k=-G_{i\bar \jmath}+R_{i\bar \jmath k}^{\ \ l} \chi_l$ 
and using (\ref{special_geom}) one gets 
\begin{equation}
[D_i,D_{\bar k}]^{\ \, l}_j=R^{\ \ \ l}_{i\bar k j}=G_{i\bar k}\delta^l_j+G_{j\bar k}\delta^l_i-C_{ijm} {\bar C}_{\bar k}^{ml}\ ,
\label{eq:special}
\end{equation}
where we abbreviated  
\begin{equation} \label{def-Cud}
{\bar C}_{\bar k}^{(0) ml} = 
e^{2K}\bar C^{(0)}_{\bar k\bar \imath\bar \jmath} 
G^{m\bar\imath} G^{l \bar \jmath}\ ,\qquad \qquad {\bar C}_{\bar k}^{ml}=i{\bar C}_{\bar k}^{(0) ml}\ .
\end{equation}

Let us also summarize some relations obeyed by $\tau_{IJ}$ and $C_{IJK}$. 
One first notes that by homogeneity and \eqref{def-CC} and \eqref{special_geom} one has
\beq
\label{vielbein}
 C_{IJK}X^K=0\ ,\qquad \qquad  C_{ijk} = C_{IJK}  \chi^I_i  \chi^J_j  \chi^K_k\ .
\eeq
Using the above definitions and the degree two 
homogeneity of $\cF$ one also shows that
\beq
\label{transversality}
2 e^{K} X^I\I \tau_{IJ} \bar X^J=1\ ,\qquad \bar X^I \I \tau_{IJ} \chi_i^J=0\ ,
\qquad   2 e^K \chi^I_i \I \tau_{IJ} \bar \chi^J_{\bar j}=G_{i \bar j}\ .  
\eeq
Denoting by $\I \tau^{IJ}$ the inverse of $\I \tau_{IJ}$ it follows from these conditions that
\beq \label{G-tau}
\chi^I_i G^{i \bar j} \bar \chi^J_{\bar j} e^K  = \tfrac{1}{2}  \I \tau^{IJ} + X^I \bar X^J e^K\ .
\eeq

\section{Theta functions and modular forms}\label{theta}

Our conventions for the Jacobi theta functions are:
\be
\begin{aligned}
\vartheta_1(\nu|\tau)&=\vartheta [^1_1](\nu|\tau)=
i \sum_{n \in {\bf Z}} (-1)^n q^{{1\over 2}(n+1/2)^2} e^{i \pi (2n +1) \nu},\\
\vartheta_2(\nu|\tau)&= \vartheta [^1_0](\nu|\tau)=
\sum_{n \in {\bf Z}} q^{{1\over 2}(n+1/2)^2} e^{i \pi (2n +1) \nu},\\
\vartheta_3(\nu|\tau)&= \vartheta [^0_0](\nu|\tau)=
\sum_{n \in {\bf Z}} q^{{1\over 2} n^2} e^{i \pi 2n  \nu},\\
\vartheta_4(\nu|\tau)&= \vartheta [^0_1](\nu|\tau) =
\sum_{n \in {\bf Z}} (-1)^n q^{{1\over 2}n^2} e^{i \pi 2n  \nu},
\end{aligned}
\ee
where $q=e^{2\pi i \tau}$. When $\nu=0$ we will simply denote $\vartheta_2(\tau)=
\vartheta_2(0|\tau)$ (notice that $\vartheta_1(0|\tau)=0$).
The theta functions $\vartheta_2(\tau)$, $\vartheta_3(\tau)$ and $\vartheta_4(\tau)$ have the
following product representation:
\be
\begin{aligned}
\vartheta_2(\tau)&  =2 q^{1/8}\prod_{n=1}^{\infty} (1-q^n)(1+q^n)^2,\\
\vartheta_3(\tau)&  = \prod_{n=1}^{\infty} (1-q^n)(1+q^{n-\half} )^2,\\
\vartheta_4(\tau)& = \prod_{n=1}^{\infty} (1-q^n)(1-q^{n-\half} )^2
\end{aligned}
\ee
and under modular transformations they behave as:
\be
\begin{aligned}
\vartheta_2 (-1/\tau)= &{\sqrt { \tau \over i}} \vartheta_4 (\tau),\\
\vartheta_3 (-1/\tau)= &{\sqrt { \tau \over i}} \vartheta_3 (\tau),\\
\vartheta_4 (-1/\tau)= &{\sqrt { \tau \over i}} \vartheta_2 (\tau),
\end{aligned}
\quad\quad
\begin{aligned}
\vartheta_2 (\tau +1)= &{\rm e}^{i\pi/4} \vartheta_2 (\tau),\\
\vartheta_3 (\tau +1)= &\vartheta_4 (\tau),\\
\vartheta_4 (\tau +1)= & \vartheta_3 (\tau).
\end{aligned}
\label{thetatransformation} 
\ee
The theta function $\vartheta_1(\nu|\tau)$ has the product representation
\be
\label{prodone}
\vartheta_1(\nu|\tau)=-2 q^{1\over 8} \sin (\pi \nu) \prod_{n=1}^{\infty} (1-q^n) (1-2 \cos (2 \pi \nu) q^n + q^{2n}).
\ee
We also have the following useful identities:
\be
\label{sumthet}
\vartheta_3^4 (\tau) = \vartheta_2^4 (\tau) + \vartheta_4^4 (\tau),
\ee
and
\be
\label{prodtheta}
\vartheta_2 (\tau)\vartheta_3 (\tau)\vartheta_4 (\tau) = 2\, \eta^{3}(\tau),
\ee
where
\be
\label{dede}
\eta(\tau)= q^{1/24} \prod_{n=1}^{\infty} (1- q^n)
\ee
is the Dedekind eta function. One has the following doubling formulae,
\be
\label{deta}
\begin{aligned}
\eta(2\tau)=&{\sqrt {\eta(\tau)\vartheta_2(\tau) \over 2}}, \qquad
\vartheta_2(2 \tau) ={\sqrt { {\vartheta_3^2(\tau) -\vartheta_4^2(\tau) \over 2}}}, \\
\vartheta_3(2 \tau) =&{\sqrt { {\vartheta_3^2(\tau) +\vartheta_4^2(\tau) \over 2}}},\qquad
\vartheta_4(2 \tau) ={\sqrt { \vartheta_3(\tau) \vartheta_4(\tau)}},\\
\eta(\tau/2)=&{\sqrt {\eta(\tau) \vartheta_4(\tau)}}.
\end{aligned}
\ee
The Eisenstein series are defined by
\be
\label{geneis}
E_{2n}(q)=1-{4n \over B_{2n}}\sum_{k=1}^{\infty}{k^{2n-1} q^{k}\over 1-q^{k}},
\ee
where $B_{m}$ are the Bernoulli numbers. The covariant version of $E_2$ is
\be
\widehat E_2(\tau, \bar \tau)=E_2(\tau) -{3\over \pi {\rm Im}\, \tau}=E_2(\tau)-{6\ri \over \pi (\tau-\bar \tau)}.
\ee

The formulae for the derivatives of the theta functions are also useful:
\be
\begin{aligned}
q {d \over dq}\log\, \vartheta_4=&{1\over 24}\biggl( E_2 -\vartheta_2^4 -\vartheta_3^4\biggr),\\
q {d \over dq}\log\, \vartheta_3=&{1\over 24}\biggl( E_2 +\vartheta_2^4 -\vartheta_3^4\biggr),\\
q {d \over dq}\log\, \vartheta_2=&{1\over 24}\biggl( E_2 +\vartheta_3^4 +\vartheta_4^4\biggr),
\end{aligned}
\ee
and from these one finds 
\be
q {d \over dq}\log\, \eta ={1\over 24}E_2(\tau)
\label{Eta-der}
\ee
and the Ramanujan identities 
\be\label{E-der}
\ba
q {d \over dq}\ E_2(q) =& {1\over 12}(E_2^2(q)-E_4(q)),\\
q {d \over dq}\ E_4(q) =& {1\over 3}(E_2(q)E_4(q)-E_6(q)),\\
q {d \over dq}\ E_6(q) =& {1\over 2}(E_2(q)E_6(q)-E_4^2(q)).
\ea
\ee
These can be used to compute the $q$-derivatives of the generators $K_2$, $K_4$ introduced in (\ref{gamma2generators}):
\be
\ba
q \partial_{q} K_2 &= {1\over 6} E_2(q) K_2(q) +{1\over 4} K_4(q) -{1\over 12} K_2^2(q),\\
q \partial_q K_4 &= {1\over 3} K_4(q) (E_2(q) + K_2(q)).
\ea
\ee
The doubling formulae for $E_2(\tau), E_4(\tau)$ are
\be
\label{deis}
\begin{aligned}
E_2(2 \tau)= &{1\over 2} E_2(\tau) + {1\over 4}(\vartheta_3^4(\tau) +\vartheta_4^4(\tau)),\\
E_4(2 \tau)= & {1\over 16} E_4(2\tau) + {15\over 16}\vartheta_3^4 (\tau)\vartheta_4^4 (\tau).
\end{aligned}
\ee
%


\section{The antiholomorphic dependence of the heterotic $F^{(g)}$}\label{heteroticFg}
In this Appendix we find the antiholomorphic dependence of $F^{(g)}(t,\bar t)$ in the heterotic theory. In section \ref{stu}, we show how the complicated 
result of the heterotic computation of the $F^{(g)}$ in the STU-model given in \cite{mm} can be simplified, along the lines of \cite{borcherds}. In section \ref{enr} we 
write down the result for $F^{(g)}_E$ in the Enriques Calabi-Yau and derive \eqref{Fgantihol}.

\subsection{A simple form for $F^{(g)}$ in the STU-model}\label{stu}

In \cite{mm}, an explicit expression for the holomorphic and antiholomorphic dependence of the topological amplitudes in the fiber limit of the STU-model was found. 
This expression is obtained from a one--loop computation in the dual heterotic theory, given by the integral \eqref{hetint}, which is then 
performed by using the technique of lattice reduction \cite{borcherds}. One finds that $F^{(g)}=F^{(g)}_{\rm deg}+F^{(g)}_{\rm ndeg}$, where \cite{mm}
\be\label{fgdeg}
F^{(g)}_{\rm deg}= 4 \pi^2 U_1 \delta_{g,1} +{2^{2g-1} \pi^{4g-3} \over
T_1^{2g-3}} \sum_{l=0}^{g} c_g (0,l) {l!\over \pi^{l+3} }
\biggl( {T_1 \over
U_1} \biggr)^l \zeta (2(2+l-g)),
\ee
\be\label{fgnondeg}
\ba
&F^{(g>1)}_{\rm ndeg} = 4\pi^{2g-2}(-1)^{g-1}\sum_{r\not=
0 }
 \sum_{l=0}^{g}
 \sum_{h=0}^{2g-2} \sum_{j=0}^{[g-1-h/2]}
\sum_{a=0}^s c_g ( r^2/2, l)
{(2\pi)^l(2g-2)! \over  j! h! (2g-h-2j-2)!}\\ &  \times  {(-1)^{j+h} \over  2^{j+a} }{(s+a)! \over  a! (s-a)!}({\rm
sgn} \left({\rm Re}(r\cdot y)\right)^h
{1\over (T_1U_1)^l} \left({\rm Re}(r \hat{\cdot} y)\right)^{l-j-a}
\Li _{3+a+j+l-2g}(\re^{-r \hat{\cdot} y} )
\\
& +{2\pi^{3g-3}c_g(0,g-1)\over (T_1 U_1)^{g-1}}
\sum_{s=0}^{g-1}  (-1)^s { (2g-2)!  \over s! (g-1-s)!} \psi ({1\over 2}+s)  \\
& + \sum_{\begin{subarray}{c}l=0\\l\neq g-1\end{subarray}}^{g} 4^{l+g}\pi^{2g+l-5/2}c_g(0,l){\zeta (3+2(l-g)) \over
(T_1U_1)^l} \\
& \times \sum_{s=0}^{g-1}(-1)^s 2^{2(s-2g)+5} { (2g-2)! \over (2s)!
(g-1-s)!}   \Gamma \Bigl({3\over 2}+  s+l -g\Bigr).
\ea
\ee 
We refer to $F^{(g)}_{\rm deg}$, $F^{(g)}_{\rm ndeg}$ as the degenerate and nondegenerate contributions, respectively. 
Also, $s:=|2g-2-h-j-l-1/2|-{1/2};\quad y=(T,U),\quad$ the complex norm is defined as $r^2=2r_1r_2$, and
\be
r\hat{\cdot}y\equiv |{\rm Re}(r\cdot y)|+{\rm i}{\rm Im}(r\cdot y)\nonumber.
\ee
The coefficients $c_g (m,l) $ can be obtained from the expansion
\be\label{expex}
{E_4 E_6 \over \eta^{24}} \widehat{\CP}_{g}= \sum_{m\in \mathbb{Q}}
\sum_{l\ge 0} c_g (m,l) q^m \tau_2^{-l},
\ee
where $\widehat{\CP}_{g}$ are defined by
\be
\biggl( { 2\pi  \eta^3 \lambda \over \vartheta_1(\lambda|\tau)}\biggr)^2\re^{-{\pi\lambda^2\over \tau_2}}=
\sum_{g=0}^{\infty} (2 \pi \lambda)^{2g} {\widehat{\cal P}}_{g}(\tau, \bar \tau).
\ee
Note that these $\widehat{\CP}_{g}(\tau,\bar{\tau})$ are the modular, almost holomorphic extensions of the $\CP_{g}(\tau)$ defined in \eqref{defpg}, that is,  $\widehat{\CP}_{g}$ is obtained from $\CP_{g}$ by replacing $E_2\rightarrow \widehat{E}_2$.
The only antiholomorphic dependence in $\widehat{\CP}_{g}$ thus lies in the $\widehat{E}_2(\tau,\bar{\tau})$. Using the explicit expressions for $\widehat{\CP}_{g}$ given in \cite{km}, one can show that independently of the specific model,
\be\label{rec}
c_g(m,l)={(-1)^l\over l!(4\pi)^l}c_{g-l}(m),
\ee
where $c_g(m)$ are defined analogously to \eqref{geomrmod}, that is
\be
\sum_n c_g(n) q^n ={\cal P}_{g}(q){E_4E_6\over \eta^{24}}.
\ee
In what follows, we will systematically express everything in terms of the coefficients $c_g(m)$.

It turns out that \eqref{fgnondeg} can be dramatically simplified. We will need the identity:
\be
\label{blemma}
\sum_j(-1)^j\binom{C}{j}\binom{A-2j+C-B-1}{A-2j}=\sum_j(-1)^j\binom{C}{A-j}\binom{B}{j}
\ee
This is valid for any $A,B,C \in\mathbb{Z}$, see \cite{borcherds} for a proof. A special case of the above formula is the following. 
Let $C$, $l$ and $m^+-h^+$ be integers such that $0\leq l<C<m^+-h^+-l$. Then,
\be
\label{cor2}
\sum_{\begin{subarray}{c}j\end{subarray}}{(-1)^j(m^+-h^++C-2j-1-l)!\over j!(m^+-h^+-2j)!(C-j)!}=0.\nonumber
\ee
The proof of this statement is easy. Set $B=l, A=m^+-h^+$ in (\ref{blemma}) to obtain
\be
\sum_{\begin{subarray}{c}j\end{subarray}}(-1)^j{(m^+-h^++C-2j-1-l)!C!\over j!(m^+-h^+-2j)!(C-j)!(C-1-l)!}=\sum_j\binom{C}{m^+-h^+-j}\binom{l}{j}.
\ee
Since $C > l\geq 0$, any non-vanishing term on the right-hand side must fulfill $m^+-h^+-C\leq j\leq l$, in contradiction with the assumption $C<m^+-h^+-l$. 

We also have the following three additional nontrivial identities. First of all, let $s:=|2g-2-h-j-l-1/2|-{1/2}$. Then, 
\ben
&&\sum_{h=0}^{2g-2}\sum_{j=0}^{C}{(2g-2)!\over 2^{2g-2}}{(s+C-j)!(-1)^{C-j}\over l!h!j!(2g-2-h-2j)!(s-C+j)!(C-j)!}\\
&&\hspace{4cm}=\left\{\begin{array}{ll}\binom{2g-3-l}{C}{1\over (l-C)!} & C\leq{\rm min}(l,2g-3-l)\\
0&{\rm otherwise.}\end{array}\right.
\label{cor3}
\een
This is valid for any pair of positive integers $g,l$. The second identity reads, 
\be
\label{cor4}
\sum_{s=0}^{g-1}(-1)^s{(2g-2)!\over s!(g-1-s)!}\psi\Bigl(s+{1\over 2}\Bigr)=-2^{(2g-2)}(g-2)!.
\ee
The final identity we will need is
\be
\label{cor5}
\sum_{s=0}^{g-1}(-1)^{l+s}2^{2(s-2g)+5}{(2g-2)!\over (2s)!(g-1-s)!}\Gamma\Bigl({3\over 2}+s+l-g \Bigr)={(-1)^{g-1}(2g-3-l)!\over (2g-3-2l)!}\sqrt{\pi}4^{-l}.
\ee
which is valid for any $l\in \mathbb{N}, l<g-1$. Making use of (\ref{cor2}) and (\ref{cor3}), we can convert the sums over $h,j,a$ in \eqref{fgnondeg} into a single one over $C=j+a=\{0,\cdots,l\}$. Then, 
(\ref{cor4}) and (\ref{cor5}) can be used to simplify the second respectively third term in \eqref{fgnondeg}. The sum over $r$ can be restricted for all $g\geq 3$ to a sum over $r$ for which ${\rm Re}(r\cdot y)<0$, or equivalently to a sum over positive $r$ and a finite number of boundary cases. At genus 2, however, there is a contribution from ${\rm Re}(r\cdot y)>0$, it reads \cite{mm}
\be\label{negr}
{c_0(r^2/2)\over 16 T_1U_1}\Li_3(\re^{-r\cdot y}).
\ee
We can then write down a simplified expression for the nondegenerate part of $F^{(g)}$ in the STU model:
\be\label{Fgndfin1}
\ba
&F^{(g>1)}_{\rm nd,STU}\\
&=
\sum_{l=0}^{g-1}\sum_{C=0}^{{\rm min}(l,2g-3-l)}\sum_{r>0}{\binom{2g-l-3}{C}\over(l-C)! 2^C}{(-{\rm Re}(r\cdot y))^{l-C}\over (2T_1U_1)^l}c_{g-l}({r^2\over 2})\Li_{3-2g+l+C}(\re^{-r\cdot y})\\
&+{22\over 2^g(g-1)}{1\over (2T_1U_1)^{g-1}}+\sum_{l=0}^{g-2}{c_{g-l}(0)\over l!(4T_1U_1)^l}\zeta(3+2(l-g)){(2g-3-l)!\over (2g-3-2l)!},
\ea
\ee
where we also have used the fact that in the STU model, \mbox{$c_1(0)=-22$}, and we have removed an overall prefactor of $4(2\pi \ri)^{2g-2}$ to agree with the normalization of the 
topological string amplitudes. 
 
\subsection{Application to the Enriques Calabi-Yau}\label{enr}
The above expressions have to be adapted slightly for the Enriques Calabi-Yau. We only consider here the geometric reduction suited to the large radius limit. As shown in \cite{km}, 
the polylogarithm is replaced by $\Li_m(x)\rightarrow 2^m\Li_m(x^{1\over 2})-\Li_m(x)$, and the norm of the reduced lattice is doubled. We also replace the quantity $2T_1U_1$ appearing in the STU-model by $Y=\re^{-K}$ as in \eqref{def-KY}, and the coefficients $c_g(m)$ are now defined by \eqref{geomrmod}. There is a new important simplification: $c_{0}(r^2)$ and $c_{g>1}(0)$ vanish, 
and thus there is no contribution from negative $r$ at any genus $g>1$, since \eqref{negr} becomes
\be
{c_0(r^2)\over 8 Y}\left(8\Li_3(\re^{-r\cdot y})-\Li_3(\re^{-2r\cdot y})\right)=0.
\ee
Furthermore, the degenerate contribution \eqref{fgdeg} and the last term in \eqref{fgnondeg} vanish for all $g>1$,  while $c_1(0)=4$, and the full $F^{(g)}_E(t,\bar{t})$ for the Enriques reads
\be
\ba
&\hspace{-1cm}F^{(g>1)}_E(t,\bar{t})=\sum_{l=0}^{g-1}\sum_{C=0}^{{\rm min}(l,2g-3-l)}\sum_{r>0}{\binom{2g-l-3}{C}\over(l-C)! 2^C}{(-2{\rm Re}(r\cdot t))^{l-C}\over Y^l}c_{g-l}(r^2)\\
&\cdot\left(2^{3-2g+l+C}\Li_{3-2g+l+C}(\re^{-r\cdot t})-\Li_{3-2g+l+C}(\re^{-2r\cdot t})\right)-{1\over 2^{g-2}(g-1)}{1\over Y^{g-1}}.
\ea
\ee
Using
\be
{\rm Re}(t^a)\partial_{t^a}\Li_n(\re^{-r\cdot t})=-{\rm Re}(r\cdot t)\Li_{n-1}(\re^{-r\cdot t}),
\ee
this can be cast into the following recursive form: 
\be
\ba
&F^{(g)}_E(t,\bar{t})\\
=&\sum_{l=0}^{g-1}\sum_{C=0}^{\begin{subarray}{c}{\rm min}\\(l,2g-3-l)\end{subarray}}{(2g-3-l)!\over (2g-3-l-C)!(l-C)!C!2^l}{(t^{a_1}+\bar{t}^{a_1})\cdots(t^{a_{l-C}}+\bar{t}^{a_{l-C}})\partial_{a_1}\cdots\partial_{a_{l-C}}\CF^{(g-l)}(t)\over Y^l}\\
&\hspace{2cm}-{1\over 2^{g-2}(g-1)Y^{g-1}}.\\
\ea
\ee
Notice that this exhibits the structure of the antiholomorphic amplitudes written down in \cite{abk}. 

\section{Anomaly equations for $F^{(g)}$ on the big moduli space \label{Cal_big_Fg}}

\subsection{Anomaly equation for $F^{(g)}\ (g>1)$ \label{Cal_big_Fg1}}
Here we provide some details on the calculation of the recursive anomaly equations on the 
big moduli space. 
We like to rewrite the equation \eqref{rec_Fg} in terms of the variables $Y^K = \lambda^{-1} X^K(t)$ and
$\bar Y^K$. First note that 
\beq
   \frac{\partial}{\partial t^i} - K_i \lambda \frac{\partial}{\partial \lambda} = \lambda^{-1} \chi^I_i \frac{\partial}{\partial Y^K}\ ,
\eeq
where $\chi_i^I$ is defined in \eqref{exp_Omega}.
This implies that the first derivative of $F^{(g)}$ can be written as
\beq \label{D_iF}
  D_i F^{(g)} = \lambda^{-2g+1} \chi^I_i \, \partial_{ Y^I} F^{(g)}(Y)\ .
\eeq
where we have used the fact that $\lambda \partial_\lambda F^{(g)}(Y) = (2g-2) F^{(g)}(Y)$ 
due to \eqref{Fg_Y}. Moreover, one derives that the second derivative reads
\bea
   D_i D_j F^{(g-1)} &=&   \lambda^{-2g+3} (D_i \chi^I_j) \, \partial_{ Y^I} F^{(g-1)}(Y) +  \lambda^{-2g+3}  \chi^I_j D_i \, \partial_{Y^I} F^{(g-1)}(Y) \nn \\
  &=& i \lambda^{-2g+3} C_{ijk} G^{k\bar k} \bar \chi^I_{\bar k}\, \partial_{Y^I} F^{(g-1)}(Y) + \lambda^{-2g+2}  \chi^I_i \chi^J_j\,  \partial_{Y^I} \partial_{Y^J} F^{(g-1)}(Y) \nn \\
  &=& \lambda^{-2g+2}  \chi^I_i \chi^J_j  \Big[ \tfrac{i}2 C_{IJ}^{(Y)\, K}  \partial_{Y^K} F^{(g-1)}(Y)  +   \partial_{Y^I} \partial_{Y^J}  F^{(g-1)}(Y) \Big]\ .
\eea
In order to evaluate the second identity we have used the special geometry relation \eqref{special_geom} and \eqref{D_iF} 
while for the third identity we have used
\eqref{G-tau}.
Also notice that from \eqref{D_iF} one infers that
\beq
  \sum_{r=1}^{g-1} D_i F^{(r)} D_j F^{(g-r)} =  \lambda^{-2g+2} \chi^I_i \chi^J_j  \sum_{r=1}^{g-1} \partial_{Y^I} F^{(r)}(Y)\, \partial_{Y^J} F^{(g-r)}(Y)
\eeq
Finally, we need the identity 
\beq
  \tfrac{i}2 e^{2K} \bar C_{\bar i \bar j \bar k} G^{\bar j j} G^{\bar k k} \chi^I_j \chi^J_k =  \tfrac{i}8 \lambda^{-1}\bar \chi^K_{\bar i} \bar C_{\ K}^{(Y)\, IJ}\ .
\eeq
Hence, we conclude that 
\beq
  \partial_{\bar Y^I} F^{(g)} =  \tfrac{i}8 \bar C_{K}^{\ IJ}  \Big[ \partial_{Y^I} \partial_{Y^J}  F^{(g-1)} + \sum_{r=1}^{g-1} \partial_{Y^I} F^{(r)}\, \partial_{Y^J} F^{(g-r)} \Big]
                                                       -  \tfrac{1}{16} \bar C_{K}^{\ IJ}  C_{IJ}^{\ \ K}  \partial_{Y^K} F^{(g-1)}\ ,
\eeq
where $C_{IJK}$ and $F^{(r)}$ are functions of $Y^K,\bar Y^K$. This equation is precisely the recursive anomaly equation 
given in \eqref{big_Fg}.

Let us also present the derivation of the simplest solution to \eqref{big_Fg}. In other words, we 
calculate $F^{(2)}$ by using  the integration by parts method of \cite{bcov}.
To do that we use the definition \eqref{d_prop} of the propagator to
replace $\bar C_K^{\ IJ}$ in \eqref{F_2big}. We pull the derivative $\partial_{\bar I}$ in front 
of all the terms and evaluate
\bea
  &&\partial_{\bar I} \Big[ F^{(2)} + \tfrac{1}{2} \Delta^{JK} \big( 
   D_J \partial_{K} F^{(1)} + \partial_{J} F^{(1)} \partial_{K} F^{(1)} \big)\Big]  
  =-\big(\tfrac{\chi}{24} -1 \big) \partial_{\bar I} \big[\Delta^{JK} K_{J}\big] \partial_K F^{(1)} \nn \\
  &&-
  \tfrac{1}{8} \partial_{\bar I} \big[\Delta^{JK} \Delta^{LM}\big] \big(C^{(0)}_{KLMJ} + 4 C^{(0)}_{JLM} \partial_{K} F^{(1)} \big)
  -\tfrac12 \big(\tfrac{\chi}{24} -1 \big) \partial_{\bar I} \big[\Delta^{JK} K_{J} K_K\big]
      \ ,
\eea
where $C^{(0)}_{IJK}=iC_{IJK}$ as defined in \eqref{def-C^g}.
In performing the derivative we used the equation \eqref{F_2big} to eliminate the 
terms arising when $\partial_{\bar I}$ hits the propagator. Furthermore 
we commuted $\partial_{\bar I}$ with the covariant derivative $D_J$ by using the identity 
\beq
    \big[ \partial_{\bar I},D_J\big] V_K= \tfrac{1}{4} C_{JK}^{\ \ \, P} \bar C_{IP}^{\ \ \, M} V_M\ .
\eeq
One can then eliminate the second derivative $\partial_{\bar I} \partial_K F^{(1)}$ by inserting the 
equation \eqref{big_F1} and applying the useful identities
\beq
   D_I \widehat K_{I \bar J}=D_I  \widehat K_J = 0 \ , \qquad \qquad \Delta^{IJ} D_I K_{J \bar K} = 2\Delta^{IJ} K_{I\bar K} K_J\ ,
   \qquad K_J \partial_{\bar I}  \Delta^{JK} =0\ .  
\eeq
In the next step we once again pull the derivative $\partial_{\bar I}$ in front of all terms
and evaluate 
\bea
  &&\partial_{\bar I} \Big[ F^{(2)} + \tfrac{1}{2} \Delta^{JK} \big( 
   D_J \partial_{K} F^{(1)} + \partial_{J} F^{(1)} \partial_{K} F^{(1)} \big)
   +\tfrac12 \big(\tfrac{\chi}{24} -1 \big) \Delta^{JK} K_{J} K_K \\
  && +\tfrac{1}{8} \Delta^{JK}\Delta^{LM} \big(C^{(0)}_{KLMJ} + 4 C^{(0)}_{JLM} \partial_{K} F^{(1)} \big)
     + \big(\tfrac{\chi}{24} -1 \big) \Delta^{JK} K_{J} \partial_K F^{(1)}\Big] \nn \\
  &&=  -\partial_{\bar I} \Big[ \tfrac{1}2 \big(\tfrac{\chi}{24} -1 \big)  C^{(0)}_{JLM} \Delta^{JK} \Delta^{LM} K_J 
         + \tfrac{1}2 \big(\tfrac{\chi}{24} -1 \big)^2 \Delta^{JK} K_J K_K \nn\\
         &&\ \ + \Delta^{JK} \Delta^{LM} \Delta^{QP}\big(\tfrac1{12} C^{(0)}_{KMQ} C^{(0)}_{PLJ} +\tfrac{1}{8}  C^{(0)}_{KJQ} C^{(0)}_{PML} \big) \Big]\ . \nn
\eea
We are now in the position to read off $F^{(2)}(Y,\bar Y)$ up to a holomorphic ambiguity $f^{(2)}(Y)$.
The corresponding solution can be found in \eqref{F_2_big}.

\subsection{Anomaly equation for $F^{(1)}$ on big phase space \label{Cal_big_F1}}

In this appendix we discuss the lift of the holomorphic anomaly equation \ref{anomaly_F1} for $F^{(1)}$ 
to the big moduli space $\widehat \cM$. To begin with let us first note that 
\beq
   |\lambda|^{-2} \chi^I_i \bar \chi^J_{\bar j} \partial_{Y^I} \partial_{\bar Y^J} K= G_{i\bar \jmath}\ , \qquad  
   Y^I \partial_{Y^I} \partial_{\bar Y^J} K= 0\ ,\qquad 
   \bar Y^J \partial_{Y^I} \partial_{\bar Y^J} K= 0\ .
\eeq
where $K$ is the K\"ahler potential \eqref{KpotII} and $G_{i\bar \jmath}$ is the Weil-Petersson metric.
We also evaluate the first derivative $K_I$ of $K$ and show that it satisfies
\beq \label{K_Iproj}
   K_I \chi^I_i = 0\ , \qquad \qquad K_I Y^I = -1 \ .
\eeq
With these identities at hand we now lift the holomorphic anomaly equation \eqref{anomaly_F1}.
Using the homogeneity condition \eqref{F1_hom} one derives 
\beq \label{F1rewrite1}
   \partial_{i} \partial_{\bar j} F^{(1)} = |\lambda|^{-2} \chi^I_i \bar \chi^J_{\bar j} \partial_{Y^I} \partial_{\bar Y^J} F^{(1)}(Y)  
\eeq
Moreover, one shows that 
\beq
  \tfrac{1}{2} e^{2K} G^{k\bar k} G^{l \bar l} C_{i k l } \bar C_{\bar j \bar k \bar l} = |\lambda|^{-2} \chi^I_i \bar \chi^J_{\bar j} \tfrac18 C_{ILM} \bar C_{J}^{\ LM}\ ,
\eeq
as well as
\beq \label{F1rewrite3}
  \Big(\frac{\chi}{24} -1 \Big)G_{i\bar j} = |\lambda|^{-2}\chi^I_i \bar \chi^J_{\bar j} \Big(\frac{\chi}{12} -2 \Big) e^{K(Y,\bar Y)} \I \tau_{IJ}\ .
\eeq
Inserting \eqref{F1rewrite1}-\eqref{F1rewrite3} into the anomaly equation \eqref{anomaly_F1}
we verify its big moduli space counterpart \eqref{big_F1}. It is straightforward to 
integrate \eqref{big_F1} to find the local solution \eqref{F_1solution} for $F^{(1)}$ by 
applying the identity
\beq
  R_{IJ} =  \partial_{Y^I}  \partial_{\bar Y^J} \log \det \I \tau = - \tfrac14 C_{I KL} \bar C_J^{\ KL}\ .
\eeq
It is however instructive to also recall a second alternative approach which integrates 
\eqref{anomaly_F1} rather then  \eqref{big_F1}.

Let us end this appendix by recalling the direct integration of \eqref{anomaly_F1}.
First note that the Riemann tensor on a special K\"ahler manifold is given by 
\beq
   R_{i\bar j l \bar m} = G_{i \bar j} G_{l \bar m} + G_{i \bar m} G_{l \bar j} - e^{2K} C_{ilp} \bar C_{\bar j \bar m \bar p} G^{p\bar p}\ .   
\eeq
The Ricci tensor takes the form
\beq
  R_{i\bar j} = \partial_i \partial_{\bar j} \log \det G = G_{i \bar j} (h^{2,1}+1) - e^{2K} C_{ilp} \bar C_{\bar j \bar m \bar p} G^{l\bar m} G^{p\bar p}\ .\ , 
\eeq
such that 
\beq
   \tfrac{1}{2} e^{2K} G^{k\bar k} G^{l \bar l} C_{i k l } \bar C_{\bar j \bar k \bar l} = G_{i \bar j} (h^{2,1}+1) -  
   \partial_i \partial_{\bar j} \log \det G\ . 
\eeq
Using this equation we solve \eqref{anomaly_F1} as \footnote{The derivative of the determinate of the matrix $A$ is given by
 $\partial_x \det A = \det A \cdot A^{-1\, IJ} \partial_x A_{IJ}$ }
\beq
  F^{(1)} = -\tfrac{1}{2} \log \det G + \big(\tfrac{1}{2} (h^{2,1} + 1) -\frac{\chi}{24} + 1 \big)K + h(t) + \bar h(\bar t)
\eeq
where $h(t)$ is a holomorphic function arising as integration constant.
Now note that it follows from \eqref{G-tau} that \cite{N=2rev} 
\beq
  \det(2\I\tau_{IJ}) = - \det( G_{i\bar j}) e^{-(h^{(2,1)}+1)K} |\det(\chi^I_i, X^I)|^{-2}\ .
\eeq
This equation can be used to rewrite $F^{(1)}$ as
\beq \label{F1_somlutionapp}
  F^{(1)} = - \tfrac{1}{2}\log \det(2\I\tau_{IJ})+ \Big(1-\frac{\chi}{24}\Big)K  +\tfrac{1}{2} \log\Big( -|\det(\chi^I_i, X^I)|^{-2}\Big)  + h(t) + \bar h(\bar t)
\eeq
One can evaluate the determinate of the coordinate change and shows \cite{N=2rev}
\beq
 |\det(\chi^I_i, X^I)|^{-2} =|X^0|^{-2(h^{2,1} + 1)}|\det e^i_j|^{-2}\ .
\eeq
where $e^i_j = \partial_{t^i} (X^i / X^0)$. But since $X^0$ and $e^i_j$ are holomorphic in the coordinates $t^i$
they can be absorbed into $h$ such that \eqref{F1_somlutionapp} becomes \eqref{F_1solution}.


\end{document}